%% file: RandomPowerOptimization.tex
\documentclass[journal,twoside,nofonttune]{IEEEtran}
\usepackage{amsmath}
\usepackage{amssymb}
\usepackage{amsfonts}
\usepackage{amsthm}

\usepackage{pdfcomment}
\usepackage{soul}
\setulcolor{Blue}
\setstcolor{Red}
\sethlcolor{LightBlue}

\usepackage[T1]{fontenc}
\usepackage{txfonts}

\usepackage[svgnames]{xcolor}
\usepackage{subfigure}
\usepackage{tikz}
\usepackage{acronym}
\usepackage{balance}
\usepackage{latexsym}
\usepackage{paralist}
\usepackage{xspace}

\usepackage[numbers,sort&compress]{natbib}

\usepackage{hyperref}
\hypersetup{
colorlinks=true,
linktocpage=true,
pdfstartview=FitH,
breaklinks=true,
pdfpagemode=UseNone,
pageanchor=true,
pdfpagemode=UseOutlines,
plainpages=false,
bookmarksopen=false,
hypertexnames=true,
pdfhighlight=/O,
urlcolor=Maroon, linkcolor=SteelBlue, citecolor=SeaGreen, % <--- for screen
pdfcreator={pdfLaTeX},
pdfproducer={LaTeX with hyperref}
}

\newcommand{\R}{\varmathbb{R}}
\newcommand{\Q}{\varmathbb{Q}}
\newcommand{\Z}{\varmathbb{Z}}
\newcommand{\N}{\varmathbb{N}}

\DeclareMathOperator{\bigoh}{\mathcal O}

\DeclareMathOperator{\diag}{diag}

\newcommand{\dom}{D}

\DeclareMathOperator{\ex}{\varmathbb{E}}

\DeclareMathOperator{\prob}{\varmathbb{P}}
\DeclareMathOperator{\rank}{rank}

\DeclareMathOperator{\supp}{supp}
\DeclareMathOperator{\tr}{tr}
\DeclareMathOperator{\var}{Var}

\newcommand{\dd}{\:d}

\newcommand{\eps}{\varepsilon}

\newcommand{\mgeq}{\succcurlyeq}

\newcommand{\pd}{\partial}

\newcommand{\wilde}{\widetilde}

\newcommand{\txs}{\textstyle}

\newcommand{\insum}{\sum\nolimits}
\newcommand{\inprod}{\prod\nolimits}

\theoremstyle{plain}
\newtheorem{theorem}{Theorem}
\newtheorem{corollary}{Corollary}
\newtheorem*{corollary*}{Corollary}
\newtheorem{lemma}{Lemma}
\newtheorem{proposition}{Proposition}

\theoremstyle{definition}

\newtheorem*{definition*}{Definition}

\theoremstyle{remark}
\newtheorem{remark}{Remark}
\newtheorem*{remark*}{Remark}

\newcommand{\bA}{\mathbf{A}}
\newcommand{\bB}{\mathbf{B}}
\newcommand{\bC}{\mathbf{C}}
\newcommand{\bD}{\mathbf{D}}
\newcommand{\bE}{\mathbf{E}}

\newcommand{\bG}{\mathbf{G}}
\newcommand{\bH}{\mathbf{H}}
\newcommand{\bI}{\mathbf{I}}

\newcommand{\bM}{\mathbf{M}}

\newcommand{\bT}{\mathbf{T}}

\newcommand{\bV}{\mathbf{V}}

\newcommand{\be}{\mathbf{e}}

\newcommand{\bh}{\mathbf{h}}
\newcommand{\bk}{\mathbf{k}}
\newcommand{\bm}{\mathbf{m}}

\newcommand{\bp}{\mathbf{p}}
\newcommand{\bq}{\mathbf{q}}

\newcommand{\bu}{\mathbf{u}}
\newcommand{\bv}{\mathbf{v}}

\newcommand{\bx}{\mathbf{x}}
\newcommand{\by}{\mathbf{y}}
\newcommand{\bz}{\mathbf{z}}

\newcommand{\bxi}{\boldsymbol{\xi}}

\newcommand{\bPsi}{\mathbf{\Psi}}

\def\b0{{\mathbf 0}}

\DeclareMathOperator{\arcosh}{arcosh}
\DeclareMathOperator{\one}{\varmathbb{I}}
\DeclareMathOperator{\spec}{spec}
\DeclareMathOperator{\sinr}{SINR}

\newcommand{\avg}{\mathrm{avg}}
\newcommand{\eff}{\mathrm{eff}}
\newcommand{\eigmin}{\eps_{0}}
\newcommand{\epr}{\mathtt{e}}
\newcommand{\ids}{\mathcal{N}}
\newcommand{\lattice}{\Lambda}
\newcommand{\latsize}{L}
\newcommand{\ball}{B}

\newcommand{\pdist}{\mathcal{P}}
\newcommand{\scale}{a}
\DeclareMathOperator{\K}{\mathsf{K}}
\DeclareMathOperator{\gen}{\mathcal{L}}

\defineavatar{AM}{author=Aris,color=red,markup=Higlight,icon=Star,voffset=5pt}

\newfont{\bb}{msbm10 scaled 1100}

\defineavatar{PM}{author=Panayotis,color=ForestGreen,markup=Highlight,icon=Help,hoffset=-1em,voffset=5pt}

\newcommand{\noisevar}{\sigma^{2}}
\newcommand{\noisevard}{\sigma^{4}}

\begin{document}
\bstctlcite{BSTcontrol}

%*************************************************************
%*****    FRONT MATTER
%*************************************************************

%----------------------------------------------------------------------
%%% TITLE & AUTHORS
%----------------------------------------------------------------------
\title{Power Optimization in  Random Wireless Networks}

\author{%
Aris~L.~Moustakas,
%\IEEEmembership{Member,~IEEE},
\and
Panayotis Mertikopoulos
%\IEEEmembership{Member,~IEEE},
and
Nicholas~Bambos
%\IEEEmembership{???,~IEEE}
\thanks{%
A.~L.~Moustakas is with the Physics Department of the University of Athens, Greece, and the \'Ecole Sup\'erieure d'\'Electricit\'e (Sup\'elec), Gif-sur-Yvette, France;
P.~Mertikopoulos is with the French National Center for Scientific Research (CNRS) and the Laboratoire d'Informatique de Grenoble, Grenoble, France;
N.~Bambos is with the Department of Electrical Engineering of Stanford University, Palo Alto, CA 94305. This research has been co-financed by the European Union (European Social Fund - ESF) and Greek national funds through the Operational Program "Education and Lifelong Learning" of the National Strategic Reference Framework (NSRF) Research Funding Program: ``THALES: Investing in knowledge society through the European Social Fund.''. This research was presented in part at SpaSWIN 2012, IEEE ISIT 2012 and IEEE ISIT 2013.}
}

%\author{\IEEEauthorblockN{Aris L. Moustakas$^1$ and Nicholas Bambos $^2$\thanks{$^1$ ALM is with the Physics Dept., Univ. of Athens, 157 84  Athens, Greece; email: arislm@phys.uoa.gr}
%\and
%\IEEEauthorblockN{Nicholas Bambos}\\
%\IEEEauthorblockA{Stanford University, Palo Alto, CA 94305\\
%Email: bambos@stanford.edu}}
%\author{Aris L. Moustakas$^{1,2}$,  Panayotis Mertikopoulos$^{3}$ and Nicholas Bambos$^{4}$
%\thanks{{1}:Physics Dept., Univ. of Athens, 157 84  Athens, Greece; {2}: Supelec, Gif-sur-Yvette, CEDEX France; ; {3}: CNRS, Grenoble, France; {4} Stanford University, Palo Alto, CA 94305. This work was funded by the research program `CROWN', through the Operational Program `Education and Lifelong Learning 2007-2013' of NSRF, which has been co-financed by EU and Greek national funds. }}
%\date{\today}

%\pubid{0000--0000/03\$17.00~\copyright~2003 IEEE}

\maketitle

%----------------------------------------------------------------------
%%% ABSTRACT
%----------------------------------------------------------------------
\begin{abstract}
Consider a wireless network of transmitter-receiver pairs where the transmitters adjust their powers to maintain a target SINR level in the presence of interference.
In this paper, we analyze the optimal power vector that achieves this target in large, random networks obtained by ``erasing'' a finite fraction of nodes from a regular lattice of transmitter-receiver pairs.
We show that this problem is equivalent to the so-called Anderson model of electron motion in dirty metals which has been used extensively in the analysis of diffusion in random environments.
A standard approximation to this model
so-called coherent potential approximation (CPA) method which we apply to evaluate the first and second order intra-sample statistics of the optimal power vector in one- and two-dimensional systems.
This approach is equivalent to traditional techniques from random matrix theory and free probability, but while generally accurate (and in agreement with numerical simulations), it fails to fully describe the system:
in particular, results obtained in this way fail to predict when power control becomes infeasible.
In this regard, we find that the infinite system is always unstable beyond a certain value of the target SINR, but any finite system only has a small probability of becoming unstable.
This instability probability is proportional to the tails of the eigenvalue distribution of the system which are calculated to exponential accuracy using methodologies developed within the Anderson model and
its ties with random walks in random media.
Finally, using these techniques, we also calculate the tails of the system's power distribution under power control and the rate of convergence of the Foschini\textendash Miljanic power control algorithm in the presence of random erasures. Overall, in the paper we try to strike a balance between intuitive arguments and formal proofs.
\end{abstract}

%----------------------------------------------------------------------
%%% KEYWORDS, ACRONYMS ETC.
%----------------------------------------------------------------------
%\begin{IEEEkeywords}
%These;
%are;
%the;
%keywords.
%\end{IEEEkeywords}

\newacro{CDF}{cumulative distribution function}
\newacro{CDMA}{code division multiple access}
\newacro{CPA}{coherent potential approximation}
\newacro{FPA}{free probability assumption}
\newacro{IDS}{integrated density of states}
\newacro{RMT}{random matrix theory}
\newacro{SINR}{signal-to-interference-and-noise ratio}
\newacro{iid}[i.i.d.]{independent and identically distributed}
\acused{iid}

%*************************************************************
%*****    BODY TEXT
%*************************************************************

%----------------------------------------------------------------------
%%% INTRODUCTION
%----------------------------------------------------------------------
\section{Introduction}
\label{sec:introduction}
\input{Introduction}

%----------------------------------------------------------------------
%%% MODEL
%----------------------------------------------------------------------
\section{Model description}
\label{sec:model}
\input{Model}

%----------------------------------------------------------------------
%%% WYNER MODEL
%----------------------------------------------------------------------
\section{The Wyner model: Exact results}
\label{sec:WynerModel}
\input{WynerModel}

%----------------------------------------------------------------------
%%% CPA
%----------------------------------------------------------------------
\section{Average power via the coherent potential approximation}
\label{sec:CPA}
\input{CPA}
\section{Stability analysis}
\label{sec:stability_analysis}
\input{Outage}

%----------------------------------------------------------------------
%%% POWER DISTRIBUTION
%----------------------------------------------------------------------
\section{Tails of the power distribution}
\label{sec:PowerDistribution}
\input{PowerDistribution}

%----------------------------------------------------------------------
%%% DYNAMICS
%----------------------------------------------------------------------
\section{Long-term behavior of the power control dynamics}
\label{sec:Dynamics}
\input{Dynamics}

%%----------------------------------------------------------------------
%%%% APPLICATIONS
%%----------------------------------------------------------------------
%\section{Applications}
%\label{sec:Applications}
%\input{Applications}

%----------------------------------------------------------------------
%%% CONCLUSIONS
%----------------------------------------------------------------------
\section{Conclusions}
\label{sec:conclusions}
\input{Conclusions}

\appendices

%%----------------------------------------------------------------------
%%%% APPENDIX: MODELS
%%----------------------------------------------------------------------
%\section{Random Erasures and the Anderson Model}
%\label{app:Anderson}
%\input{App-Anderson}

%----------------------------------------------------------------------
%%% APPENDIX: CPA
%----------------------------------------------------------------------
\section{Derivation of the CPA equations}
\label{app:cpa}
\input{App-CPA}

%----------------------------------------------------------------------
%%% APPENDIX: OUTAGE
%----------------------------------------------------------------------
\section{Derivation of the integrated density of states}
\label{app:lifshitz_proofs}
\input{App-Outage}

%----------------------------------------------------------------------
%%% APPENDIX: CONTINUOUS APPROXIMATION
%----------------------------------------------------------------------
\section{Continuous Approximation of $\bH_0$}
\label{app:cont_approx}
\input{App-ContApprox}

%----------------------------------------------------------------------
%%% APPENDIX: TAILS
%----------------------------------------------------------------------
\section{Details for the bounds of the power distribution}
\label{app:power_tails}
\input{App-Tails}
\footnotesize
%\balance
\bibliographystyle{IEEEtran}
%%%\bibliography{IEEEabrv,../bibliography/wireless}
\bibliography{IEEEabrv,C:/Users/ARISLM/ALMDocuments/Dropbox/Work/CurrentWork/bibliography/wireless}

\end{document}

%% file: Introduction.tex
%----------------------------------------------------------------------
%%% INTRODUCTION
%----------------------------------------------------------------------
% !TEX root = ./RandomPowerOptimization.tex

\IEEEPARstart{T}{he} importance of transmitted power has made power control an essential component of network design ever since the early development stages of legacy wireless networks. Power control allows wireless links to achieve their required throughputs,
minimizing the power used in the process and, hence, the interference induced
on other links. This increases the spatial spectrum reuse, as a result, the network
capacity, and prolongs the battery life of mobile users. For example, the introduction of efficient power control algorithms (both closed- and open-loop), was one of the main improvements that were brought about in third generation CDMA-based cellular networks.
Likewise, substantial effort has been made to optimize the performance of future and emerging network paradigms (such as ad hoc networks) by analyzing connectivity and transport capacity under power control
\cite{Franceschetti_book_RandomNetworks,
Baccelli2009_StocGeometryWirelessNetsTheory,
Baccelli2009_StocGeometryWirelessNetsApplications,
Krunz2004_TransmissionPowerControlWAHN}.
As a result, several algorithms have been developed that provably allow receivers to meet  \ac{SINR} requirements of the form $\sinr\geq\gamma$ (where the threshold value $\gamma$ is determined by the requested rate $r = \log_{2}(1+\gamma)$ of each link) while minimizing
%the total power or the power per user
power
subject to feasibility constraints \cite{Foschini1991_FMAlgorithm_PowerControl, Bambos2000_PowerControl_ActiveLinkProtection}.
However, while the benefits of such algorithms are easy to evaluate in small networks or networks with simple geometries (e.g. with transmitters and receivers located on a grid), their behavior in large-scale random networks has not been quantified analytically.

The conditions for the feasibility of power control have been discussed extensively under general assumptions \cite{Baccelli2006_OptimalPowerThruputRouting, Baccelli2009_StocGeometryWirelessNetsApplications} but without characterizing the properties of the optimal power vector in a quantitative way. In contrast, using the Laplace transform method, the authors of \cite{Haenggi2008_InterferenceLargeWirelessNets, Krunz2004_TransmissionPowerControlWAHN} calculated the effects  of fading, pathloss and random erasures on the interference to a random receiver in both regular and Poisson random networks;
in addition, the authors also analyzed therein the effects of power control by inverting the pathloss and/or the fading coefficient of the direct link of a given transmitter-receiver pair.
That said, interference from neighboring transmitters is
%only taken into account by considering them as an
modeled as an effective medium without any feedback:
as a result, the impact that increasing power in a given link has on its neighbors (that also control their power in order to meet a target \ac{SINR} value) is ignored.

A similar approach is taken by the authors of \cite{Jindal2008_FractionalPowerControl} who introduce a scheme to compensate for the fading coefficient of the direct link between transmitter and receiver (but, again, without addressing the effects on neighboring links).
Such effects were partially included in the context of percolating networks in \cite{Paschos2009_ExtendingPercolationWithPowerControl};
there however, the network was initially assumed to percolate with all users transmitting at maximum power, and then reducing their power while maintaining connectivity.
In this way, only the links that are already connected transmit at their optimal power level, without any guarantees to others.

Interference is a serious problem in dense WiFi networks, and it is also expected to remain a major issue in the recently proposed femto-cell paradigm when such cells are deployed at a massive scale \cite{Chandrasekhar2012_FemtoCells}.
Due to their close proximity, neighboring femto-cells may create interference to one another, so when a transmitter increases its power to compensate for interference, it may precipitate a cascade of power increases which needs to be kept in check.
As a result, power optimization is crucial in the above scenarios;
nonetheless, little progress has been made in finding analytic performance estimates for random, interference-limited networks under power control \cite{Chiang2007_NOW_PowerControl}.

In this paper, we present an analytical framework to quantify the optimal power characteristics of large random networks in the presence of interference by introducing a number of methods from statistical mechanics.
We begin with a pure, ordered network in the form of an equally spaced square lattice of $N$ transmitters, each with a receiver located at a fixed distance in its (Voronoi) neighborhood.
Randomness is then introduced in the network by removing (``erasing'') each transmitter-receiver pair with probability $\epr$, leading to a network of (roughly) $N(1-\epr)$ transceiver pairs that are placed randomly on the original lattice.
This thinned network is a plausible model for a cellular network with random transmitter locations;
it is also a reasonable model for a wireless network with intermittent activity where a fraction $\epr$ of the transmitters are inactive at any given time.

To derive an expression for the average transmitted power in a random network of this type, we employ the so-called \ac{CPA} approach, an approximate self-consistent method which was first introduced in the study of disordered metals \cite{Davies1963_CPA, Elliott1974_CPA_RevModPhys}.
The expressions obtained in this way turn out to be identical to those obtained using \ac{RMT} \cite{Tulino2007_GaussianErasureChannel, Tulino2010_GaussianErasureChannel, Moustakas2012_PowerOptimizationErasures} and they agree with numerical results when power control is feasible. However, they fail to account for the fact that an infinite system is \emph{always} infeasible while a finite network only becomes infeasible with increasing probability for larger values of the target \ac{SINR} value $\gamma$.

As a result, even though the problem of determining the average transmit power under power control can be reduced to the analysis of a large random matrix, traditional \ac{RMT} methods are only approximately correct.
The shortcomings of such methods can be traced to the fact that the interference that each receiver observes is mostly due to nearby sites, so it exhibits sizable spatial fluctuations.
Consequently, the interference fluctuations at each site do \emph{not} vanish in the large system limit (as posited by \ac{RMT});
in fact, these fluctuations persist and, in some cases, end up dominating the behavior of the system.
Instead, by modeling power control as a random walk in a random medium,
we show that the problem is equivalent to the so-called \emph{Anderson impurity model} which was originally introduced to describe the motion of electrons in random crystal lattices \cite{Anderson1958_AbsenceDiffusionRandomLattices} and was later applied to the study of
%random walks and
diffusion processes in disordered media \cite{Haus1987_DiffusionRegularIrregularLattices}.
%In this model, the randomness appears only on the main diagonal of the matrix.
Using this equivalence, we obtain analytic results for the probability that the system becomes infeasible and we are also able to estimate the tails of the distribution of power in the system under power control.

Even though we work with a specific network model, we will argue throughout the paper that this paradigm is generic for power controlled networks when interference and randomness both play a significant part.
In fact, one of the main contributions of the paper is the introduction of tools and methodologies from the physics of disordered metals and the theory of random walks in random media to analyze such networks.

\subsection{Summary of results}
\label{sec:SummaryOfResults}

We will now provide an outline of the paper, while at the same time summarizing our main contributions. In the main text of the paper, we try to use intuitive arguments \textendash as opposed to strictly mathematics based ones, trying to bring out the important connections between the physics of
disordered systems and the power control dynamics of random wireless networks. Most appendices, in contrast, are more rigorous and there we try to elucidate the details of the proofs.

Our random network model is introduced in Section \ref{sec:model}, where we also establish the connection between the erasure channel model of \cite{Tulino2007_GaussianErasureChannel, Tulino2010_GaussianErasureChannel},
random walks in random media, and the Anderson impurity model.  In Section \ref{sec:WynerModel}, we then focus on a specific one-dimensional network where only adjacent transmitters interfere with each other \textendash\ the so-called Wyner model \cite{Wyner1994_WynerModel}.
In this simple, yet insightful framework, we are then able to compute all relevant quantities exactly:
in particular, we calculate
\begin{enumerate}
\item
The eigenvalue distribution of the system's pathloss matrix, which determines its feasibility (Section \ref{sec:Wyner_eig_cdf});
\item
The system's probability of infeasibility \textendash\ which, for large but finite systems, turns out to be asymptotically proportional to the tails of the system's eigenvalue distribution (Section \ref{sec:Wyner_model_outage_prob});
\item
The  tails of the empirical distribution of powers in the optimal power vector (Section \ref{sec:Wyner_model_power distribution}).
\end{enumerate}
Accordingly, the Wyner model will serve as a reference point throughout the paper, and will motivate the results of later sections:
for example, the failure of traditional \ac{RMT} techniques will be established by comparing the exact density of states of the Wyner model (a distribution with a countable dense set of atoms) to that derived by \ac{RMT} methods.

In Section \ref{sec:CPA}, we introduce the so-called \acf{CPA} method and we show that it is equivalent to \ac{RMT} (although more general in scope).
Despite its approximate nature, we demonstrate numerically that it is an extremely accurate predictor of both the average optimal power and the average variance of the power vector of the network when power control is feasible.
That said, \ac{CPA} exhibits a fundamental shortcoming in that it fails to predict the probability of instability of the network when operated beyond the stability region of the pure, deterministic system \textendash\ an instability which stems from the infrequent appearance of small eigenvalues in the random, disordered system.

In Section \ref{sec:stability_analysis} we show that power control in the network is \emph{always} infeasible in the infinite system regime beyond a particular value $\gamma_{c}$ of the \ac{SINR} target, irrespective of the degree of randomness in the network.
Nevertheless, for large (but finite) networks, this instability can be described by the so-called \emph{Lifshitz tails} of the network's cumulative density of eigenvalues.
In Section \ref{sec:outage_probability}, we show that the probability that power control becomes infeasible in a finite (but large) network is proportional to the cumulative density of eigenvalues $\ids(\lambda)$ of the corresponding infinite system, thus providing an infeasibility criterion for network operation.
%In this section we also obtain qualitatively the low-valued tails of the distribution of eigenvalues, deferring the details of the calculation to Appendix \ref{app:lifshitz_proofs}.
In particular, we find that the tails of this distribution scale as $\ids(\lambda)\sim (1-\epr)^{k_\lambda \lambda^{-q_\lambda}}$ (to exponential accuracy), where both $k_\lambda$ and $q_\lambda$ depend on the system's dimensionality $d$ and the pathloss exponent $\alpha$ in an explicit way (that we also calculate).
%Then in Section \ref{sec:NumericsOutage} we demonstrate the validity of our results for both one and two dimensional networks.

Even though the average variance of the power vector calculated in Section \ref{sec:CPA} provides an indication of how large the optimal transmitting powers of the systems can become, it is also important to have an understanding of how often much higher powers occur.
In Section \ref{sec:PowerDistribution}, we obtain a lower bound for the tails of the empirical distribution of the optimal power vector, and we find that the cumulative power distribution $\pdist(p)$ scales as $(1-\epr)^{R(p)}$ for $\alpha>d+2$ (where $R(p)$ is a power law which depends on $\alpha$ and $d$);
in particular, in the near-critical limit $\gamma\to\gamma_c^-$, we find that $\pdist(p)$ scales as $(1-\epr)^{k_p p^{d/2}}$ for $\epr\geq1/2$.
We argue that this bound appears to be tight, but we have not been able to prove this;
that said, in Appendix \ref{app:power_tails} we \emph{do} establish a tight upper bound for $2$-dimensional systems where only adjacent transmitters interfere.
%In this Appendix we also conjecture for the value of the exponent for pathloss exponents between $d<\alpha<d+2$.

Finally, in Section \ref{sec:Dynamics}, we analyze the long-term behavior of the Foschini\textendash Miljanic power control algorithm \cite{Foschini1991_FMAlgorithm_PowerControl} and we examine its rate of convergence to the optimal power vector in the presence of random erasures.

\subsection{Notational conventions}

Throughout the paper, we will use the asymptotic equality notation ``$f(x) \sim g(x)$ near $x_{0}$'' to mean $\lim_{x \to x_{0}} f(x)/g(x) = 1$;
when $x_{0} = +\infty$, we will write more simply ``$f(x) \sim g(x)$ for large $x$''.
To maintain the intuitive flow of the discussion, we will sometimes not distinguish between finite- and infinite--dimensional operators in the main text;
whenever such a distinction is important, it will be detailed in a series of appendices at the end of the paper.
%Also, by $\ex$ we will usually assume the expectation over randomness ($\bE$), and in the main text we will not specify this.
%In Appendix \ref{app:lifshitz_proofs}, we also have a different meaning for the expectation, namely over random walks. In that section, we will be more diligent in specifying the type of expectation.
Also, if $\lattice$ is a discrete set, the real space spanned by $\lattice$ will be denoted by $\R^{\lattice}$ and the basis vector of $\R^{\lattice}$ corresponding to $\bm\in\lattice$ will be denoted by $\be_{\bm}$.
Finally, we will use $\one$ to denote the indicator function which takes the value $1$ if its argument is true and zero otherwise.

%% file: Model.tex
%----------------------------------------------------------------------
%%% MODEL
%----------------------------------------------------------------------
% !TEX root = ./RandomPowerOptimization.tex

\subsection{Definitions and connection to random walks}
\label{sec:ConnectionRW}

We start by defining the basic quantities of the problem and establishing a deep connection between power control and random walks of a particle in a random medium (a connection which will be crucial for later sections).

Consider a general network with $N$ transmit-receive pairs.
Let $f_{ij}$ be the channel coefficient, or power gain, between transmitter $i$ and receiver $j$, and let $p_{i}$ denote the transmit power of transmitter $i$.
We then assume that every transmitter adjusts their power to meet the target \ac{SINR} criterion
\begin{equation}
\label{eq:SINRk_def}
\sinr_{k}
	\equiv \frac{p_{k} f_{kk}}{\noisevar + \sum_{j\neq k} p_{j} f_{jk}}
	\geq \gamma_k,
\end{equation}
where $\gamma_{k}$ denotes the threshold SINR of the $k$-th transmitter\textendash receiver pair and $\noisevar$ is the thermal noise level at the receiver.
This inequality can then be written in linear form as
\begin{equation}
\label{eq:SINRk_ineq}
\gamma_k^{-1} f_{kk} p_{k} - \sum_{j\neq k} p_{j} f_{jk} \geq \noisevar,
\end{equation}
or, more concisely, as
\begin{equation}
\label{eq:SINR_vec_ineq}
\bM\bp
	\mgeq \noisevar \bu,
\end{equation}
where $\bu = (1,\dotsc,1)$ is a vector of ones, $\bp =(p_{1},\dotsc,p_{N})$ is the network's power vector, and the matrix $\bM \equiv \bM(\gamma)$ is defined in components as:
\begin{equation}
\label{eq:Mmat_def}
M_{ij} =
\begin{cases}
\gamma_i^{-1} f_{ii}
	&\quad
	\text{if $i=j$},
	\\
-f_{ji}
	&\quad
	\text{if $i\neq j$}.
\end{cases}
\end{equation}
We will then say that power control in the network is \emph{feasible} \cite{Baccelli2009_StocGeometryWirelessNetsApplications}, if there exists a finite positive vector $\bp^{\ast}$ which saturates the constraints \eqref{eq:SINR_vec_ineq};
in particular, if $\bM$ is invertible, we will have:
\begin{equation}
\label{eq:M_eq_def}
\bp^{\ast}
	= \noisevar \bM^{-1} \bu.
\end{equation}
%exists and has positive and finite entries then this will be the optimal power vector, which minimizes the total power of each user, while maintaining all constraints.

In the seminal paper \cite{Foschini1991_FMAlgorithm_PowerControl}, it was shown  that
the power control dynamics
\begin{equation}
\label{eq:FoschiniM_algo_def}
\frac{dp_{k}}{dt}
	= \noisevar + \sum_{j\neq k} p_{j} f_{jk} - \gamma_k^{-1} f_{kk} p_{k}
%	 + \sigma^2
\end{equation}
converge to the power vector $\bp^{\ast}$ (if it exists), which saturates the inequalities in \eqref{eq:SINRk_ineq} \textendash\ assuming of course that their feasible set is not empty. In matrix form we can simply write
\begin{equation}
\label{eq:FoschiniM_algo_matrix_form}
\dot \bp = -\bM \bp + \noisevar \bu,
\end{equation}
so the corresponding stationary solution is simply $\bp^\ast$. In this way, \eqref{eq:FoschiniM_algo_def} may be viewed as the evolution of a population of particles spread over a point lattice (indexed by $i=1,\dotsc,N$) with constant birth rate equal to $\noisevar$, where $f_{ij}$ is the particle transition rate from site $j$ to site $i$ and $-f_{ii}\gamma_{i}^{-1}$ represents the rate of absorption at each site $i=1\dotsc,N$.
The optimal power vector $\bp^*$ describes the stationary distribution of the process.
This interpretation will allow us to view power control as a random walk process, and will be crucial in what follows.

\subsection{Networks without disorder}
\label{sec:NetworkWithoutDisorder}

\begin{figure}[t]
\includegraphics[width=\columnwidth]{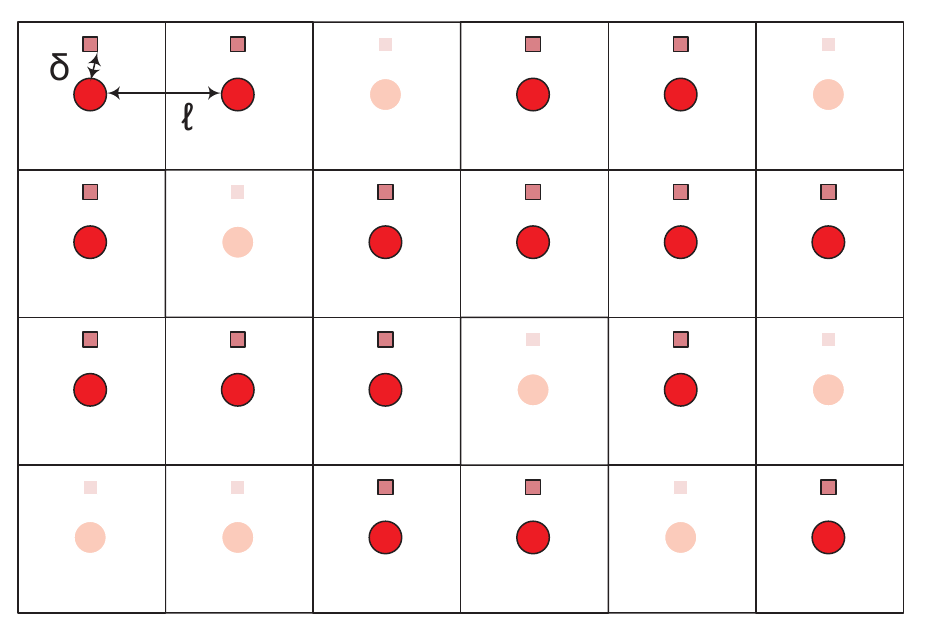}
\caption{%
Schematic of a random wireless network:
circles correspond to transmitters, while squares to receivers;
the faded squares represent transmitter-receiver pairs that have been ``erased''  and are thus inactive.}
\label{fig:net_pic}
\end{figure}

We begin with our model of an \emph{ordered} network, namely a regular, deterministic network consisting of $N$ transmitters situated on the nodes of a regular $d$-dimensional lattice ($d=1,2$).
For concreteness, in two dimensions, we will focus on square lattices with inter-neighbor distance $\ell$ and we will assume that each receiver is located at distance $\delta$ from the corresponding transmitter (see {Fig.~\ref{fig:net_pic}}).

More precisely, let $\latsize$ be a positive integer and let $\lattice \equiv \Z_{\latsize}\times\dotsm\times \Z_{\latsize} = \Z_{\latsize}^{d}$ denote the $d$-fold product of the cyclic group $\Z_{\latsize} = \{0,1,\dotsc,\latsize-1\}$ of integers modulo $\latsize$.
The elements of $\lattice$ will be indexed by $i=1,\dotsc,N \equiv \vert\lattice\vert = \latsize^{d}$ so that $\bm_{i} = (m_{i,1},\dotsc,m_{i,d})$ denotes the position of the $i$-th transmitter on $\lattice$ and addition or subtraction of $\bm_{i},\bm_{j} \in \lattice$ is taken modulo $\latsize$.

In this context, our model for the network's channel coefficients (averaged for fading) will be:
\begin{equation}
\label{eq:g_ij_def}
 f_{ij} = f(\bm_i-\bm_j)
 	\equiv \frac{\delta^{\alpha}}{(|\bm_i-\bm_j|^2\ell^2+\delta^2)^{\alpha/2}},
\end{equation}
where
\begin{enumerate}
[\itshape a\upshape)]
\item
The \emph{pathloss exponent} $\alpha>d$($=1,2$) expresses how fast the channel strength decays as a function of distance.
\item
$\delta$ denotes the distance between a transmitter and its intended receiver.
\item
$\ell$ represents the physical distance between elements of $\lattice$.
\item
The function $f(\bm) = \delta^{\alpha}(|\bm|^{2} \ell^{2} + \delta^{2})^{-\alpha/2}$ describes the pathloss between a transmitter and a receiver located at the points in the lattice $\lattice$ with distance  $\bm$ apart.
\end{enumerate}

This ordered network model will be crucial to our analysis, so a few remarks are in order:

\begin{remark}
A simplifying assumption in \eqref{eq:g_ij_def} is the dependence on the distances between transmitters and receivers:
indeed, \eqref{eq:g_ij_def} is technically correct only when each receiver is positioned vertically to the space spanned by the transmitter lattice $\lattice$ (a line for $d=1$ and a plane for $d=2$) (see for example \cite{Baccelli2006_OptimalPowerThruputRouting}).
Nevertheless, \eqref{eq:g_ij_def} exhibits the correct behavior for $\bm_{i}=\bm_{j}$ as well as for $|\bm_{i}-\bm_{j}|\gg \delta/\ell$
thus, given that we will be focusing on the case where interference is relevant, the exact location of interferers far away is not important.
Moreover, when the pathloss exponent $\alpha$ has to be estimated by curve-fitting large amounts of data with sizable errors, the error induced by the perpendicularity assumption in \eqref{eq:g_ij_def} becomes negligible when compared to the estimation error for $\alpha$ \cite{COST259_book}, so this approximation is harmless in the large system limit.
\end{remark}

\begin{remark}
It should be mentioned here that the periodicity assumption of taking addition modulo $\latsize$ in $\lattice$ was introduced in \eqref{eq:g_ij_def} purely for convenience:
in the large system limit that we will focus on, boundary effects that would occur from embedding $\lattice$ in $\R^{d}$ instead of a $d$-dimensional torus can be effectively ignored when $\alpha> d$ because each line of the matrix is absolutely summable so the approximation error from \eqref{eq:g_ij_def} becomes negligible \cite{Gray2005_ToeplitzMatrices}.
%Since the integers $m_{ia}$ fully specify the position $\bm_i$, we will drop the index $i$ when not necessary. For concreteness, we have normalized the channel gains to unity for $i=j$, i.e. $f_{ii}=f(\b0)=1$.
\end{remark}

\subsection{Average power and feasibility}

The first metric that we will consider for the optimal power vector $\bp^{\ast}$ is the average power per node, which can be expressed \eqref{eq:M_eq_def} as
\begin{eqnarray}
\label{eq:P_ave_def}
p_{\avg} &=& N^{-1}\sum_{i=1}^N \bp^\ast_i = \noisevar N^{-1} \sum_{i=1}^N \left[\bM^{-1} \bu\right]_i \\ \nonumber
&=&\noisevar N^{-1} \bu^{\top}\bM^{-1} \bu
\end{eqnarray}
where $\bu=[1,\ldots,1]^\top$. Clearly, for $p_{\avg}$ to be well-defined, the eigenvalues of the inverse matrix must be themselves positive, so it will be important to analyze the eigenvalue structure of $\bM$.
To that end, note first that the eigenvalues of $\bM$ will be real on account of $\bM$ being real and symmetric.%
\footnote{This is actually one of the main reasons for choosing the model \eqref{eq:g_ij_def}.}
Furthermore, given that the modular arithmetic of $\lattice$ allows us to view $\bM$ as a generalized circulant matrix indexed by $\bm\in\lattice$ \cite{CohenTannoudji_book}, the eigenvectors of $\bM$ will be Fourier modes indexed by the row vector
\begin{equation}
\label{eq:q_vec_def}
\bq = \frac{2\pi}{\latsize} \left(k_{1},\dotsc, k_{d}\right),
	\quad
	k_{1},\dotsc, k_{d}\in\Z_{\latsize}.
\end{equation}
The eigenvalue $\mu(\bq)$ corresponding to the index vector $\bq$ will then be the associated Fourier transform of any
line of $\bM$, i.e.
\begin{flalign}
\label{eq:eigvalueM_def}
\mu(\bq)
	&= \gamma^{-1} - \sum_{\bm\in\lattice\setminus\{0\}} \frac{\delta^{\alpha} e^{i\bq\cdot\bm}}{(|\bm|^{2}\ell^{2}+\delta^{2})^{\alpha/2}}.
\end{flalign}
Accordingly, the minimum eigenvalue of $\bM$ (corresponding to the eigenvector $\bu$) will be:
\begin{equation}
\label{eq:min-eigenvalue}
z_{\gamma}
	= M_{ii} + \insum_{j\neq i} M_{ij}
	= \gamma^{-1} - \gamma_{c}^{-1},
\end{equation}
where
\begin{equation}
\label{eq:lambda0_def}
\gamma_c^{-1}
	= \sum_{j \neq i} f_{ij}
	= \sum_{\bm\in\lattice\setminus\{0\}} \frac{\delta^{\alpha}}{(|\bm|^{2}\ell^{2}+\delta^{2})^{\alpha/2}}.
\end{equation}
In a network of infinite size, $\gamma_{c}$ is finite if and only if $\alpha>d$;
%which we will assume henceforth in the paper.
the optimal power vector of the system will then be
\begin{equation}
\label{eq:opt_P_for H0}
\bp^{\ast}
	=z_{\gamma}^{-1} \noisevar \bu
\end{equation}
leading to average power
\begin{equation}
p_{\avg} = z_{\gamma}^{-1}\noisevar.
\end{equation}
Hence, for the system to be well-defined and feasible we need $z_\gamma>0$ or, equivalently:
\begin{equation}
\label{eq:finiteP_ave_no_erasurecondition}
\gamma < \gamma_c.
\end{equation}

For simplicity, it will be convenient to shift the spectrum of $\bM$ to positive values by introducing the positive-semidefinite matrix $\bH_{0}$ via the equation
\begin{equation}
\label{eq:H0_def}
\bM =\bH_{0} + z_{\gamma} \bI.
\end{equation}
In view of \eqref{eq:eigvalueM_def}, the eigenvalues of $\bH_0$ will then be
\begin{equation}
\label{eq:eigvalue_def}
\eps(\bq)
	= \sum_{\bm\in\lattice} \frac{\delta^{\alpha} \left(1-e^{i\bq^T\bm}\right)}{(|\bm|^{2}\ell^{2}+\delta^{2})^{\alpha/2}},
\end{equation}
so the feasibility of the optimal power vector $\bp^{\ast}$ will be determined by the behavior of $\eps(\bq)$ for small $|\bq|$ (i.e. by the lowest eigenvalues of $\bH_{0}$).

For $\alpha \geq d+2$, an asymptotic expansion of $\eps(\bq)$ yields
\begin{equation}
\label{eq:low_q_eig_a>d+2}
\eps(\bq)
	= t_{2} \vert\bq\vert^2 + \bigoh(\vert\bq\vert^{4})
\end{equation}
with
\begin{equation}
\label{eq:low_q_t2}
t_{2}
	= \frac{1}{2d} \sum_{\bm} \frac{\delta^\alpha |\bm|^2}{(\vert\bm\vert^{2}\ell^{2}+\delta^{2})^{\alpha/2}}.
\end{equation}
On the other hand, for $d<\alpha<d+2$, the series \eqref{eq:low_q_t2} for $t_{2}$ is no longer summable;
instead, using the Poisson summation formula, it can be shown that the leading order asymptotic expression for $\eps(\bq)$ will be of the form
\begin{equation}
\label{eq:low_q_eig_a<d+2}
\eps(\bq)
	= t_{\alpha-d} \vert\bq\vert^{\alpha-d} + \bigoh(\vert\bq\vert^2)
\end{equation}
where $t_{\alpha-d}$ is a computable constant.
%
%%% FOR d=2
%\begin{equation}
%\label{eq:low_q_t_{a-d}}
%t_{\alpha,d=2} = \frac{1}{\sin\left[\left(\frac{\alpha-2}{2}\right)\pi\right]} \frac{\pi^2 2^{2-\alpha}}{\Gamma\left(\frac{\alpha}{2}\right)^2} \left(\frac{\delta}{\ell}\right)^\alpha
%\end{equation}
%%% FOR d=1
%\begin{equation}
%\label{eq:low_q_t_{a-d}}
%t_{\alpha,d=1} = \left(\frac{\delta}{\ell}\right)^\alpha \frac{\pi}{\Gamma(\alpha)\left|\cos\frac{\pi\alpha}{2}\right|}
%\end{equation}
%
Thus, with a fair degree of hindsight, it will be convenient to introduce here the \emph{effective pathloss exponent}
\begin{equation}
\label{eq:alpha-eff}
\alpha_{\eff}
	\equiv \min\{\alpha, d+2\}
\end{equation}
and the corresponding leading order coefficient
\begin{equation}
\label{eq:t-eff}
t_{\eff}
	\equiv
	\begin{cases}
	t_{\alpha-d}
	&
	\text{if $\alpha\in(d,d+2)$},
	\\
	t_{2}
	&
	\text{if $\alpha \geq d+2$}.
	\end{cases}
\end{equation}
In this way, \eqref{eq:low_q_eig_a>d+2} and \eqref{eq:low_q_eig_a<d+2} may be written more simply as:
\begin{equation}\label{eq:eps-eff}
\eps(\bq) \sim t_{\eff} \vert \bq \vert^{\alpha_{\eff}-d}
	\quad
	\text{for small $\vert\bq\vert$.}
\end{equation}

\subsection{Random networks: disorder and erasures}
\label{sec:randomly_thinned_model}

There are two ways of introducing randomness (disorder) in the network model of the previous section.
First, the target \ac{SINR} $\gamma_{k}$ of each user (and hence, the corresponding rate) may be random at each site;
second, a random fraction of the transmitters could be turned off (``erased'') at any given time.
The former type of randomness can be analyzed in conjunction with the latter but, due to space limitations, we will defer this analysis for the future.
In the present paper, we will only focus on erasures, which will be introduced in two different (but equivalent) ways.
%which will be introduced in two equivalent ways.
%For simplicity, we will focus on the second type of randomness, but we stress that the first can also be analyzed in conjunction with the second.
%We will analyze this type of disorder in the future.

\subsubsection{The Anderson model}
The first ``erasure'' procedure that we will consider may be described as follows:
first, the sites to be turned off are chosen at random with a fixed \emph{erasure probability} $\epr\in[0,1]$.
Then,  the optimal transmitting power $p_k$ of a transmitter which is to be switched off is set to $0$ by setting $f_{kk} = +\infty$ for the corresponding channel strength between the $k$-th transmitter and its intended receiver. Indeed it is not hard to see that when $f_{kk}$ becomes arbitrarily large in \eqref{eq:SINRk_ineq}, the \ac{SINR} target constraint for the $k$-th link may be met with arbitrarily small power $p_k$.
Formally, consider the random diagonal matrix
\begin{equation}
\bE = \diag(e_{1},\dotsc,e_{N})
\end{equation}
with random \ac{iid} entries $e_{i}\in\{0,1\}$ such that
\begin{equation}
\label{eq:P(e)_def}
\begin{aligned}
\prob(e_{i} = 1)
	&=\epr,
	\\
\prob(e_{i} = 0)
	&=1-\epr.
\end{aligned}
\end{equation}
Erasures are then introduced by replacing $\bH_0$ in \eqref{eq:H0_def} with
\begin{equation}
\label{eq:H_def}
\bH_{V}
	= \bH_{0} + V\bE
	\equiv \bH_{0} + \bV,
%	\equiv \bH_{0} + V\bE,
\end{equation}
with matrix elements $\bE_{V,ij}=\bH_{0,ij}+Ve_i\delta_{ij}$, where $\bV = V \bE$ and $V>0$ is a large positive parameter which turns off the sites determined by $\bE$ in the limit $V\to\infty$.
In particular, the quantity $V$ plays the role of the excess channel gain of a given transmitter to its intended receiver:
since we are interested only in optimal power solutions which assign finite positive transmitting power to each site, the limit $V\to+\infty$ can then be taken in the end of the calculation of the inverse matrix $\bM^{-1} = (z_{\gamma}\bI + \bH_{V})^{-1}$.

The case of spatially random $\gamma_k$ can be treated in a similar fashion, by including $\gamma_k^{-1}$ in $\bV$.
 
\begin{remark*}
The matrix $\bH_{V}$ above has deterministic off-diagonal elements and diagonal disorder and it is known in the physics literature as the \emph{Anderson model}.
This model was introduced by P.~W.~Anderson to explain localization of particles (and waves) in random media \cite{Anderson1958_AbsenceDiffusionRandomLattices}, and it has since been extended to study random walks in random media \cite{Grassberger1986_DiffusionStaticTraps}.
\end{remark*}

In this context, the optimal power vector $\bp^{\ast}$ will be given by
\begin{equation}\label{eq:P_vec_eq_random}
\bp^{\ast}
	= \noisevar \big[\bH_{V} + z_\gamma\bI\big]^{-1} \bu,
\end{equation}
so its intra-sample average over non-erased sites can be derived by multiplying from the left by $\bu^{\top}$ and dividing with the expected number of non-erased sites $N(1-\epr)$, producing
\begin{equation}
\label{eq:P_ave_Anderson_def}
p_{\avg}
	= \frac{\noisevar}{N(1-\epr)} \lim_{V\to\infty} \bu^{\top} \big[\bH_{V} + z_\gamma \bI\big]^{-1} \bu.
\end{equation}
As a result, via spectral decomposition, $p_{\avg}$ may be expressed directly in terms of the eigenvalues and eigenvectors
of the random matrix $\bH_{V}$ as
\begin{equation}
\label{eq:P_ave_Anderson_eigenval_expansion}
p_{\avg} = \frac{\noisevar}{N(1-\epr)} \lim_{V\to\infty} \insum_{s} \frac{\vert\bu_{s}^{\top} \bu\vert^{2}}{\lambda_{s} + z_\gamma},
\end{equation}
where $\lambda_{s} \equiv \lambda_{s}(\bH_{V})$ denotes the $s$-th eigenvalue of $\bH_{V}$ and $\bu_{s}$ is the corresponding eigenvector (note that $\bu$ is itself an eigenvector in the absence of erasures). The effect of the $V\to\infty$ limit above can be appreciated by invoking Gershgorin’s circle theorem, which tells us that for large $V$ and a given realization of the randomness $\bE$ with $K\approx N\epr$ ones in $\bE$, the spectrum of $\bH_V$ will consist of $K$ large eigenvalues of order $\bigoh(V)$ and and the remaining ones are $\bigoh(1)$ in $V$. Hence the former will not play any role in the power vector above.

In view of the above, the average optimal power will be finite and positive as long as the eigenvalues of the matrix $\bH_{V}$ are large enough, i.e. $\lambda_s+z_\gamma>0$.
More importantly, the analysis of \cite{Baccelli2006_OptimalPowerThruputRouting, Baccelli2009_StocGeometryWirelessNetsApplications} (see Lemma 18.2.4 in \cite{Baccelli2009_StocGeometryWirelessNetsApplications}) readily yields the following stronger statement for the feasibility of power control:

\begin{theorem}
\label{thm:relation_Pout_P(Emin)}
Power control is feasible
%i.e. there exists a $\bp$ with positive and finite elements such that $(\bH_V+z_\gamma\bI)\bp\mgeq \noisevar \bu$  \eqref{eq:SINR_vec_ineq},
if and only if
$\bH_V+z_\gamma\bI \succ 0$.
Consequently, the probability of instability (or infeasibility) for the network will be:
\begin{equation}
\label{eq:thm:p_outage=prob_min}
 P_{\mathrm{inst}}(\gamma)
 	=\prob\big[\lambda_{\min}(\bH_{V})<-z_{\gamma}\big],
\end{equation}
where $\lambda_{\min}(\bH_{V})$ denotes the minimum eigenvalue of $\bH_{V}$.
\end{theorem}

This result provides a close connection between the feasibility of the system and the lower part of the spectrum of $\bH_V$.
In fact, as an immediate corollary of Theorem \ref{thm:relation_Pout_P(Emin)}, we obtain:
\begin{corollary}
\label{cor:feasibility_gamma<gamma_c}
The system is always feasible for $\gamma < \gamma_{c}$.
\end{corollary}

Despite their apparent simplicity, the results above do not provide any intuition on what happens in the network for $\gamma>\gamma_c$ and how feasibility breaks down for larger $\gamma$. In the next sections we will see that for $\gamma>\gamma_c$ the system becomes unstable (i.e. its powers explode) in the network configurations where the minimum eigenvalue of $\bH_V$ becomes larger than $-z_\gamma$. We will also calculate the probability for this to happen.

\subsubsection{The erasure channel model}
To make contact with previous work on the erasure channel \cite{Tulino2007_GaussianErasureChannel, Tulino2010_GaussianErasureChannel, Moustakas2012_PowerOptimizationErasures}, we will also consider a different random network model and show that it is equivalent to the large $V$ limit of \eqref{eq:H_def}.
In particular, for every transmitter-receiver pair that is to be ``switched off'', we will set the corresponding column and row elements of the channel matrix $\bM$ to zero by considering the matrix
%more precisely, working directly with the representation \eqref{eq:H0_def} of $\bM$, we will consider the matrix
\begin{equation}
\label{eq:EME_def}
\bH
	= (\bI-\bE) \bH_0 (\bI-\bE),
\end{equation}
with matrix elements $\bE_{V,ij}=\bH_{0,ij}(1-e_i)(1-e_j)$, and with $\bE$ given by \eqref{eq:P(e)_def} as before.
In this way, the multiplication with $\bI-\bE$ from the left and right, the ``erased'' sites are completely decoupled \textendash\ and, hence, switched off. This has the the effect of completely decoupling the erased sites, which are thus effectively switched off.

The previous discussion (e.g. the statement of Theorem \ref{thm:relation_Pout_P(Emin)}) obviously still applies with $\bH_{V}$ replaced by $\bH$ and with the caveat that ``minimum eigenvalue'' should be interpreted as the ``minimum eigenvalue over the range of $\bI-\bE$'' \textendash\ simply note that the zero eigenvalues contributed by the erased sites should not be counted in \eqref{eq:thm:p_outage=prob_min}.
Thus, given that the lower part of the spectrum of $\bH_{V}$ approaches that of $\bH$ for large $V$ (see Proposition \ref{prop:model-equivalence} in Appendix \ref{app:lifshitz_proofs}), the two erasure models will be equivalent in the limit $V\to\infty$.

%% file: WynerModel.tex
%----------------------------------------------------------------------
%%% WYNER MODEL
%----------------------------------------------------------------------
% !TEX root = ./RandomPowerOptimization.tex

Our goal in this section will be to analyze the so-called \emph{Wyner model} \cite{Wyner1994_WynerModel}, a simple one-dimensional random network where the asymptotic behavior of the optimal power vector can be calculated exactly.
Thanks to this simple model, we will have the opportunity to introduce several metrics for the behavior of the optimal power vector that are at the core of our considerations;
more importantly, the exact results obtained here will provide the intuition and necessary groundwork to understand the asymptotic behavior of more general network models that require significantly more sophisticated tools.

The Wyner model consists of a circular array $\lattice$ of $\vert\lattice\vert = N$ transmitters,%
\footnote{Again, the effects of the geometry may safely be ignored for large $N$, so the system may be considered linear in the large $N$ limit.}
located a fixed distance apart so that only neighboring transmitters interfere with each others' transmissions.
Accordingly, the matrix $\bH_0$ describing the system in the sense of \eqref{eq:H0_def} will be a tridiagonal matrix with elements
\begin{equation}
\label{eq:H0_wyner_def}
H_{ij}^{0}
	= 2t\left[\delta_{ij} - \tfrac{1}{2}\big(\delta_{i,j+1} + \delta_{i,j-1}\big)\right],
%	- \frac{t}{2}\left(\delta_{i,1}\delta_{j,N} + \delta_{i,N} \delta_{j,1}\right),
\end{equation}
where
%$\vert i \rangle$ denotes the basis vector corresponding to the $i$-th transmitter,
addition in $i$ and $j$ is taken modulo $N$,
%where the last two terms are necessary to keep the matrix circulant.
and the parameter $t$ determines the interference level between users.

Comparing the above with \eqref{eq:g_ij_def}, \eqref{eq:lambda0_def} and \eqref{eq:low_q_t2}, it follows that the Wyner model \eqref{eq:H0_wyner_def} will have
\begin{equation}
\gamma_c = 1/(2t),
\quad
%\text{ and }
t_{2}=t.
\end{equation}
Furthermore, since the system is one-dimensional and interference only comes from a site's nearest neighbors, erasures will simply partition the system into independent blocks of different (random) lengths, separated by sites with zero power.
In particular, in the infinite system limit, the distribution $\pi_{r}$ of the cluster length $r\geq 0$ can be shown to be exponential, i.e.
\begin{equation}
\label{eq:3diag_exp_distribution}
\pi_{r} = \epr (1-\epr)^{r}.
\end{equation}
Thanks to this partition, we will calculate
\begin{inparaenum}[\itshape a\upshape)]
\item
the eigenvalue distribution of $\bH_{0}$ in the presence of erasures;
\item
the resulting optimal power vector;
\item
the system's instability probability (i.e. the probability of the optimal power vector being infeasible);
and
\item
the tails of the power distribution when power control is feasible.
\end{inparaenum}

\subsection{Eigenvalue distribution}
\label{sec:Wyner_eig_cdf}

As we indicated in the previous section, the feasibility of the optimal power vector $\bp^{\ast}$ for a given erasure matrix $\bE$ will be determined by the spectrum of $\bH = (\bI - \bE) \bH_{0} (\bI - \bE)$.
Accordingly, our aim here will be to determine the system's \acf{IDS}, i.e. the number of eigenvalues not exceeding a given level divided by the size $N = \vert\lattice\vert$ of the system;
formally, we let:
\begin{equation}
\label{eq:IDS-Wyner}
\ids(\lambda)
	= \lim_{N\to\infty} N^{-1} \vert\{\lambda'\in\spec(\bH): 0<\lambda'\leq\lambda\}\vert,
\end{equation}
where $\spec(\bH)$ is the set of eigenvalues of the $N\times N$ matrix $\bH$ (see Appendix \ref{app:lifshitz_proofs} for a more detailed discussion).
Clearly, each realization of $\bE$ partitions $\bH_{0}$ into disjoint tridiagonal T\oe plitz blocks of varying lengths, so the eigenvalues corresponding to a block of length $r$ will be:
\begin{equation}
\label{eq:segment_L_eigs}
\txs
\eps_{r}(k)
	= \gamma_c^{-1} \left[1 - \cos\left(\frac{k\pi}{r+1}\right)\right],
	\quad
	k =1,\dotsc,r.
\end{equation}
In view of the above,
the probability of observing a given eigenvalue may be calculated by averaging over the possible block lengths $r$ for which this eigenvalue may occur.
To that end, since the probability of observing a segment of length $r$ in the infinite system limit follows the geometric distribution \eqref{eq:3diag_exp_distribution},
some algebra yields the following expression for the integrated eigenvalue density $\ids(\lambda)$:
\begin{flalign}
\label{eq:3diag_eig_distr}
  \ids(\lambda)
  	&= \sum_{r=1}^\infty  \frac{\epr\pi_{r}}{1-\epr} \sum_{k=1}^{r}
	\one\left[\lambda \geq \gamma_c^{-1} \left(1-\cos\left(\tfrac{k\pi}{r+1}\right)\right)\right]
	\notag\\
	&= \sum_{q\in\Q\cap(0,1)}\,
	\hat{\pi}_{\ell(q)}   \one\left[\lambda \geq \gamma_c^{-1}\left(1-\cos(q\pi)\right)\right],
\end{flalign}
where
\begin{equation}
\hat{\pi}_\ell
	= \frac{\epr^2(1-\epr)^{\ell-2}}{1-(1-\epr)^{\ell}},
\end{equation}
and $\ell(q)$ denotes the denominator of $q$ in lowest terms.

To understand this expression, we note that the second sum in the first line of \eqref{eq:3diag_eig_distr} counts the number of non-zero eigenvalues that do not exceed $\lambda$ in a block of length $r$.
One then needs to normalize the expression with the average number of eigenvalues, or, equivalently, the average block size $\epr/(1-\epr)$;
finally, the expression for $\hat{\pi}_{\ell(q)}$ results from summing over all rationals of the form $q = k/(r+1)$ that correspond to an eigenvalue occurring in blocks of different length.
%summing $\pi_{r}$ over all multiples of $\ell(q)$ which correspond to segments with the same eigenvalue.

The cumulative eigenvalue density \eqref{eq:3diag_eig_distr} above has two interesting properties:
First, the set of discontinuities of $\ids(\lambda)$ (corresponding to the atoms of the underlying eigenvalue distribution) is \emph{dense} in $[0,+\infty)$:
in particular, $\ids(\lambda)$ is discontinuous at all points of the form $\gamma_c^{-1}(1 - \cos(q\pi))$, $q\in\Q\cap(0,1)$, and is continuous otherwise.
This is consistent with the prediction that the cumulative density of eigenvalues is discontinuous for the one-dimensional Bernoulli-distributed random potential above \cite{Halperin1967_PropertiesParticle1DRandomPotential, Kirsch2007_IDOS_RandomSchroedingerOperators}.

Second, the infimum of the support of $\ids(\lambda)$ is zero for all $\epr>0$, a behavior which is intimately connected with the infeasibility of power control in the system.
However, very small eigenvalues correspond to very large (and very rare) clusters that occur with probability of the order of $(1-\epr)^{r_{\lambda}}$ where
\begin{equation}
r_{\lambda} \sim \frac{\pi}{\sqrt{2\gamma_c\lambda}}
\end{equation}
denotes the inverse of \eqref{eq:segment_L_eigs} for $k=1$ (i.e. $r_{\lambda}$ is the size of the smallest cluster which supports the eigenvalue $\lambda$).
%$r_{\lambda}$ is the inverse of $\lambda=t(1-\cos(\pi/(L+1)))$,
%and for small $\lambda$ it becomes $L(\lambda)\sim\pi\sqrt{t/(2\lambda)}$
As a result, for small $\lambda$, the integrated density of eigenvalues becomes
\begin{equation}
\label{eq:3diag_CDF_asympt}
\ids(\lambda) \sim \left(1-\epr\right)^{\pi\left(2\gamma_c\lambda\right)^{-1/2}}.
\end{equation}
The importance of this expression will become clear below, where we show that $\ids(\lambda)$ is proportional to the instability probability for large (but finite) systems.

\begin{figure}[t]
\includegraphics[width=\columnwidth]{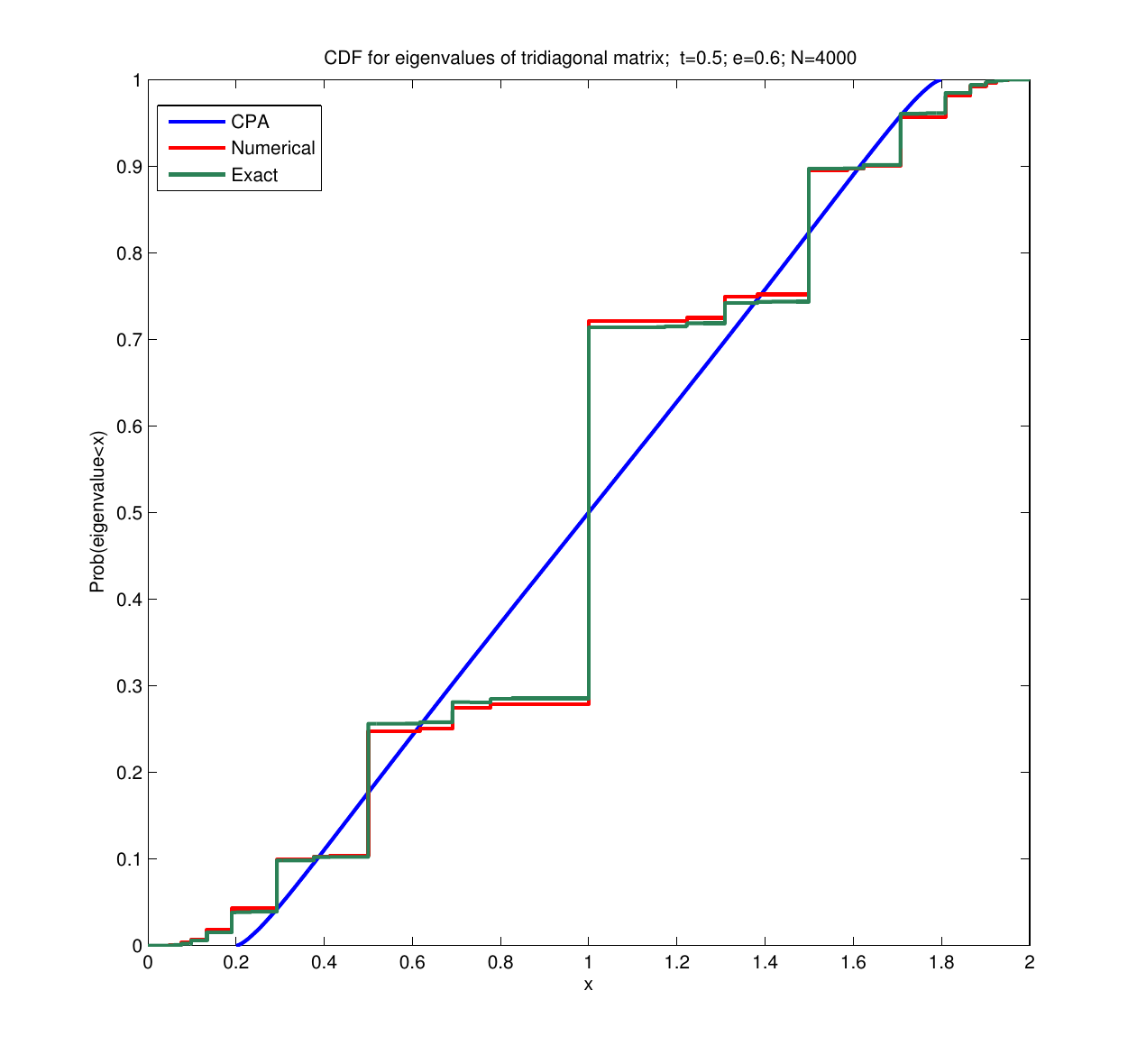}
\caption{Cumulative distribution of eigenvalues for the Wyner model.
The system's eigenvalues form a dense countable subset of $[0,2]$ but, given the size of the simulated system, these discontinuities cannot be represented graphically.
This figure demonstrates the failure of the \ac{CPA} expression obtained in Section \ref{sec:CPA} (continuous line) to capture the exact structure of the distribution of eigenvalues in the system.
For our purposes, the inconsistency is most important in the lower left tails of the distributions.}
\vspace{-5pt}
\label{fig:asympt_cdf_3diag}
\end{figure}

\subsection{The optimal power vector}
\label{sec:Wyner_power_evaluation}

Owing to the partition of the system into erasure-free blocks, the optimal power at each point may be calculated by noting that, in any given block of length $r$, the power control equations \eqref{eq:SINRk_ineq} may be rewritten more suggestively as:
\begin{equation}
\label{eq:pc_eqs_wyner}
-\frac{1}{2}\gamma_c^{-1} \, \Delta^{(2)} p_{k} + \big(\gamma^{-1} - \gamma_{c}^{-1}\big) p_{k}
%\gamma^{-1} p_{k} - t (p_{k-1} + p_{k+1})/2
	= \noisevar,
	\quad
	k = 1,\dotsc,r,
\end{equation}
where $\Delta^{(2)}$ denotes the second-order difference operator $\Delta^{(2)} p_{k} = p_{k+1} + p_{k-1} - 2 p_{k}$,
and we are taking boundary conditions $p_{0} = p_{r+1} =0$ (recall that each end of the block is erased).
Depending on the value of the target \ac{SINR} $\gamma$, we thus obtain three different solutions:
\begin{enumerate}
\item
For subcritical $\gamma < \gamma_{c}$, we get the hyperbolic expression:
\begin{subequations}
\label{eq:pk_stable_wyner}
\begin{flalign}
p_{k}	
	&= \frac{\noisevar}{z_{\gamma}} \left[
	1-\frac{\cosh\left[\kappa\left(k-\frac{r+1}{2}\right)\right]}{\cosh\left(\kappa \frac{r+1}{2}\right)}
	\right],
	\\
\kappa
% 	&= \log\left(\frac{\gamma_c-\sqrt{\gamma_c^2-\gamma^2}}{\gamma}\right).
	&= -\arcosh(\gamma_{c}/\gamma).
\end{flalign}
\end{subequations}

\item
At the critical value $\gamma = \gamma_{c}$, we get the quadratic solution:
\begin{equation}
\label{eq:pk_critical_wyner}
p_{k}
	= \gamma_c\noisevar k(r + 1 - k).
%	\quad
%	\text{for $1\leq k\leq r$}.
\end{equation}
\item
Finally, supercritical $\gamma>\gamma_{c}$ leads to the elliptic solution:
\begin{subequations}
\label{eq:pk_unstable_wyner}
\begin{flalign}
p_{k}	
	&= \frac{\noisevar}{z_{\gamma}}\left[
	1-\frac{\cos\left[\left(k-\frac{r+1}{2}\right)\phi\right]}{\cos\left(\frac{r+1}{2}\phi\right)}
	\right],
	\\
%  	&= \frac{2n}{z_\gamma} \frac{\sin\left[\left(r+1-k\right)\phi\right]\sin(k\phi)}{\cos\left((r+1)\phi\right)}
%	\\
 \phi
	&= \arccos(\gamma_c/\gamma),
\end{flalign}
\end{subequations}
\noindent
which is obviously equivalent to the hyperbolic solution \eqref{eq:pk_stable_wyner} with $\kappa = -i\phi$.
\end{enumerate}

\begin{remark*}
We note here that the solutions \eqref{eq:pk_stable_wyner}\textendash\eqref{eq:pk_unstable_wyner} of the finite difference equation \eqref{eq:pc_eqs_wyner} may be mapped to the solutions of the continuous \emph{differential} equation
\begin{equation}
\label{eq:pc_eqs_wyner-continuous}
-\frac{1}{2}\gamma_c^{-1} \cdot p''(x) + (\gamma^{-1} - \gamma_{c}^{-1}) p(x)
	= \noisevar,
	\quad
	x\in[0,r+1],
\end{equation}
with boundary conditions $p(0) = p(r+1) = 0$.
As we shall see in the next section, this last equation may be viewed as a ``large $r$'' limit of \eqref{eq:pc_eqs_wyner} where the sites $k=0,\dotsc,r+1$ are approximated by a continuum of sites $x\in[0,r+1]$ and the power vector $p_{k}$ by the power distribution $p(x)$ (see also Appendix \ref{app:cont_approx}).
This approximation will be key to the analysis of more general problems, so it is worth keeping in mind even in the exactly solvable Wyner model.
\end{remark*}

%With these expressions at hand, one can then calculate the total power per segment and then average over each segment's length using \eqref{eq:3diag_exp_distribution} to obtain expressions for the average power per transmitter.

\subsection{Feasibility analysis and probability of instability}
\label{sec:Wyner_model_outage_prob}

Obviously, for power control to be feasible, the components of the system's optimal power vector $\bp^{\ast}$ (given by \eqref{eq:pk_stable_wyner} and \eqref{eq:pk_unstable_wyner} for the subcritical and supercritical regime respectively) must be finite and nonnegative.
Since \eqref{eq:pk_stable_wyner} is positive for all $k=1,\dotsc,r$,%
\footnote{Simply note that $\cosh\left(\kappa \left(k - \frac{r+1}{2}\right) \right) \leq \cosh\left(\kappa \frac{r+1}{2}\right)$ for $k = 0,\dotsc,r+1$.}
the optimal power vector $\bp^{\ast}$ will always be feasible if $\gamma<\gamma_{c}$ (cf. Corollary \ref{cor:feasibility_gamma<gamma_c}).
On the other hand, for $\gamma > \gamma_{c}$, $p_{k}$ may take on negative values if $\phi > \phi_{c} \equiv \frac{\pi}{r+1}$:
indeed, the denominator of \eqref{eq:pk_unstable_wyner} vanishes for $\phi=\phi_{c}$, so the optimal power vector $\bp^{\ast}$ will start becoming infeasible beyond the critical value $\phi_{c}$.

The above criterion may be reformulated in terms of the length of each erasure-free block as follows:
by \eqref{eq:segment_L_eigs}, the minimum eigenvalue of a block of length $r$ such that $\frac{\pi}{r+1} < \arccos(\gamma_{c}/\gamma)$ will satisfy the inequality:
\begin{equation}
\label{eq:instability_criterion_wyner}
\eps_{\min}(r)
	= \gamma_c^{-1} \left(1 - \cos\frac{\pi}{r+1}\right)
	< \gamma_{c}^{-1} (1 - \gamma_{c}/\gamma)
	= -z_{\gamma}.
\end{equation}
As a result, for $\gamma > \gamma_{c}$, for a given realization of the erasure matrix $\bE$, power control will be feasible only if the system's largest erasure-free region (where the system's smallest eigenvalue is encountered) satisfies the criterion \eqref{eq:instability_criterion_wyner}.
Hence, in view of Theorem \ref{thm:relation_Pout_P(Emin)}, the \emph{instability probability} for a finite system of size $N$ and target \ac{SINR} $\gamma>\gamma_c$ will be:
\begin{equation}
\label{eq:Lc_wyner}
P_{\mathrm{inst}}(\gamma)
	=\prob(r_{\max} > r_c(\gamma)),
\end{equation}
where $r_{\max}$ is the maximum realized cluster size and
\begin{equation}
\label{eq:Rcrit}
r_{c}(\gamma)
	\equiv \left\lfloor \pi \big/ \arccos\left(\gamma_c\big/\gamma\right) \right\rfloor
\end{equation}
denotes the minimum cluster size for which the infeasibility criterion \eqref{eq:instability_criterion_wyner} is satisfied.

The RHS of \eqref{eq:Lc_wyner} may be evaluated explicitly to yield
\begin{equation}
\label{eq:outage_wyner}
\begin{aligned}
P_{\mathrm{inst}}
	&= (1-\epr)^N \, \one\big[N > r_{c} + 1\big]
	\\
	&+ N\sum_{a=1}^N (-\epr)^{a-1} (1-\epr)^{a r_{c}} \frac{\Gamma(N-ar_{c})}{\Gamma(N-ar_{c} + 1 - a)\Gamma(a+1)},
\end{aligned}
\end{equation}
where each term counts the number of ways that $a$ blocks of $r_{c}$ non-erased sites
%\st{separated by an erasure}
can appear in a circle of length $N$.%
\footnote{\eqref{eq:outage_wyner} was obtained by expressing $P_{\mathrm{inst}}$ as a sum over the possible positions of erasure-free regions, taking the $z$-transform, averaging over the corresponding probabilities and taking the inverse $z$-transform of the result.}

\setcounter{remark}{0}

\begin{remark}
As $N\to\infty$, the probability of encountering arbitrarily large clusters approaches $1$, so very large Wyner networks will be infeasible almost surely.
This prediction is consistent with \eqref{eq:outage_wyner} where, with a little algebra, one can show that $P_{\mathrm{inst}} \to 1$ as $N\to\infty$.
Importantly, even though this result seems to depend crucially on the specific structure of the Wyner model, we will see in Section \ref{sec:stability_analysis} that this property remains true in a significantly more general class of random networks.
\end{remark}

\begin{remark}
For the instability probability to be small, $r_{c}(\gamma)$ has to be large and hence $\gamma$ must be close to $\gamma_{c}$.
In this case, \eqref{eq:outage_wyner} may be expressed to leading order as
\begin{equation}
\label{eq:outage_wyner_lifshitz}
P_{\mathrm{inst}}
	\sim N\epr \left(1-\epr\right)^{r_{c}(\gamma)}
%	\notag\\
	\sim N\epr \left(1-\epr\right)^{\pi/\sqrt{2\gamma_c \vert z_{\gamma}\vert}},
\end{equation}
%with $r_{c} \gg -\log N/\log(1-\epr)$
with the approximation being valid for $r_{c} \gg -\log N/\log(1-\epr)$ or, equivalently:
\begin{equation}
\label{eq:outage_criterion_z_g_wyner}
\vert \frac{\gamma_c}{\gamma} -1\vert \ll \frac{1}{2} \left(\frac{\pi\log(1-\epr)}{\log N}\right)^2.
\end{equation}
This shows that the instability probability in a network of size $N$ is small whenever the target \ac{SINR} value $\gamma$ lies within $\bigoh((\log N)^{-2})$ of the network's critical threshold $\gamma_{c}$;
in other words, if $N$ is not too large, the parameter range of $\gamma$ for which power control remains feasible can be itself fairly large.
\end{remark}

\begin{remark}
It is also important to note that the instability probability \eqref{eq:outage_wyner_lifshitz} is proportional to the tails of the integrated density of eigenvalues $\ids(-z_\gamma)$ in \eqref{eq:3diag_CDF_asympt}.
This is no coincidence:
the instability probability is given by the cumulative distribution function of the minimum eigenvalue of the system, which is in turn proportional to $\ids(\lambda)$.
This important point will be made more precise in Section \ref{sec:outage_probability}.
\end{remark}

%\begin{remark}
%The previous analysis reveals that an outage occurs when a region of the network of size $\bigoh(r_{c})$ has no erasures.
%However atypical, such events are local and hence have a probability of occurrence which does not scale exponentially with the size of the system
%\textendash\ for instance, as would be the case if $r_{c} = \bigoh(N^\eps)$ for some $\eps>0$.
%\PMcomment{Do not understand this comment\dots}
%In Section \ref{sec:stability_analysis}, we will see that this is a general phenomenon that does not depend on the specific simple structure of the Wyner model, but holds for a much more general class of Hamiltonian matrices $\bH_0$.
%\end{remark}

\subsection{Power distribution in the Wyner model}
\label{sec:Wyner_model_power distribution}

Thanks to the simplicity of the Wyner network model, we may also calculate the tails of the empirical distribution of powers in the optimal power vector $\bp^{\ast}$, or equivalently the fraction $\pdist(p)$ of sites with power exceeding some large value $p$.
%This distribution $\pdist(p)$ can be seen as the fraction of sites with power greater than a fixed power $p$ in the infinite system limit.
Since all sites are statistically equivalent, this distribution may be viewed as the probability that the optimal power $p_{0}$ at the origin exceeds $p$, i.e.:
\begin{equation}
\pdist(p) = \prob[p_{0}\geq p].
\end{equation}

Now, given that the fraction of clusters of size $r$ follows the geometric distribution \eqref{eq:3diag_exp_distribution} for large $N$, the distribution of powers over the network may be written similarly to \eqref{eq:3diag_eig_distr} as
\begin{equation}
\label{eq:Prob(p)_Wyner_def}
\pdist(p)
	= \sum_{r=1}^\infty \frac{\epr\pi_r}{1-\epr} \sum_{k=1}^r \one[p_{k}>p].
\end{equation}
The above expression is derived in a similar fashion as \eqref{eq:3diag_eig_distr}: We have taken into account the geometric distribution of segment lengths and have normalized over the average segment length $\epr^{-1}-1$. In addition, the second sum in the above expression corresponds to the possible positions $k=1,\cdots,r$ of the site located at the origin of the lattice within a segment of length $r$.

As we saw in the previous section, in the supercritical regime $\gamma>\gamma_c$, there is a finite probability that the system will be infeasible, so it only makes sense to analyze the distribution of powers for $\gamma\leq \gamma_c$.
To that end, we will first consider the critical \ac{SINR} target value $\gamma=\gamma_{c}$ with $p_k$ given by \eqref{eq:pk_critical_wyner}.

Obviously, if we focus on the tails of the distribution (i.e. for powers $p\gg \noisevar \gamma_c$), only the terms with sufficiently large $r$ will contribute to the sum \eqref{eq:Prob(p)_Wyner_def}:
in fact, since the maximum power for a segment of size $r$ is roughly $\noisevar \gamma_c r^{2}/4$, \eqref{eq:Prob(p)_Wyner_def} will only count the terms with $r > r_{c}(p)\equiv \sqrt{4p/(\noisevar \gamma_c)}$.
Hence, using the Euler-MacLauren formula \cite{Bender_Orszag_book} to replace sums by integrals, we get
\begin{equation}
\label{eq:Prob(p)_Wyner1}
\pdist(p)
	\sim \int_{r_{c}(p)}^{\infty} \epr^2 (1-\epr)^{r-1} \sqrt{r^{2} - r_{c}^{2}(p)} \dd r,
\end{equation}
where $r_{c}(p) = \sqrt{4p/(\noisevar \gamma_c)}$ and $\sqrt{r^{2} - r_{c}^{2}(p)}$ is the number of sites in a segment of length $r$ with power greater than $p$.
This yields
\begin{equation}
\label{eq:Prob(p)_Wyner2}
\pdist(p)
	\sim A \sqrt{r_{c}(p)} \, (1-\epr)^{r_{c}(p)}
\end{equation}
for some constant $A>0$ (independent of $p$), so the tails of the power distribution $\pdist(p)$ are again determined by the rare event of observing an erasure-free region of size exceeding $r_{c}(p)$.

The subcritical regime $\gamma<\gamma_c$ can be treated in the same way, the only difference being that the power in the system will always be bounded by $p_{\max} = z_{\gamma}^{-1} \noisevar$.
When $p_{\max}$ is small there is no point in discussing the tails of the distribution.
However, the situation becomes quite interesting in the near-critical regime $\gamma_{c}/\gamma-1\ll 1$ where powers $p\gg \noisevar\gamma_{c}$ are allowed.
As before, $p$ introduces a characteristic length $r_{\gamma}(p)$ which corresponds to the minimal segment supporting power equal to $p$ at its midpoint (i.e. the point of highest power in the segment);
then, by inverting \eqref{eq:pk_stable_wyner} for $k = (r+1)/2$, we obtain:
\begin{equation}
\label{eq:Lhatp_def}
r_{\gamma}(p)
	= \sqrt{\frac{2}{\gamma_c z_{\gamma}}}
	\cdot
	\log\frac{\noisevar +\sqrt{z_\gamma p(2\noisevar - z_\gamma p)}}{\noisevar - z_\gamma p},
\end{equation}
and hence:
\begin{equation}
\label{eq:Prob(p)_Wyner3}
\pdist(p)
	\sim A' \sqrt{r_{\gamma}(p)}
	\,(1-\epr)^{r_{\gamma}(p)}
\end{equation}
This formula is quite interesting, because the exponent $r_\gamma(p)$ interpolates between $r_\gamma(p)\sim \sqrt{4p/(\noisevar\gamma_c)}$ for $\noisevar/z_\gamma \ll p\gg \sigma^2\gamma_c$, and $r_\gamma(p)\sim \sqrt{2/(z_\gamma\gamma_c)} |\log(1-p/p_{\max})| \to +\infty$ when $p\to p_{\max}^-$.

\subsection{Bird's eye view of the Wyner model}

To sum up, it is worth pointing out here that the simple (but not simplistic) one-dimensional Wyner model carries all the qualitative properties of the more general models that we will encounter in the following sections.

On the one hand, power control is feasible for all $\epr\geq0$ when the users' \ac{SINR} target $\gamma$ is below the critical feasibility \ac{SINR} threshold $\gamma_{c}$ of the pure, ordered Wyner network ($\epr= 0$).
In this case, one obtains an explicit expression for the average power per node, simply by summing over the distribution of erasure-free segments.
On the other hand, in the supercritical regime $\gamma>\gamma_{c}$, the infinite Wyner network becomes infeasible almost surely;
nonetheless, networks of finite size exhibit a finite instability probability, and this probability becomes exponentially small when $\gamma\to\gamma_{c}^{-}$.
This instability is due to the occurrence of large, erasure-free regions, and the probability of this rare event is proportional to the integrated density of states evaluated at $\lambda=- z_\gamma=\gamma_c^{-1}-\gamma^{-1}$ (in fact, these rare, erasure-free regions are also responsible for the occurrence of atypically large powers in the optimal power vector).
In Sections \ref{sec:outage_probability} and \ref{sec:PowerDistribution}, we will see that these mechanisms are responsible for the instability and large power characteristics of more general networks as well.

%% file: CPA.tex
%----------------------------------------------------------------------
%%% CPA
%----------------------------------------------------------------------
% !TEX root = ./RandomPowerOptimization.tex

In this section, we will focus on the ``bulk'' characteristics of the network in the presence of randomness;
in particular, we will calculate the (intra-sample) average power per node and its variance by means of the so-called \acf{CPA} approach, an approximative methodology which has been applied extensively in the physics literature to study the movement of electrons in disordered alloys
\cite{Davies1963_CPA,
Elliott1974_CPA_RevModPhys,
Kroha1990_SelfConsistentTheoryAndersonLocalization}.
For simplicity, we will only show the intuition and the end results of the \ac{CPA} method here;
a more detailed discussion of the derivation will be given in Appendix \ref{app:cpa} where we also provide further pointers to the extensive literature on the \ac{CPA} method.

Importantly, even though \ac{CPA} is not an exact method, it has enjoyed considerable success in calculating the energy spectrum of systems with diagonal disorder, and its predictions become increasingly accurate when the number of connections between different sites increases.
It should also be mentioned that results obtained by the \ac{CPA} method turn out to be identical with those predicted in \cite{Tulino2007_GaussianErasureChannel, Tulino2010_GaussianErasureChannel} using tools and techniques from \acl{RMT} and free probability theory:
essentially, the self-energy $\Sigma$ that is the cornerstone of the \ac{CPA} method corresponds to the $R$-transform in \ac{RMT}, so \ac{CPA} may be viewed as an approximative way of applying \ac{RMT} methods.

To proceed, let $\bG_{V}$ be the Green's function operator (often called the \emph{resolvent} in \ac{RMT}) associated to the matrix $\bH_{V} = \bH_{0} + V\bE$ of \eqref{eq:H_def}, namely:
\begin{equation}
\label{eq:G_function_def}
\bG_{V}(\lambda)
	= \big[\lambda\bI-\bH_{V}\big]^{-1}.
\end{equation}
In this notation, the intra-sample average optimal power of the system becomes
\begin{equation}
\label{eq:p_avg_def_usingG}
p_{\avg} = -\frac{1}{N(1-\epr)}\bu^{\top} \bG_{V}(-z_\gamma) \bu,
\end{equation}
so we will calculate $p_{\avg}$ by taking the expectation $\ex[\bG_{V}(\lambda)]$ of $\bG_{V}$ over all realizations of $\bH_{V}$ and then letting $V\to\infty$.

The first implicit assumption of the \ac{CPA} method is that $p_{\avg}$ becomes deterministic in the large $N$ limit, i.e. $p_{\avg} \to \ex[p_{\avg}]$ as $N\to\infty$ (a.s.).
With this in mind, we will replace each random diagonal element of $V\bE$ in $\bH_{V}$ with a so-called ``self-energy'' term $\Sigma(\lambda)$ capturing the effects of all other sites in the network in a self-consistent fashion (see Appendix \ref{app:cpa} for a more detailed discussion of what ``self-consistency'' means here).
In other words, \ac{CPA} is essentially a ``mean-field'' solution to the problem where interactions across different sites are replaced by a ``mean field'' which measures the average effect of these interactions.

Apart from  these caveats, we are now in a position to state the \ac{CPA} equations (see Appendix \ref{app:cpa} for details on their derivation).
To begin with, the average Green's function operator in the \ac{CPA} regime will be:
\begin{equation}
\label{eq:G_fun_CPA}
\ex[\bG(\lambda)]
	= \bG_{0}(\lambda)\big[\bI - \Sigma(\lambda)\bG_{0}(\lambda)\big]^{-1}
\end{equation}
where
\begin{equation}
\label{eq:G0_fun_def}
\bG_0(\lambda) = \big[\lambda\bI - \bH_{0}\big]^{-1}.
\end{equation}
is the resolvent (Green's function) operator in the absence of randomness and
\begin{equation}
\label{eq:S_fun_CPA_eq}
\Sigma(\lambda)
	= \ex\left[ \frac{Ve_i}{1 - g(\lambda) [Ve_{i} - \Sigma(\lambda)]}\right]
	= \frac{\epr V}{1 - g(\lambda) [V-\Sigma(\lambda)]}
\end{equation}
is the system's \emph{self energy}.
Strictly speaking, this self energy corresponds to site $i$, hence the \ac{CPA} recipe requires only an averaging over the randomness of the given site.
The implicit assumption is that all other sites have been taken into account self-consistently and have been lumped into the diagonal element $g(\lambda)$ of $\ex[\bG_{V}(\lambda)]$ (see Appendix \ref{app:cpa}), given by
\begin{equation}
\label{eq:g_fun_CPA_def}
g(\lambda)
	= \frac{1}{N} \tr\bG_0(\lambda-\Sigma(\lambda))
	= \int \frac{d\bq}{(2\pi)^d}\frac{1}{\lambda-\Sigma(\lambda)-\eps(\bq)}.
\end{equation}

In this way, letting $V\to\infty$ in \eqref{eq:S_fun_CPA_eq} readily gives
\begin{equation}
\label{eq:Sigma_CPA2_eq}
\Sigma(\lambda)
%	= \frac{\epr V}{1-(V-\Sigma(\lambda))g(\lambda)}
	= -\frac{\epr}{g(\lambda)},
\end{equation}
and hence, for $\lambda = - z_{\gamma}$, \eqref{eq:g_fun_CPA_def} leads to the following implicit expression for the self-energy $\Sigma\equiv\Sigma(-z_{\gamma})$:
\begin{equation}
\label{eq:fixed_point_equation}
\epr
	= \int \frac{\Sigma}{\Sigma+\eps(\bq)+z_\gamma} \frac{\dd\bq}{(2\pi)^d}.
\end{equation}
The above equation can be solved numerically to yield $\Sigma(-z_\gamma)$.
Then, to evaluate the average optimal power we may use \eqref{eq:P_ave_Anderson_def} and the fact that $\bu$ is proportional to the $\bq=0$ eigenvector;
doing just that, we get:
\begin{equation}
\label{eq:p_ave_result}
p_{\avg}
	= \frac{\noisevar}{1-\epr}\frac{1}{\Sigma+z_\gamma}.
\end{equation}
Importantly, this equation is identical to the one derived in \cite{Moustakas2012_PowerOptimizationErasures} under the (false) conjecture that the matrices $\bE$ and $\bM$ are asymptotically free.
Moreover, it is easy to see that the above result reduces to $p_{\avg} = \noisevar/z_\gamma$ in the limit $\epr\to 0$:
for $z_\gamma>0$, the RHS of \eqref{eq:fixed_point_equation} vanishes only when $\Sigma=0$, so \eqref{eq:p_ave_result} yields $p_{\avg} = \noisevar/z_{\gamma}$.
Similarly, for $\epr\to 1$, \eqref{eq:fixed_point_equation} gives $(1-\epr)\Sigma\approx \gamma^{-1}$ leading to the non-interference value $p_{\avg} = \noisevar \gamma$.

Remarkably, the \ac{CPA} approach also allows us to describe the fluctuations of the optimal power vector from its average value.
Indeed, for large $N$, the (intra-sample) variance of the optimal power vector will be
\begin{equation}
\label{eq:VarP_def}
\var(\bp^{\ast})
	= \frac{1}{N(1-\epr)} \insum_{k} (p_{k} - p_{\avg})^2,
%	= \frac{1}{N(1-\epr)} \sum_{\bm\in\lattice} p_{\bm}^{2} - p_{\avg}^{2},
\end{equation}
so, by employing \eqref{eq:P_vec_eq_random}, we will have
\begin{equation}
\label{eq:p_var_as_a_derivative}
\insum_{k} p_{k}^{2}
	= \noisevard\bu^{\top}\big[z_\gamma \bI + \bH_{V}\big]^{-2} \bu
	= -N (1-\epr) \frac{dp_{\avg}}{d z_\gamma}.
\end{equation}
By differentiating \eqref{eq:p_ave_result} with respect to $z_\gamma$, we then obtain
\begin{flalign}
\label{eq:var_min_result}
\var(\bp^{\ast})
	&= \frac{\noisevard}{1-\epr}
	\frac{1}{\big(\Sigma+z_\gamma\big)^2}
	\frac
	{\int \frac{\Sigma}{\left(\Sigma+z_\gamma+\eps(\bq)\right)^2} \dd \bq}
	{\int\frac{\eps(\bq)+z_\gamma}{\left(\Sigma+\eps(\bq)+z_\gamma\right)^2} \dd \bq}
	\notag\\
	&- \frac{\epr\noisevard}{(1-\epr)^2} \frac{1}{(z_\gamma+\Sigma)^2},
\end{flalign}
with $\Sigma$ given by \eqref{eq:fixed_point_equation}.

\begin{figure*}[t]
\centering
\subfigure
[Average power in a $1$-dimensional network]
{\label{fig:p_ave_vs_sinr_pl4_n1_d05_num}
\includegraphics[width=.955\columnwidth]{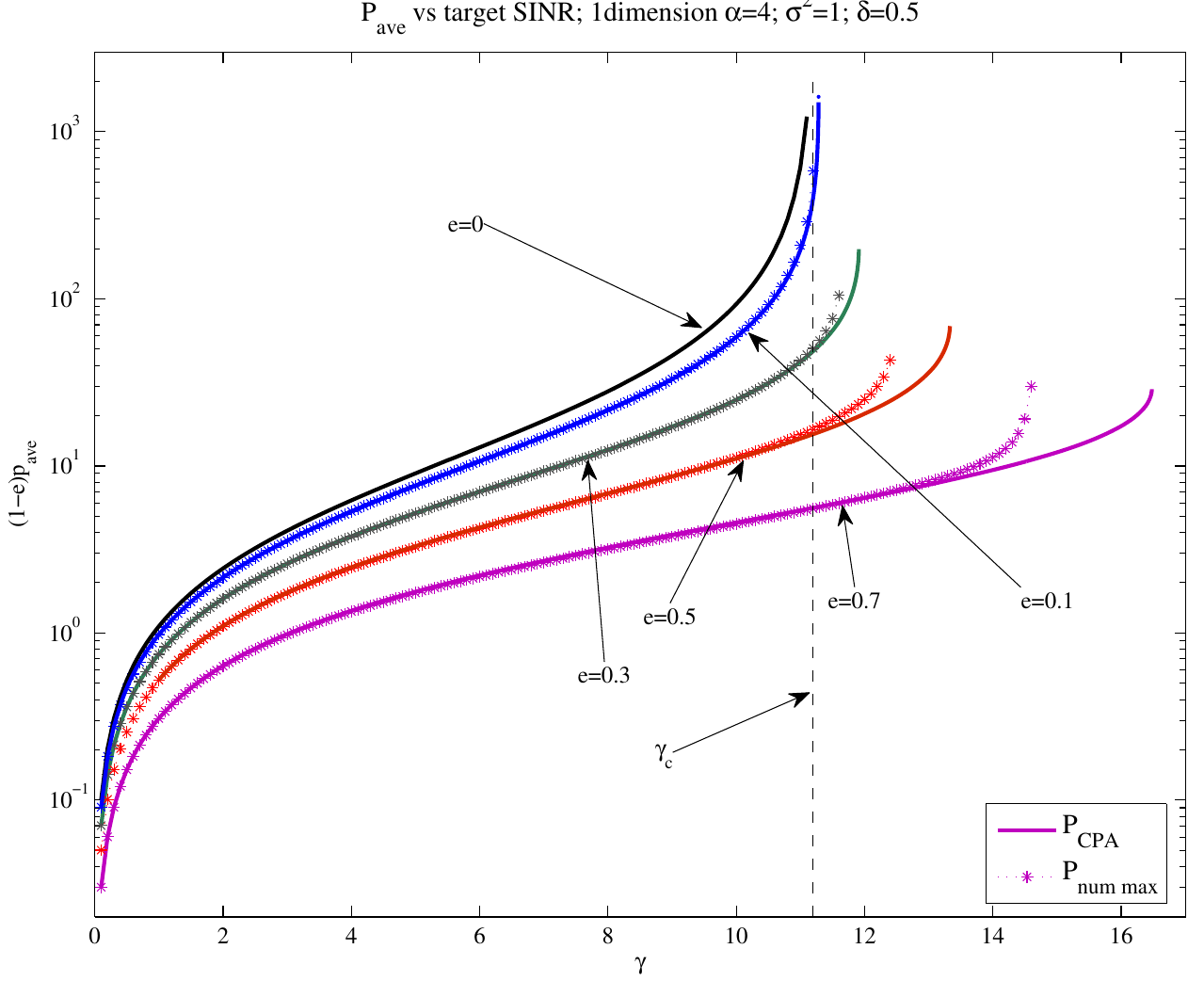}}
\hfill
\subfigure
[Average power in a $2$-dimensional network]
{\label{fig:p_ave_vs_sinr_2dim_pl5_n1_e030507holes}
\includegraphics[width=\columnwidth]{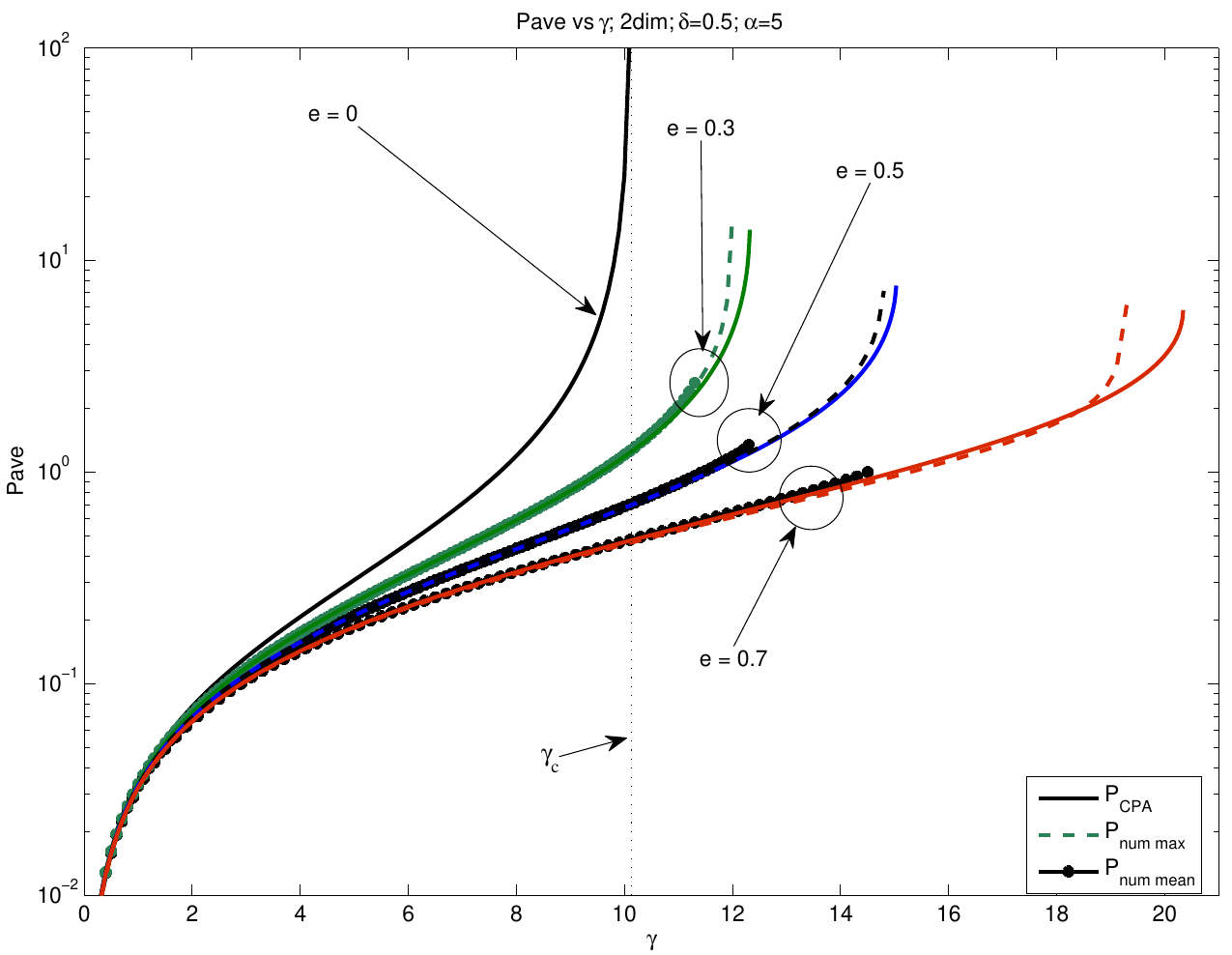}}
\caption{Plot of average power for a $1$- and $2$-dimensional network as a function of the target \ac{SINR} value $\gamma$ for different values of the erasure probability $\epr$.
In both cases, we took a pathloss exponent $\alpha = d+3$, noise level $\sigma=1$, and ratio $s = \delta/\ell =0.5$.
The solid curves represent our theoretical predictions, while the numerical datasets (starred and dashed curves) were generated from $10^{3}$ and $500$ random erasure instantiations in $1$- and $2$-dimensional networks respectively;
in the $1$-dimensional case we only plot the realization which has remained feasible the longest (indicated as ``num max''), while for $2$-dimensional networks we also plot an average over all realizations that remain feasible at any given $\gamma$.
In both cases, the vertical line corresponds to the critical threshold \ac{SINR} value $\gamma_{c}$ where the ordered network ($\epr=0$) becomes infeasible.}
\vspace{-1ex}
%one-dimensional chain of transmitters as a function of the target SINR for various values of $\epr$. The solid curves are analytically obtained, while the starred curves are numerically generated with $10^3$ independent realizations and correspond to the realizations which have remained feasible the longest possible. $\gamma_c$ is shown using a dashed vertical line. The path-loss exponent is set to $\alpha=4$, the noise level equal to $1$, and the ratio $s=\delta/\ell=0.5$ and the value of $\gamma_c$ is depicted with a dashed vertical line.}
%\label{fig:p_ave_vs_sinr_pl4_n1_d05_num}
\end{figure*}

\subsection{Numerical analysis and validation}
\label{sec:Analysis}

To study the accuracy of the \ac{CPA} appraoch, we will analyze here the validity of \eqref{eq:p_ave_result} and \eqref{eq:var_min_result} for the average optimal power and its variance via numerical simulations.
In Figs. \ref{fig:p_ave_vs_sinr_pl4_n1_d05_num} and \ref{fig:p_ave_vs_sinr_2dim_pl5_n1_e030507holes}, we plot $p_\avg$ for one and two dimensional systems respectively, as calculated from \eqref{eq:p_ave_result} and as obtained by generating instances for $\bE$ in \eqref{eq:P_ave_Anderson_def} versus the \ac{SINR} threshold $\gamma$.
As it turns out, the analytically calculated value of $p_{\avg}$ is finite not only for $\gamma<\gamma_{c}$,
but also for a range of \ac{SINR} target values $\gamma>\gamma_c$ for which the erasure-free network ($\epr=0$) is infeasible.
Nonetheless, \eqref{eq:Sigma_CPA2_eq} and \eqref{eq:fixed_point_equation} show that the \ac{CPA} solution cannot be extended indefinitely:
it eventually reaches a value of $\gamma$ beyond which the optimal power vector becomes infeasible.

\begin{figure}[htb]
\includegraphics[width=\columnwidth]{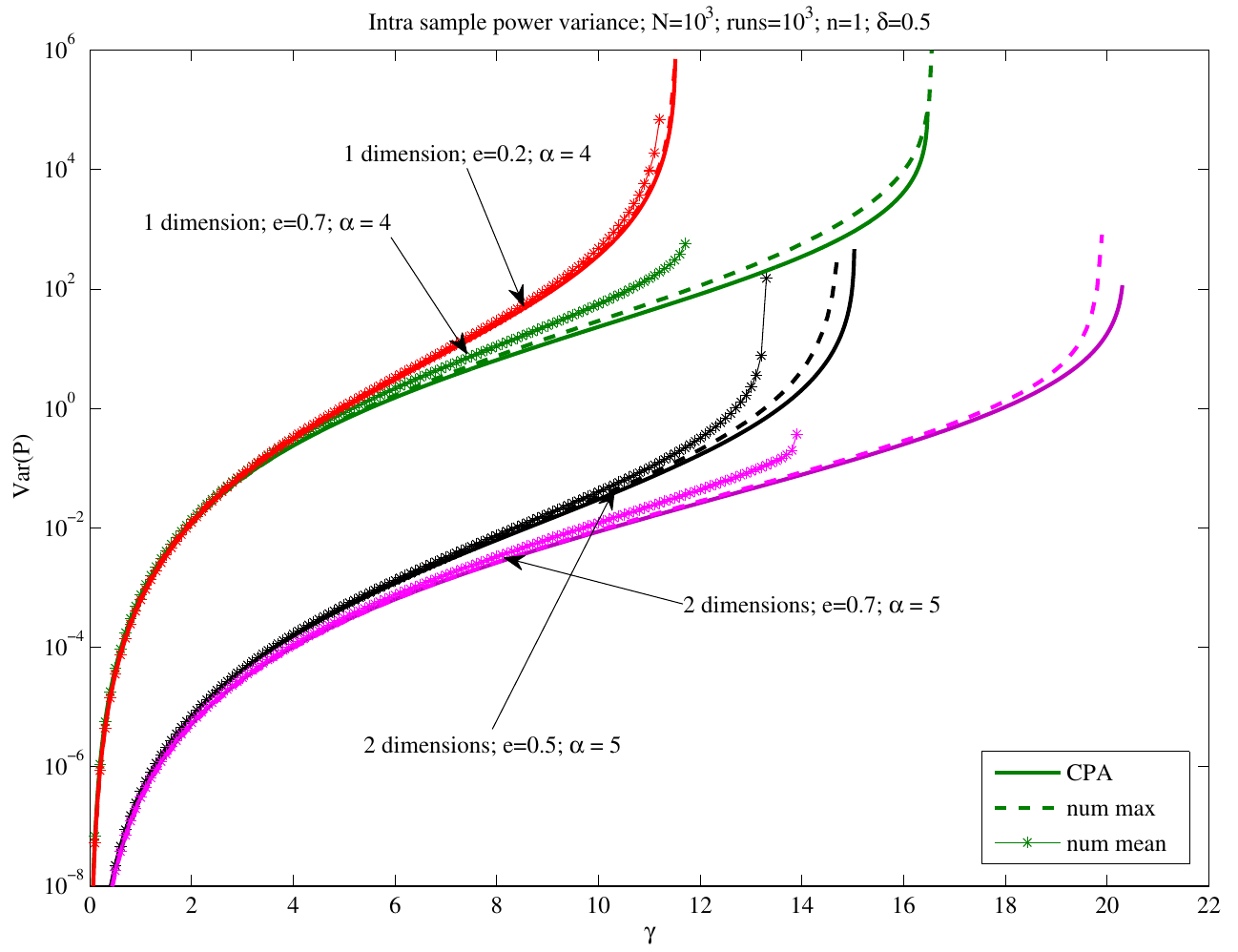}
\caption{Plot of the intra sample variance of the power for one- and two- dimensional networks of transmitters of size $N=10^3$, for various values of $\epr$. The solid curves are analytically obtained from \eqref{eq:var_min_result}. The meanings of the dashed and starred curves are the same as in Fig. \ref{fig:p_ave_vs_sinr_2dim_pl5_n1_e030507holes}} \label{fig:p_var_vs_sinr_dim12_num}
\end{figure}

The agreement between the \ac{CPA} solutions and the Monte Carlo data is excellent over a wide range of $\gamma$. Nevertheless, for $\gamma>\gamma_{c}$, the behavior of the simulated system becomes sample-dependent:
in particular, for any given realization of $\bE$, the graph of $p_{\avg}$ versus $\gamma$ follows the \ac{CPA} curve very closely until a certain random $\gamma > \gamma_{c}$ beyond which the two curves start to diverge, with the simulated network becoming infeasible soon after.
We illustrate this phenomenon from two different points of view in both Fig. \ref{fig:p_ave_vs_sinr_pl4_n1_d05_num} and Fig. \ref{fig:p_ave_vs_sinr_2dim_pl5_n1_e030507holes}. In Fig. \ref{fig:p_ave_vs_sinr_pl4_n1_d05_num}, and for each value of $\epr$, we plotted the curve $p_\avg$ vs. $\gamma$ that became infeasible at the \emph{largest} value of $\gamma$ from a sample of $10^3$ random realizations.
In Fig. \ref{fig:p_ave_vs_sinr_2dim_pl5_n1_e030507holes} we also plot the curve corresponding to the average value of $p_{\avg}$ over all realizations generated.
This last curve terminates at the minimum value of $\gamma$ at which some realization became infeasible.
Although both curves look identical, what is striking is the significant gap in the value of $\gamma$ where the first realization became infeasible, compared to the last.
The good agreement between numerics and \ac{CPA} appears also in the case of the variance \eqref{eq:var_min_result}, which is plotted for both one- and two-dimensional networks in Fig. \ref{fig:p_var_vs_sinr_dim12_num}.

\subsection{The breakdown of the \ac{CPA} approach}

Remarkably, even though the \ac{CPA} expressions agree with the numerically generated data when the simulated system is feasible, there exists a significant gap between the infeasibility threshold predicted by the \ac{CPA} approach and the largest value of $\gamma$ where the simulated system breaks down.
This is strongly reminiscent of our analysis of the Wyner network model in the previous section:
indeed, for $\gamma>\gamma_{c}$, the Wyner network becomes infeasible with finite probability, related to the minimum eigenvalue of $\bH$ becoming negative.
In other words, while the \emph{bulk} behavior of the system is captured remarkably well by the \ac{CPA} method, \emph{tail events} are not.

The aim of the following sections will be to highlight this \emph{tail} behavior;
for now, we will only give an intuitive explanation of why the \ac{CPA} and \ac{RMT} equations cannot be expected to obtain a result which remains valid for all values of $\gamma$.%
\footnote{A similar version of the \ac{CPA} equations was also derived in  \cite{Wegner1979_DisorderedSystemNOrbitalsCPA,Neu1995_FreeProbabilityApproachCPA, Khorunzhy1993_LimitsInfiniteRadiusDimensionalityOffDiagonalRandomOperators}.}
Indeed, \ac{RMT} typically addresses systems described by operators (or matrices) connecting all states in a random way:
in the context of matrices, this means that the randomness permeates the whole matrix, so every site experiences the same, average environment.
By contrast, randomness in our systems appears only in the diagonal elements of the matrix, and as it turns out, this is not ``enough'' to apply an approach based on a law of large numbers.
In particular, since each site is connected to a finite number of sites, it experiences an independent realization of the randomness and hence the behavior at different parts of the system will exhibit significant fluctuations;
as a result, it may be very misleading to replace a site's local environment with an average ``mean field'' quantity.%
\footnote{This only makes sense in the large $d$ limit discussed in\cite{Khorunzhy1993_LimitsInfiniteRadiusDimensionalityOffDiagonalRandomOperators}.}

This was first exemplified by Anderson \cite{Anderson1958_AbsenceDiffusionRandomLattices} who suggested that averages may often be spurious, while the distribution of rare events can be more important.
The significance of tail events already appeared in the instability analysis for the Wyner model in the previous section and it will be made clearer in the following sections where we go beyond the \ac{CPA} regime.

%% file: Outage.tex
%----------------------------------------------------------------------
%%% OUTAGE ANALYSIS
%----------------------------------------------------------------------
% !TEX root = ./RandomPowerOptimization.tex

In the previous section, the numerical validation of the \ac{CPA} results showed that while the \ac{CPA} equations match numerical results very closely in most realizations of the network, power control becomes infeasible well before the \ac{SINR} threshold predicted by the \ac{CPA} method.
In particular, when the simulated network is large, this instability occurs for some random, sample-dependent $\gamma>\gamma_{c}$.
In this section, we will analyze the probability of such an instability occurring:
we will see that power control is always infeasible for $\gamma>\gamma_{c}$ for infinite networks (Section \ref{sec:outage_probability_infinite_nets}), and we will also calculate the instability probability for finite networks (Section \ref{sec:outage_probability}).

\subsection{Feasibility and instability in infinite networks}
\label{sec:outage_probability_infinite_nets}

Corollary \ref{cor:feasibility_gamma<gamma_c} shows that the system is always feasible if $\gamma<\gamma_{c}$, i.e. for all $\epr\geq 0$ and for all $\latsize$.
In contrast, we will now show that the infinite system is always infeasible if $\gamma>\gamma_{c}$:

\begin{theorem}
\label{thm:Feasibility_infinite_system}
In the infinite system limit, power control is
%\begin{inparaenum}
%[\itshape a\upshape)]
%\item
feasible if $\gamma < \gamma_{c}$
and
%\item
infeasible if $\gamma > \gamma_{c}$ (a.s.).
%\end{inparaenum}
\end{theorem}

\begin{proof}
In view of Corollary \ref{cor:feasibility_gamma<gamma_c}, it suffices to consider the case $\gamma>\gamma_{c}$.
To that end, consider a finite network of edge length $\latsize$ and a cubic region with $M \ll \latsize$ sites per edge.
Initially, we will ignore the surroundings of the smaller region, which corresponds to setting all sites outside this region to zero.
Let $\mu_{\min}\equiv \mu_{\min}(M)$ be the minimum eigenvalue of $\bM$ in this smaller region.
Since $\bM$ is a T\oe plitz matrix, we will have
\begin{equation}
\mu_{\min}\geq z_\gamma=\gamma^{-1} -\gamma_c^{-1},
\end{equation}
where $\gamma_{c}$ is defined as in \eqref{eq:lambda0_def}, and the RHS corresponds to the minimum eigenvalue of $\bM$ in the limit $\latsize\to\infty$ \cite{Gray2005_ToeplitzMatrices,Widom1963_ExtremeEIgenvaluesConvolutionOperators}.
Now, with $\gamma>\gamma_{c}$ and $\lim_{M\to\infty} \mu_{\min} = \gamma^{-1} - \gamma_{c}^{-1} < 0$, there exists some $M_{\gamma}$ such that $\mu_{\min}(M) < 0$ for all $M\geq M_{\gamma}$.
This means that for $M\geq M_{\gamma}$, power control in this region is infeasible, as was to be shown.

Up to this point we have neglected the effects of neighboring sites outside the region in question.
However, since the power of each transmitter inside the smaller region will grow in the presence of other transmitters outside the region, it follows that power control will be infeasible in the $M$-sized erasure-free region for $M\geq M_{\gamma}$, even in the presence of outside transmitting powers.
As a result, if there exists an erasure-free region of size $M\geq M_{\gamma}$ the whole system will be itself infeasible.

Now, let $\mathcal{E}_{M}$ be the event that there are no erasures in any region of edge length $M$.
Clearly, any \emph{fixed} region of size $M$ will be erasure-free with probability $p_{M} = (1-\epr)^{M^{d}}$;
%Clearly, for a given region this event has probability of occurrence $p_L=(1-\epr)^{L^d}$.
as a result, the network's instability probability will be bounded from below by the probability of $\mathcal{E}_{M}$, i.e.
\begin{flalign}
\label{eq:prob_infeasibility}
P_{\mathrm{inst}}(\gamma)
	\geq \prob({\cal E}_M)
%	\notag\\
	\geq 1- \left(1-p_{M}\right)^{(\latsize/M)^{d}} \to 1
	\quad
	\text{as $\latsize\to\infty$.}
\end{flalign}
We conclude that power control in an infinite network is infeasible for any target \ac{SINR} which is larger than the critical \ac{SINR} threshold $\gamma_{c}$ corresponding to an erasure-free network.
\end{proof}
\begin{remark*}
It should be pointed out that the above analysis does not deal with the case $\gamma=\gamma_c$.
Of course, any \emph{finite} network with $\gamma=\gamma_c$ is feasible, because it has finite power even if it is completely devoid of erasures.
Furthermore, in the case of the Wyner model (Section \ref{sec:Wyner_model_power distribution}), we saw that even though the support of the optimal power vector is unbounded for $\gamma = \gamma_{c}$, the probability of observing an infinite power value is zero.
We conjecture that this holds in general, and we will prove this assertion for some representative cases in Section  \ref{sec:PowerDistribution}.
\end{remark*}

\subsection{Instability probability in finite networks: Lifshitz tails}
\label{sec:outage_probability}

The instability in the supercritical regime $\gamma>\gamma_{c}$ that was established in the previous section concerns only infinite networks.
In finite networks, the numerical simulations of Section \ref{sec:CPA} show that this instability is probabilistic in nature:
the average power is close to the one that was derived analytically using the \ac{CPA} method until the system becomes infeasible at a random, sample-dependent value of $\gamma>\gamma_{c}$.
In this section we will quantify the \emph{instability probability} $P_{\mathrm{inst}}(\gamma)$ for $\gamma>\gamma_c$ by building on the insights of Section \ref{sec:WynerModel} where we saw that the system becomes unstable when rare, large erasure-free regions occur.
In this way, we will show that instability events are always local in origin, and we will characterize the associated instability probability by relating the size of these regions to $\gamma$.

We begin by recalling the relationship \eqref{eq:thm:p_outage=prob_min} between infeasibility and the distribution of the minimum eigenvalue of the matrix $\bH_V$ of the network, i.e.
\begin{equation}
\label{eq:p_outage=prob_min_outagesection}
 P_{\mathrm{inst}}(\gamma)
 	=\prob\big[\lambda_{\min}(\bH_{V}^{\lattice})<-z_{\gamma}\big],
\end{equation}
where we emphasize the dependence on the size $\vert\lattice\vert = \latsize^{d} = N$ of the network by writing $\bH_{V}^{\lattice}$ instead of $\bH_{V}$.
Of course, for finite $V$, sites are not really erased in the network \textendash\ their power is simply reduced.
Thus, to obtain the instability probability for a network with bona fide erasures, we will need to take $V\to\infty$ in \eqref{eq:p_outage=prob_min_outagesection} or, equivalently (see Appendix \ref{app:lifshitz_proofs}), to apply Theorem \ref{thm:relation_Pout_P(Emin)} to the erasure model \eqref{eq:EME_def} and write:
\begin{equation}
\label{eq:p_outage=prob_min_outagesection2}
 P_{\mathrm{inst}}(\gamma)
 	=\prob\big[\lambda_{\min}(\bH^{\lattice}) < -z_{\gamma}\big],
\end{equation}
where, again, we write $\bH^{\lattice}$ instead of $\bH$ to emphasize the dependence on the size of the network.

Of course, $P_{\mathrm{inst}}(\gamma)$ will be positive only if $z_\gamma<0$, i.e. when $\gamma>\gamma_{c}$;
in that case, we need to look at the low-end part of the spectrum of $\bH^{\lattice}$ which we will study by means of its cumulative eigenvalue distribution.
Formally, for finite networks, we define the \emph{cumulative densities}
\begin{equation}
\begin{aligned}
\ids^{\lattice}(\lambda)
	&= \latsize^{-d} \vert\{\lambda'\in\spec(\bH^{\lattice}): 0<\lambda'\leq\lambda\}\vert,
	\\
\ids_{V}^{\lattice}(\lambda)
	&= \latsize^{-d} \vert\{\lambda'\in\spec(\bH_{V}^{\lattice}): 0<\lambda'\leq\lambda\}\vert,
\end{aligned}
\end{equation}
where $\spec(\cdot)$ denotes the spectrum of the matrices $\bH_{V}^{\lattice}$ and $\bH^{\lattice}$, defined in \eqref{eq:H_def} and \eqref{eq:EME_def}, respectively.
Then, in the infinite system limit, we will have
\begin{equation}
\label{eq:IDS_infte system}
\begin{aligned}
\ids(\lambda)
	&= \lim_{L\to\infty} \ids^{\lattice}(\lambda),
	\\
\ids_{V}(\lambda)
	&= \lim_{L\to\infty} \ids_{V}^{\lattice}(\lambda),
\end{aligned}
\end{equation}
with $\ids_{V} \to \ids$ as $V\to\infty$ (see App.~\ref{app:lifshitz_proofs} for a detailed discussion).

Of the above quantities, the object of interest is $\ids^{\lattice}$ for large (but finite) networks $\lattice$;
indeed, we have:

\begin{lemma}
\label{lem:P(Emin)<NN(E)}
Let $\lambda_{\min}(\bH^{\lattice})$ be the minimum eigenvalue of the matrix $\bH^{\lattice}$.
Then:
\begin{equation}
\label{eq:lem:prob_min_<NN(E)}
 \ex_{\omega}\big[\ids^{\lattice}(\lambda)\big]
 	\leq \prob\big[\lambda_{\min}(\bH^{\lattice})<\lambda\big]
	\leq N \ex_{\omega} \big[\ids^{\lattice}(\lambda)\big],
\end{equation}
where the expectation $\ex_{\omega}$ is taken over the realizations of the erasure matrix $\bE$ of \eqref{eq:P(e)_def}.
\end{lemma}

\begin{IEEEproof}
%The above result is quite general and holds for all matrices.
%Beginning with the rightmost part, we have
With $\prob\big[\lambda_{\min}(\bH^{\lattice} < \lambda)\big] = \prob\big[\ids^{\lattice}(\lambda) > 1/N\big]$, Markov's inequality readily yields:
\begin{equation}
\label{eq:lem:num_eigs<lambda}
\prob(\ids^{\lattice}(\lambda) > 1/N)
	\leq N \ex_{\omega} \big[ \ids^{\lattice}(\lambda)\big].
\end{equation}
For the leftmost inequality, a second application of Markov's inequality then gives
\begin{equation}
\label{eq:lem:Markov_ineq2}
\prob\big[\ids^{\lattice}(\lambda)=0\big]
	\leq 1-\ex_{\omega} \left[\ids^{\lattice}(\lambda)\right],
\end{equation}
and our claim follows by noting that $\prob\big[\lambda_{\min}(\bH^{\lattice})<\lambda\big] = 1 - \prob\big[\ids^{\lattice}(\lambda)=0\big]$.
%\PMcomment{Why X? Couldn't we pick something that won't conflict with the random walk notation?}
%\begin{equation}
%\label{eq:lem:num_eigs<lambda}
%X_\lambda = \vert\{\lambda'\in\spec(\bH_V^{\lattice}): 0<\lambda'\leq\lambda\}\vert,
%\end{equation}
%as the number of eigenvalues of a given realization of $\bH_V^\lattice$ which are less than $\lambda$. Then $\prob\big[E_{\min,N}(\bH_{V})<\lambda\big]=\prob(X_\lambda\geq 1)$ and from Markov's inequality we obtain
%\begin{equation}\label{eq:lem:Markov_ineq1}
%\prob(X_\lambda\geq 1) \leq \ex\left[X_\lambda\right]=N\ex\left[\ids_V^{\lattice}(\lambda)\right]
%\end{equation}
%To prove the left inequality we once again apply Markov's inequality resulting to
%\begin{equation}
%\label{eq:lem:Markov_ineq2}
%\prob(X_\lambda=0)
%	=\prob(N-X_\lambda\geq N)\leq 1-\ex\left[\ids_V^{\lattice}(\lambda)\right].
%\end{equation}
\end{IEEEproof}

\begin{remark*}
The above inequalities provide bounds for $P_{\mathrm{inst}}(\gamma)$ in terms of the averaged \acl{IDS} $\ex_{\omega}\big[\ids^{\lattice}\big]$ of $\bH^{\lattice}$ evaluated at $\lambda=-z_\gamma$.
At first sight, these inequalities seem quite loose:
indeed, for large $N$ and \emph{fixed} $\lambda$, the RHS of \eqref{eq:lem:prob_min_<NN(E)} may exceed $1$, so the rightmost inequality becomes trivial.
Nevertheless, we will be interested in the case where $N$  and $\lambda$ are such that $N\ex_\omega[\ids^\lattice(\lambda)] \ll 1$, and we will argue at the end of the section that the rightmost inequality of \eqref{eq:lem:prob_min_<NN(E)} becomes tight in this case.
Hence, for large but finite $N$, the instability probability will be proportional to $N\ids^\lattice(-z_\gamma) \sim N\ids(-z_\gamma)$ with the proportionality constant depending only on $\lambda$.
\end{remark*}

In light of the above, we are left to calculate $\ids^\lattice(\lambda)$ for large $\lattice$, a quantity which we will approximate with $\ids_{V}(\lambda)$ for large $V$ (see Appendix \ref{app:lifshitz_proofs} for a justification of this approximation).
This last quantity has a long history in statistical physics:
in his study of the electronic properties of dirty semiconductors, Lifshitz conjectured the correct form of the density of eigenvalues close to the edge of the spectrum using a truly insightful argument based on the size of regions that are free of impurities \cite{Lifshitz1964_Tails}.
Subsequently, a large corpus of sophisticated mathematical techniques has provided a formal footing for the method (see e.g. \cite{Friedberg1975_DOSDisorderedSystems, Kac1974_BECPresenceImpurities, Pastur1978_WienerIntegralsDOSRandomPotentials, Donsker1975_AsymptoticsWienerSausage, Simon1985_LifshitzTailsAndersonModel, Kirsch2007_IDOS_RandomSchroedingerOperators} and references therein),
and our instability analysis follows from applying these techniques to our random network model with erasures viewed as impurities:
%Our main outage analysis results (Theorems \ref{thm:Lifshitz_tails} and \ref{thm:Lifshitz_tails_a<d+2} below) follow from the application of these methods to the random network model analyzed in this paper with erasures corresponding to impurities.
%Without further ado, we have:

\begin{theorem}
%[Lifshitz Tails  for $\alpha>d+2$]
\label{thm:Lifshitz_tails}
Let $\ids(\lambda)$ be the \acl{IDS} of the Hamiltonian matrix $\bH$ of the random network model \eqref{eq:EME_def}.
Then
\begin{equation}
\label{eq:thmLifsthitztails}
\lim_{\lambda\to 0^{+}}
	\lambda^{d/(\alpha_{\eff} - d)} \log \ids(\lambda)
	= \log(1-\epr) \left(t_{\eff}\eigmin\right)^{d/(\alpha_{\eff} - d)}
\end{equation}
or, equivalently:
\begin{equation}
\label{eq:thmLifshitztails}
\log \ids(\lambda)
	\sim \log(1-\epr) \left(\frac{t_{\eff}\eigmin}{\lambda}\right)^{d/(\alpha_{\eff} - d)}
\end{equation}
where
\begin{enumerate}
[\itshape a\upshape)]
\item
$d$ is the dimensionality of the network;

\item
$\epr$ is the erasure probability;

\item
$\alpha_{\eff} = \min\{\alpha, d+2\}$ denotes the system's effective pathloss exponent as given by \eqref{eq:alpha-eff};

\item
the leading order coefficient $t_{\eff}$ is given by \eqref{eq:t-eff};

\item
the quantity $\eigmin\equiv\eigmin(\alpha,d)$ is defined as
\begin{equation}
\label{eq:thm_min_eig def}
\eigmin = \inf\{ \lambda_{\min}(\dom): \dom\subseteq\R^{d},\,\mathrm{vol}(\dom) = 1\},
\end{equation}
where
%depending on $\alpha$ and $d$,
$\lambda_{\min}(D)$ is the lowest Dirichlet eigenvalue (over $\dom\subseteq\R^{d}$) of the linear operator:
\begin{equation}
\label{eq:generator}
\gen
%	= -\left\vert -\tfrac{1}{2} \nabla^{2} \right\vert^{(\alpha_{\eff} - d)/2}
	=
	\begin{cases}
	-\nabla^{2}
	&\text{if $\alpha \geq d+2$,}
	\\
	\left( - \nabla^{2} \right)^{(\alpha - d)/2}
	&\text{if $d < \alpha < d+2$},
	\end{cases}
\end{equation}
i.e. the infinitesimal generator of a standard Brownian motion on $\R^{d}$ for $\alpha \geq d+2$,
or of a symmetric stable process of order $\alpha-d$ for $\alpha\in(d,d+2)$.
In particular, for $\alpha > d+2$, we will have:
\begin{equation}
\label{eq:eigmin}
\eigmin
	=
	\begin{cases}
	\pi^{2}
	&\text{for $d=1$,}
	\\
	\pi k_{0}^{2}
	&\text{for $d=2$,}
	\end{cases}
\end{equation}
where $k_{0} \approx 2.4048$ is the first zero of the $0$-th order Bessel function $J_{0}(x)$.
\end{enumerate}
\end{theorem}

The convergence of $\ids^{\lattice}$ to $\ids$ then gives:
\begin{corollary}
With notation as in Theorem \ref{thm:Lifshitz_tails}, the integrated density of states in a random network of size $\vert\lattice\vert = N$ satisfies
\begin{equation}
\log\ids^{\lattice}(\lambda)
	\sim \log (1-\epr) \left(\frac{t_{\eff} \eigmin}{\lambda}\right)^{d/(\alpha_{\eff} - d)}
	\quad
	\text{for small $\lambda$},
\end{equation}
with probability approaching one as $N\to\infty$.
\end{corollary}

\begin{remark*}
Just as the Laplacian operator $-\nabla^{2}$ is the infinitesimal generator of a standard Brownian motion in $\R^{d}$, the operator denoted as $\left( - \nabla^{2} \right)^{(\alpha - d)/2}$ is the infinitesimal generator of a $d$-dimensional symmetric stable process of degree $\alpha-d<2$ \cite{Banuelos2010_SymmetrizationLevyProcesses}. Despite its similarity with the Laplacian, it is not a local operator and can be expressed equivalently as \cite{Donsker1979_NumberDistinctSitesRW} (see also Appendix \ref{app:cont_approx})
\begin{equation}
\label{eq:generator_sym_stable_proc}
\gen \phi = - \int_{\dom} \frac{\phi(\bx+\bh) - 2 \phi(\bx) + \phi(\bx-\bh)}{\vert \bh \vert^{\alpha}}
	\dd \bh,
\end{equation}
\end{remark*}

The proof of Theorem \ref{thm:Lifshitz_tails} is quite technical, so we defer it to Appendix \ref{app:lifshitz_proofs};
instead, in the remainder of this section, we will provide a qualitative analysis based on Lifshitz's original approach and the related analysis of Section \ref{sec:WynerModel} for $\alpha>d+2$.
Lifshitz's key insight was to realize that very low eigenvalues close to the minimum of the spectrum become exceedingly rare because they correspond to large regions without impurities (erasures) \textendash\
this is so because erasures create kinks in the corresponding eigenfunctions, and these tend to increase the eigenvalue.
In this way, the measure of eigenvalues below a given low eigenvalue $\lambda$ becomes dominated by the probability of having an erasure-free region $\dom(\lambda)$ in the system such that $\lambda$ is the minimum eigenvalue in $\dom(\lambda)$, i.e.
\begin{equation}
\label{eq:N_E_sim_1-e^V}
\ids(\lambda) \sim \left(1-\epr\right)^{\vert\dom(\lambda)\vert},
\end{equation}
where the dependence of $\vert\dom(\lambda)\vert$ on $\lambda$ is to be determined.

At the boundary of $\dom(\lambda)$, the corresponding eigenfunction vanishes due to the appearance of erasures, so the eigenvalues within this region can be evaluated by diagonalizing $\bH_{0}$ in $\dom(\lambda)$.
From \eqref{eq:low_q_eig_a>d+2}, we know that the eigenvalues of $\bH_{0}$ close to the minimum one will be
\begin{equation}
\label{eq: low_q_eig_a>d+2}
\eps(\bq) \sim t_2 |\bq|^2.
\end{equation}
Hence, by dimensional analysis, the value of $|\bq|$ for the minimum eigenvalue must be proportional to the inverse $R^{-1}$ of the (effective) radius of $\dom(\lambda)$, implying that $\lambda$ scales as $R^{-2}$.%
\footnote{The exact meaning of $R$ will become apparent later, but for simplicity we take $R$ to be the only characteristic lengthscale of the domain $\dom(\lambda)$.}

This conclusion can be reached independently by noting that the discrete operator $\bH_{0}$ can be approximated for large $\vert\dom\vert$ by the Laplacian;
indeed, for any $\bm\in\dom$, we will have:
\begin{equation}
\label{eq:H0=Laplacian}
\be_\bm^{\top}  \bH_{0} \bp
	= \insum_{\bm'\in \lattice} p_{\bm'} \be_\bm^{\top}  \bH_{0} \be_{\bm'}
	\sim -\frac{t_2}{R^{2}} \Delta \hat{p}(\bx)
\end{equation}
where $\be_\bm^{\top}  \bH_{0} \be_{\bm'}$ denotes the $(\bm,\bm')$-th element of $\bH_{0}$, $p_{\bm}$ stands for the power at transmitter located at  $\bm$ in $\lattice$, and $\hat{p}(\bm/L) = p_{\bm}$
(for more details about this continuum approximation, see Appendix {\ref{app:cont_approx}}).
We thus obtain
\begin{equation}
\label{eq:volume_vs_lambda}
|\dom(\lambda)|\propto \left(t_{2}/\lambda\right)^{d/2},
\end{equation}
with the proportionality constant depending on the shape of $\dom$.
Thus, in order to obtain the maximum of \eqref{eq:N_E_sim_1-e^V}, we need to minimize this constant.

This can be accomplished by means of the well-known Rayleigh\textendash Faber\textendash Krahn inequality \cite{Bandle1980_IsoperimetricInequalitiesApplications}, which states that the lowest Dirichlet eigenvalue of the Laplacian over a domain with fixed volume is minimized when the domain is a $d$-dimensional ball;
equivalently, for a fixed value of $\lambda$, this isoperimetric principle implies that the minimal erasure-free domain (and hence the most probable one) will be a $d$-dimensional ball.
The relationship between the minimum eigenvalue and the radius of this ball can then be evaluated by solving the eigenvalue problem $-t_{2} \nabla^{2} \phi = \lambda \phi$ with Dirichlet boundary conditions $\phi|_{\pd\dom}=0$.
By doing just that, we obtain:
\begin{equation}
\label{eq:min_energy_a>d+2}
\lambda = t_{2} \frac{\eigmin}{R^{2}}
\end{equation}
with $\eigmin$ given by \eqref{eq:eigmin} \cite{Jackson_EM_book}.
In this way, Lifshitz was able to obtain the following asymptotic expression for the cumulative density of eigenvalues (correct to exponential accuracy):
\begin{equation}
\label{eq:IDOS_Lif_a>d+2}
\ids(\lambda)
	\sim (1-\epr)^{\vert\dom(\lambda)\vert}
	=(1-\epr)^{\left(t_2\eigmin/\lambda\right)^{d/2}}
\end{equation}
This result coincides with \eqref{eq:thmLifsthitztails} for $\alpha>d+2$;
by contrast, it is worth recalling that the cumulative density of eigenvalues for the pure system vanishes asymptotically as $\ids(\lambda)\sim \lambda^{d/2}$ \textendash\ cf. \eqref{eq:low_q_t2}.

\setcounter{remark}{0}

\begin{remark}
To illustrate the exponential sensitivity of the above result to the occurrence of even a small number of erasures in the domain $\dom(\lambda)$, it is helpful to revisit the one-dimensional case of the Wyner model and estimate the probability of occurrence of the eigenvalue $\lambda$.
In the absence of erasures the minimum eigenvalue of a segment of length $R$ is $t_2 \eigmin/R^2$.
The appearance of even a single erasure, for simplicity in the center of the segment, increases the eigenvalue of this region to roughly $8 t_2 \eigmin/R^2$.
Hence, such an event with approximately the same probability will contribute to the eigenvalue density at a much higher value, where a region of size $R/\sqrt{8}$ which exponentially more probable.
Hence, such events with few erasures inside the region of interest are negligible.
\end{remark}

\begin{remark}
We can use the Wyner model to also show why $\ids(\lambda)$ is dominated (to leading exponential order) by the occurrence of an erasure-free disc with minimum eigenvalue $\lambda$ rather than its higher eigenvalues.
As we saw above, one way that this eigenvalue can occur is when  an erasure free region of length $R$ appears, where $\lambda=t_2\eigmin/R^2$.
This event occurs with probability of the order of $(1-\epr)^{R}$.
However,  $\lambda$ can also occur in a size $R'$ of the erasure-free region as the \emph{second} lowest eigenvalue such that $\lambda \sim 4 t_{2} \eigmin/R'^{2}$.
This means that $R' \sim 2 R$, so the probability of $\lambda$ occurring as the second lowest eigenvalue is exponentially small compared to the case where $\lambda$ is the lowest eigenvalue.
\end{remark}

\begin{remark}[Accuracy of the IDS approximation]
An important byproduct of this analysis is that in the low eigenvalue regime, an eigenvalue $\lambda$ appears only when an erasure-free region with volume roughly equal to $\vert\dom(\lambda)\vert$ occurs in the sample (recall that $\dom(\lambda)$ is such that the minimum eigenvalue of the Laplacian over $\dom(\lambda)$ is $\lambda/t_2$).
Also, since the eigenfunction of such an eigenvalue is localized within $\dom(\lambda)$, it will not depend on the random disorder beyond this region.
Therefore, since such erasure-free regions appear randomly and independently in the system, we may estimate the probability $\prob(\lambda_{\min}<\lambda)$ in \eqref{eq:p_outage=prob_min_outagesection} by assuming that there are $\bigoh(N/\vert\dom(\lambda)\vert)$ independent regions in the system, in each of which the probability that $\lambda$ appears is of the order of $(1-\epr)^{\vert \dom(\lambda)\vert}\sim\ids(\lambda)$.
As a result,
\begin{flalign}
\label{eq:bound_prob_min3}
\prob(\lambda_{\min}<\lambda)
	&\sim 1- \big(1-\ids(\lambda)\big)^{\frac{N}{\vert \dom(\lambda)\vert}}
	\notag\\
	&\sim A(\lambda) N\ids(\lambda),
\end{flalign}
where $A(\lambda)$ is a power-law function of $\lambda$, which does not depend on $N$.
As a result, when $N\ids(\lambda) \ll 1$, we conclude that
\begin{equation}
\label{eq:P_min=N(E)}
P_{\textrm{inst}}(\gamma)
	=\prob(\lambda_{\min} < -z_\gamma)
	\sim N \ids(-z_\gamma).
\end{equation}
corroborating the tightness of the upper bound in \eqref{eq:lem:prob_min_<NN(E)}.
\end{remark}

\subsection{Numerical validation in finite networks}
\label{sec:NumericsOutage}

We now  turn to the numerical validation of our stability analysis for finite networks.
As discussed above, the instability probability corresponds to the probability that the minimum eigenvalue of the system is less than $-z_\gamma>0$ (Theorem \ref{thm:relation_Pout_P(Emin)}).
To obtain a better comparison with our theoretical predictions, it will be convenient to introduce the parameter
%we calculate the outage probability numerically not in terms of $\gamma$ (as in Section \ref{sec:Analysis}) but in terms of the functional form of $|z_\gamma|$ which appears in the exponent of the tails of $\ids(\lambda)$,
\begin{equation}
y = -\log(1-\epr) \left(\frac{t_{\eff} \eigmin}{\vert z_{\gamma}\vert}\right)^{d/(\alpha_{\eff}-d)},
\end{equation}
which corresponds to the function of $z_{\gamma}$ that appears in \eqref{eq:thmLifsthitztails}.
Thus, for our numerical simulations to be consistent with Theorem \ref{thm:Lifshitz_tails}, the plots of $\log P_{\mathrm{inst}}$ against $y$ for different values of $\epr$ must be concentrated around parallel lines with negative unit slope (the axis intercept is irrelevant).

\begin{figure}[htb]
\includegraphics[width=1\columnwidth]{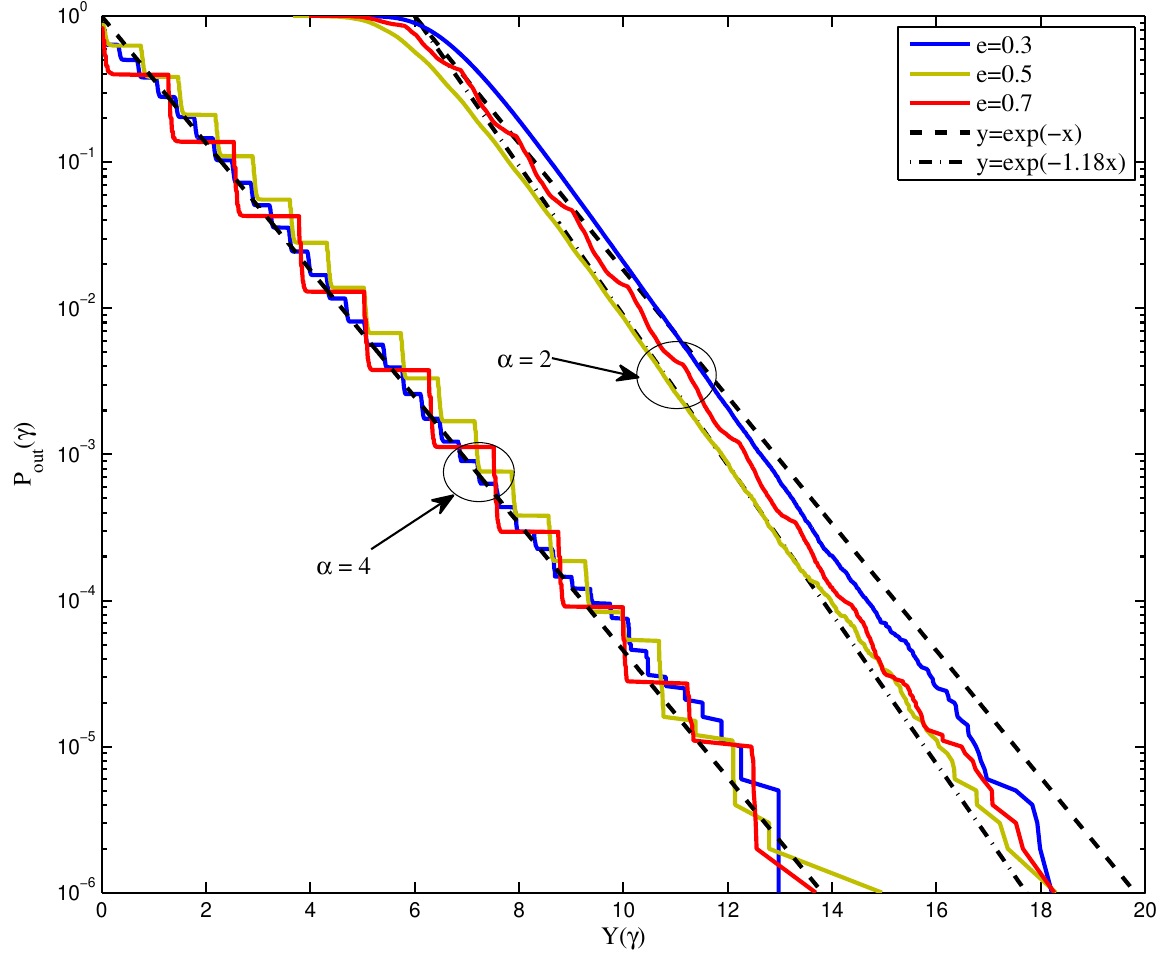}
\caption{Plot of the instability probability for a one dimensional chain versus the parameter $-\ln(1-\epr)\sqrt{t_{2}\pi^{2}/|z_\gamma|}$ for the case $\alpha=4$ and versus $y=-\ln(1-\epr)\frac{t_{eff} \eps_{0,\ell}}{|z_\gamma|}$ for $\alpha=2$, where $\eps_{0,\ell}=2$ is the lower bound for the corresponding minimum eigenvalue as discussed in Theorem \ref{thm:Lifshitz_tails}.
The curves have been shifted appropriately to allow for easy comparison.
Despite the different values of the parameters for each curve, the slope is identical for all cases with same $\alpha$.
In the case of $\alpha=2$, two straight lines have been draw for comparison, corresponding to the lower and upper bounds of $\eigmin$, i.e. $\eps_{0,h}=2.356$ and $\eps_{0,\ell}=2$ \cite{Banuelos2004_CauchyProcessSteklovProblem}.}
\label{fig:pc_1dim_outage}
\end{figure}

Fig.~\ref{fig:pc_1dim_outage} presents our simulations for one-dimensional networks and demonstrates remarkable agreement with Theorem \ref{thm:Lifshitz_tails}.
Just as in the case of the Wyner model (Fig.~\ref{fig:asympt_cdf_3diag}), the jump discontinuities that appear in the numerically calculated \ac{IDS} are due to the fact that the cumulative eigenvalue density of the system is not H\"older continuous to any order in the limit $V\to\infty$ \cite{Kirsch2007_IDOS_RandomSchroedingerOperators, Halperin1967_PropertiesParticle1DRandomPotential}.
Finally, the plots corresponding to the long-range interaction regime $\alpha=2$ also show excellent agreement with our theoretical predictions.%
\footnote{The exact value of $\eigmin$ for $d<\alpha<d+2$ has not been calculated analytically, but is known to lie between $\eps_{0,\ell}=2$ and $\eps_{0,h}=3\pi/4 \approx 2.356$ \cite{Banuelos2004_CauchyProcessSteklovProblem}.}

\begin{figure}[htb]
\includegraphics[width=\columnwidth]{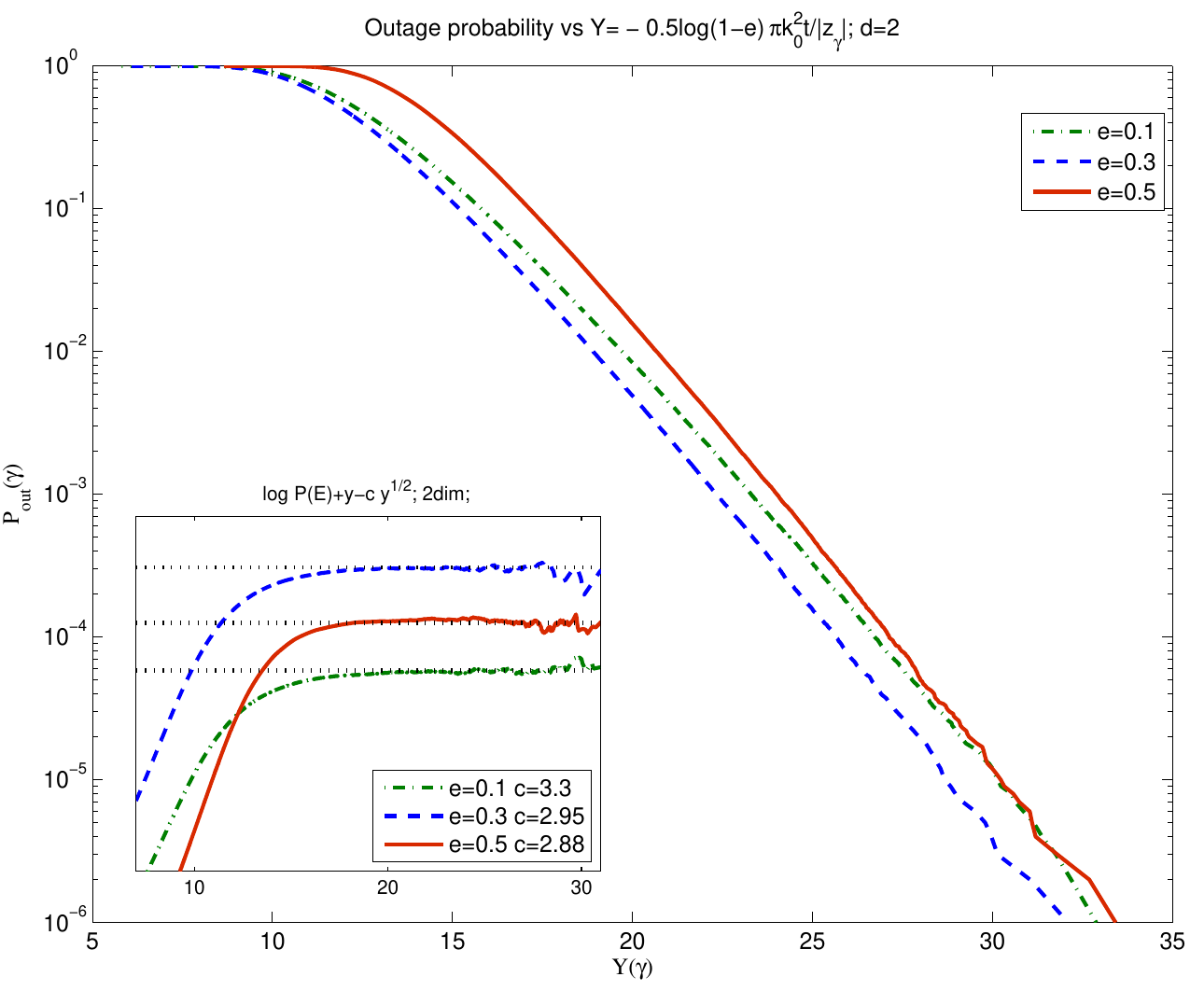}
\caption{Plot of probability of instability for two dimensional systems versus the parameter $Y(\gamma)=-\log(1-\epr)\frac{\pi t_2 k_0^2}{|z_\gamma|}$ for nearest neighbor interference. In two dimensions, the slope is not easy to extract directly from the plot, due to the sizable $\bigoh(Y^{1/2})$ correction in the exponent, as argued in the text. Instead, one can fit the curve including this term. This is done in the inset, which includes the plots of $\log(P_{inst}(\gamma))+Y(\gamma)-c Y(\gamma)^{1/2}$ for fitted values of $c$.} \label{fig:pc_2dim_outage}
\end{figure}

Fig.~\ref{fig:pc_2dim_outage} presents our simulations for $2$-dimensional networks for three different values of the erasure probability $\epr$.
For simplicity, we only simulated the case where only nearest neighbors interfere each other.
In this case, although the plots look straight, the convergence to the theoretical exponent is not so obvious.
One important reason is that the rare regions of interest are now $2$-dimensional and hence susceptible to shape fluctuations that can be significant when the radii are not sufficiently large.
In fact, based on the analysis of \cite{Bolthausen1994_Localization2dimRW}, these surface fluctuations introduce a subleading correction in the exponent of the cumulative density of eigenvalues which is of order $\bigoh(R(\lambda)^{d-1})$, where $R(\lambda)$ is given in \eqref{eq:min_energy_a>d+2}, i.e.
\begin{equation}
\label{eq:Laplace2_correction}
\ids_V(\lambda)
	\sim e^{-\frac{t_2 \eigmin}{R(\lambda)^2}} \left(1-\epr\right)^{R(\lambda)^d} e^{cR(\lambda)^{d-1}}
\end{equation}
for some constant $c>0$.
Importantly, this last term does not appear for $d=1$;
on the other hand, for $d=2$, it introduces a subleading correction of order $\bigoh(\lambda^{-1/2})$ in \eqref{eq:thmLifsthitztails} which can be significant if $\lambda$ is not sufficiently small.
In the inset of Fig. \ref{fig:pc_2dim_outage} we have subtracted such a term from the exponent and fitted the coefficient $c$, obtaining asymptotic convergence to our theoretical predictions for small $\lambda$.

%\begin{figure}[htb]
%\includegraphics[width=\columnwidth]{pc_2dim_outage_N100e20r5a100.pdf}
%\caption{Plot of probability of outage for two dimensional systems versus the parameter $-\ln(1-\epr)\frac{\pi t_2 k_0^2}{2 |z_\gamma|}$. Case 3} \label{fig:pc_2dim_outage2}
%\end{figure}
%
%\begin{figure}[htb]
%\includegraphics[width=\columnwidth]{pc_2dim_outage_N40e50r5a100.pdf}
%\caption{Outage probability for a 2\textendash dimensional system versus $-\log(1-\epr)\frac{\pi t_2 k_0^2}{2 |z_\gamma|}$. } \label{fig:pc_2dim_outage}
%\end{figure}

%% file: PowerDistribution.tex
%----------------------------------------------------------------------
%%% POWER DISTRIBUTION
%----------------------------------------------------------------------
% !TEX root = ./RandomPowerOptimization.tex

Having analyzed the instability probability for finite random networks generated by \eqref{eq:EME_def} in the supercritical regime $\gamma>\gamma_c$, we now turn to the tails of the power distribution for $\gamma \leq \gamma_{c}$.
This analysis is complementary to that of Section \ref{sec:CPA} where we calculated the ``bulk'' statistics of the optimal power vector;
indeed, the importance of analyzing the tails of the power distribution lies in that they serve as an alternative outage criterion:
since the available power at each transmitter is bounded, transmitters with assigned powers higher than their maximum power will effectively be in outage.

In this section, we will present a lower bound for the tails of the power distribution using the intuitive approach of the previous section, and we will argue that this lower bound becomes tight for large powers \textendash\ an assertion backed by our numerical simulations and the discussion of the Wyner model in Section \ref{sec:WynerModel}.
On the other hand, establishing an upper bound is significantly more difficult:
using arguments from percolation theory, we obtain a tight upper bound for the power distribution when interference is only caused by nearest neighbors and the erasure probability $\epr$ exceeds a critical value $\epr_{c}$ derived from an associated bond percolation model.
This approach however does not apply when the erasure probability is low, leaving a gap between the lower and upper bounds in this regime.
Nevertheless, we conjecture that the scaling obtained through the lower bound is tight:
in fact, as has been emphasized by Pastur for the case of the integrated density of states, the lower bounds obtained with our methodology capture the correct behavior in all known cases \cite{Kirsch2007_IDOS_RandomSchroedingerOperators}.

\subsection{A lower bound for the tails of the power distribution}

We will begin by presenting a lower bound for the tails of the power distribution in the short-range regime $\alpha \geq d+2$.
To that end, recall that the fraction $\pdist(p)$ of sites with power exceeding some value $p$ in a large network may be seen as the probability that the optimal power $p_{0}$ at the origin exceeds $p$, i.e.
\begin{equation}
\label{eq:pdist}
\pdist(p)
	= \prob(p_{0} > p).
\end{equation}
%Clearly, since the random matrix $\bE$ of \eqref{eq:EME_def} is spatially homogeneous, $0$ could be replaced by any other site in $\Z^{d}$, so $\pdist(p)$ may be interpreted as the fraction of sites with optimal power exceeding $p$.
%To estimate this probability, we will follow a line of reasoning similar to that of the previous section.
We will thus say that a connected domain $\dom \subseteq \Z^{d}$ \emph{supports power $p$ at $0$} when
\begin{inparaenum}[\itshape a\upshape)]
\item
$0\in\dom$;
and
\item
$p_{0} \geq p$ \emph{for all} realizations of the erasure matrix $\bE$ such that $\pd\dom$ is erased while $\dom$ remains erasure-free (i.e. $\bE=0$ on $\dom$ and $\bE=1$ on $\pd\dom$).
\end{inparaenum}
Of course, if $p$ is sufficiently large, arbitrarily small domains $\dom$ containing $0$ cannot support power $p$ at $0$:
if $\dom$ is small enough and every site outside $\dom$ is erased (i.e. not transmitting), no point in $\dom$ will have high optimal transmitting power.
Clearly then, if $V \equiv \vert\dom\vert$ denotes the number of sites contained in $\dom$, there exists some minimal value $V_{p}$ such that $\dom$ does not support power $p$ at $0$ if $\vert\dom\vert < V_{p}$.
Therefore, if $\dom_{p}\subseteq\Z^{d}$ is a domain supporting power $p$ at $0$ with minimal volume $\vert\dom_{p}\vert = V_{p}$, we will have
\begin{equation}
\label{eq:lower-bound}
\pdist(p)
	\geq \prob\big(\text{$\bE\vert_{\dom_{p}} \equiv 0$ and $\bE\vert_{\pd\dom_{p}} \equiv 1$}\big)
	= (1-\epr)^{V_{p}},
\end{equation}
so the problem boils down to determining the minimal volume $V_{p}$ which supports power $p$ at $0$.%
\footnote{Interestingly, even though the lower bound \eqref{eq:lower-bound} appears lax for arbitrary $p$, it tightens considerably for large $p$.
Indeed, when $p$ is large, only very large domains can support power $p$, and the minimal volume $V_{p}$ will be exponentially more probable to occur than larger erasure-free volumes; as a result, the leading contribution to $\pdist(p)$ from erasure-free domains will be coming from $\dom_{p}$.}

Since we are interested in large powers for $\alpha > d+2$, we will focus on large domains $V_{p}$. In Section \ref{sec:outage_probability} we related the volume of an erasure-free domain to the minimum eigenvalue of the Dirichlet Laplacian over the domain; here, we need to relate it instead to the maximum power that can be supported therein. In Appendix \ref{app:Tails_lower_bound} we will provide details to the proof of the following proposition:
\begin{proposition}
\label{prop:lower-bound}
Let $\alpha \geq d+2$ and $\gamma\leq\gamma_{c}$.
Then, for large $p$:
\begin{equation}
\pdist(p)
	\geq (1 - \epr)^{V_{p}}
	\sim (1-\epr)^{\Omega_{d} R_{p}^{d}},
%	\quad
%	\text{for large $p$},
\end{equation}
where $\Omega_{d} = \pi^{d/2}/\Gamma(d/2 + 1)$ is the volume of the unit $d$-dimensional ball and $R_{p}$ is given by
\begin{subequations}
\label{eq:Rp}
\begin{align}
\label{eq:Rp-1}
R_{p}
	&= \kappa^{-1} \arcosh\left(\frac{\sigma^2}{\sigma^2-z_\gamma p}\right)
	&\text{for $d=1$,}
	\\
\label{eq:Rp-2}
R_{p}
	&= \kappa^{-1} I_{0}^{-1} \left(\frac{\sigma^2}{\sigma^2-z_\gamma p}\right)
	&\text{for $d=2$,}
\end{align}
\end{subequations}
where $I_{0}$ is the $0$-th order modified Bessel function of the first kind and $\kappa^2=z_\gamma/t_2$.
In particular, for $\gamma\to\gamma_{c}^{-}$, we will have:
\begin{equation}
\label{eq:lower-bound-final}
\pdist(p)
	\geq (1 - \epr)^{V_{p}}
	\sim \exp(-c_{d} p^{d/2}),
\end{equation}
where
%$c_{d} = - (2\pi d)^{d/2} \log(1-\epr) \big/ \Gamma\left(\frac{d+2}{2}\right)$ is a positive constant depending only on the erasure probability $\epr$ and the dimensionality $d$ of the network.
%In particular, we will have
\begin{subequations}
\begin{align}
c_{1}
	&= -2\sqrt{2t_2/\sigma^2} \log(1-\epr)
	&\quad
	\text{for $d=1$,}
	\\
c_{2}
	&= - 4\pi\log(1-\epr) t_{2}/\sigma^{2}
	&\quad
	\text{for $d=2$.}
\end{align}
\end{subequations}
%$c_{1} = -2\sqrt{2t_2/\sigma^2} \log(1-\epr)$ for $d=1$ and $c_{2} = - 4\pi\log(1-\epr)t_2/\sigma^2$ for $d=2$.
\end{proposition}

\begin{remark*}
Comparing the above equations with \eqref{eq:Prob(p)_Wyner2}, \eqref{eq:Lhatp_def} and \eqref{eq:Prob(p)_Wyner3} we conclude that the above lower bound is tight in the case of the (one-dimensional) Wyner model.
\end{remark*}

\subsection{An upper bound for nearest neighbor interactions}

We now provide an upper bound for the tails of the empirical power distribution $\pdist(p)$ summarized in Proposition \ref{prop:upper-bound}. Technically, it only applies to the random network model where interference arises only from nearest neighbor interactions. In the one dimensional case, this corresponds to the Wyner model discussed in Section \ref{sec:WynerModel} for which, as mentioned above, the lower bound is indeed tight.
In the two dimensional case, we can also obtain a matching \emph{upper} bound for the tails of the power distribution by using a site percolation argument (see Appendix \ref{app:power_tails}):

\begin{proposition}
\label{prop:upper-bound}
For erasure probabilities $\epr\geq 1/2$, we have:
\begin{equation}
\label{eq:upper-bound}
\pdist(p) \leq \exp(-\eta\pi R_p^2),
\end{equation}
where $R_p$ is given by \eqref{eq:Rp-2} and $0 < \eta \leq -\log(1-\epr)$.
%is a constant.
\end{proposition}

By combining \eqref{eq:lower-bound-final} and \eqref{eq:upper-bound} for the case $\gamma=\gamma_c$, we then obtain the following growth estimate for the tails of $\pdist(p)$:

\begin{corollary}
With notation as in Propositions \ref{prop:lower-bound} and \ref{prop:upper-bound}, we have
\begin{equation}
\label{eq:asymptotic-subcritical}
c_{d}'p \lesssim -\log \pdist(p) \lesssim c_{d} p,
\end{equation}
for a constant $c_{d}'\leq c_d$, whenever $\epr \geq \epr_{c}(d)$
and $p$ is large enough.
\end{corollary}

\begin{figure}[t]
\includegraphics[width=.95\columnwidth]{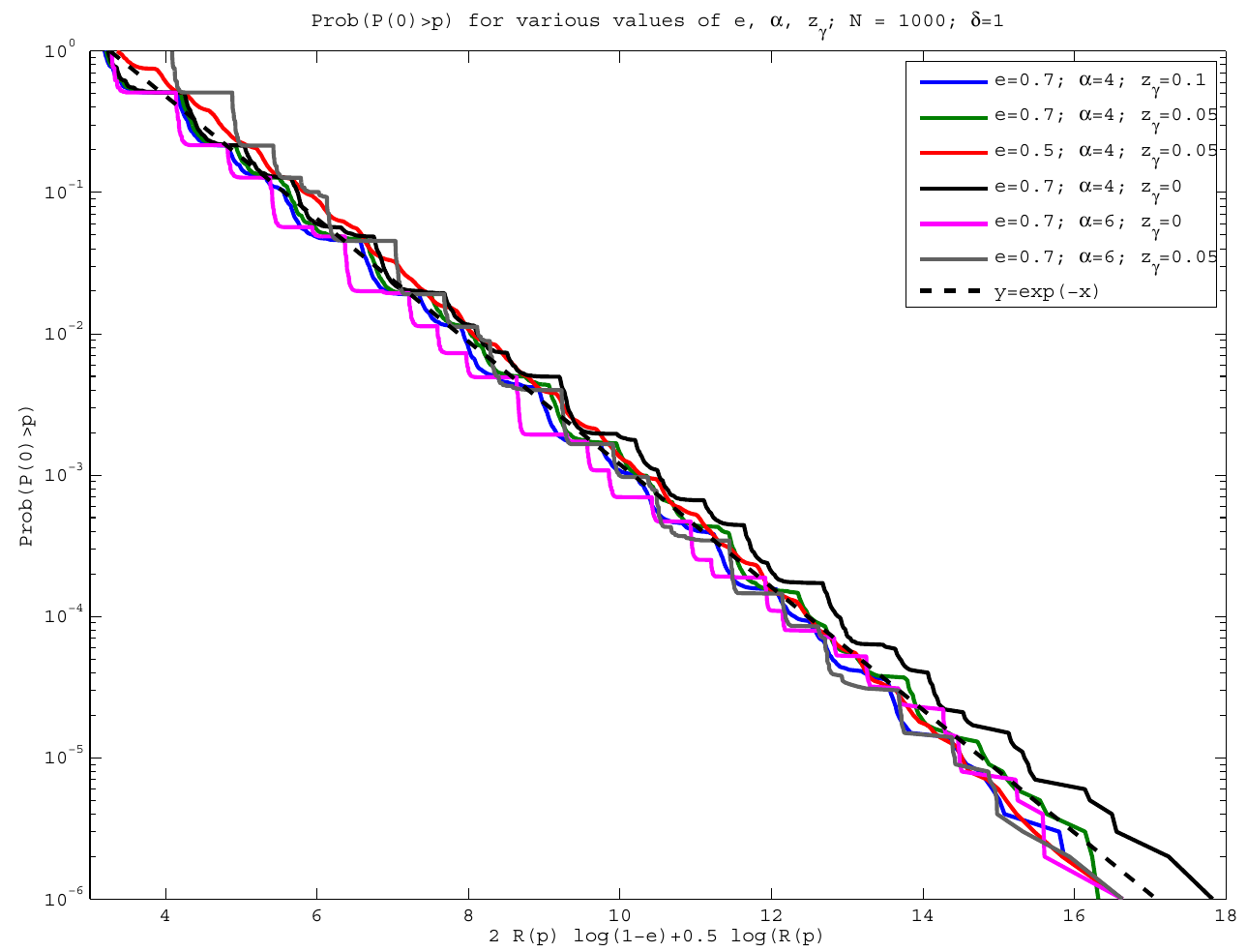}
\caption{Plot of the distribution of power for one-dimensional networks. The $x$-axis for each curve represents the {\em scaled} quantity related to the power, i.e. $X(p)=-2\log(1-e)R(p)-0.5\log(R(p))$, where $R(p)$ is defined in \eqref{eq:Rp-1}, so that ${\cal P}(p) \sim \exp(-x)$. The extra correction $-0.5\log(R_p(p))$ comes from the analysis of the Wyner model \eqref{eq:Prob(p)_Wyner3}. We include plots for different values of $\epr$, $\alpha$ and (small) $z_\gamma\geq 0$. Consistent with the calculation, we also plot the function $Y=\exp(-X)$ and shift all curves so that they coincide at one point. The curves follow a straight line, irrespective of whether $z_\gamma$ is zero or not (i.e. whether there is a maximum in the power itself).} \label{fig:pc_1dim_power_tails}
\end{figure}

\begin{figure}[htb]
\includegraphics[width=.95\columnwidth]{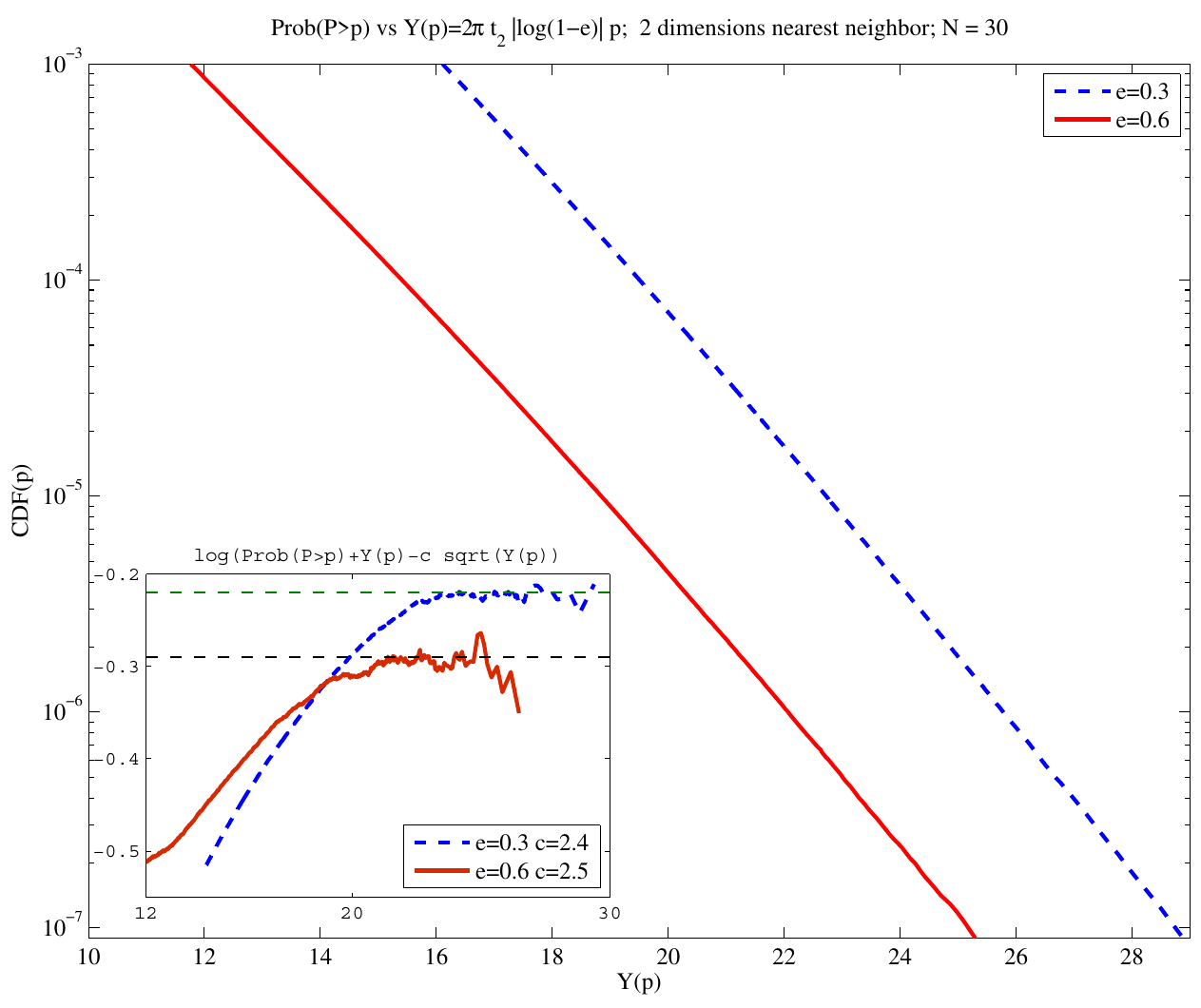}
\caption{Plot of the power distribution for two-dimensional square networks with $30\times 30$ sites in the case of nearest neighbor interferers.
The $x$-axis for each curve represents the \emph{scaled} power $Y(p) = -\log(1-\epr)2\pi t_2 p$ so that to leading $\pdist(p) \sim \exp(-Y(p))$. In two dimensions, and just as in the outage case (see Fig. \ref{fig:pc_2dim_outage}), the leading correction is proportional to $R_{p}^{d-1}\propto p^{(d-1)/2}$ and hence is expected to play an important role.
In the inset we fit the parameter $c$ in $\log \pdist(p) + Y(p)-c Y(p)^{1/2}$ to show that we indeed obtain a straight line for large $Y$ modulo numerical fluctuations due to arithmetic precision errors).}
\label{fig:pc_2dim_power_tails}
\end{figure}

%% file: Dynamics.tex
%----------------------------------------------------------------------
%%% DYNAMICS
%----------------------------------------------------------------------
% !TEX root = ./RandomPowerOptimization.tex

So far, our analysis has focused on the statistical properties of the optimal power vector $\bp^\ast$ in \eqref{eq:M_eq_def}, as well as the conditions under which this vector (and power control in general) is feasible.
In Section \ref{sec:ConnectionRW} we also discussed the Foschini\textendash Miljanic power control algorithm \eqref{eq:FoschiniM_algo_def} which provably converges to $\bp^\ast$ \textendash\ assuming that $\bp^{\ast}$ is itself feasible \textendash\ i.e. that $\gamma<\gamma_{c}$.
Two related obvious questions which arise are the following:
\begin{enumerate}
[(a)]
\item
If the system is feasible (i.e. $\gamma<\gamma_{c}$), what is the rate of convergence of the power control dynamics \eqref{eq:FoschiniM_algo_def} in the presence of random erasures?
\item
On the other hand, if $\gamma>\gamma_c$, how long does it take for the powers in the network to start becoming very large?
\end{enumerate}

To answer these questions, we will first analyze the solution of the Foschini\textendash Miljanic power control dynamics with $\bM = \bH_{V} + z_{\gamma} \bI$ for finite $V$:
in the limit $V\to\infty$, this solution will converge to the actual vector $\bp(t)$ with erasures at the sites with $\bE_{ii}=1$ in \eqref{eq:P(e)_def}.
Beginning with the subcritical case $\gamma<\gamma_c$, let $\bxi_{V}(t)=\bp_{V}(t)-\bp_{V}^{\ast}$;
then:
\begin{equation}
\label{eq:app:solutionFoschiniM_algo_matrix_form}
\bxi_{V}(t) = e^{-z_\gamma t} e^{-\left(\bH_0+V\bE\right)t} \bxi_{V}(0).
\end{equation}
Without loss of generality, we may focus on the origin $\bm=0$;
thus, projecting to the $0$-th element $\xi_{V}(0,t)$ of $\bxi_V(t)$, we obtain:
\begin{equation}
\label{eq:app:solutionFoschiniM_algo_matrix_form2}
\xi_{V}(0,t) = e^{-z_\gamma t} \, \be^{\top}_0 e^{-\left(\bH_0+V\bE\right)t} \bxi_{V}(0)
\end{equation}
In the limit $V\to\infty$, the LHS of this expression converges to $\xi(0,t)= p(0,t)- p^\ast_0$ which is the quantity we are interested in.
Since $z_\gamma>0$, if the initial powers of the sites are bounded, the elements of $\bxi_V(0)$ will also be bounded for all $V$;
hence, since $\exp(-\bH_Vt)\be_0\succ 0$, we will also have $|\xi_{V}(0,0)|\leq\delta_{+}$ for some $\delta_{+}>0$, leading to
\begin{equation}
\label{eq:app:solutionFoschiniM_algo_bound}
\left\vert \xi_{V}(0,t) \right\vert
	\leq \delta_{+} e^{-z_\gamma t} \be_{0}^{\top} e^{-\left(\bH_0+V\bE\right)t} \bu
\end{equation}
Taking the average of the above expression and the limit $V\to\infty$ we get
\begin{eqnarray}
\label{eq:app:solutionFoschiniM_algo_bound2}
\ex\left[\left|\xi(0,t)\right|\right] & = & \delta_{+} e^{-z_\gamma t} \wilde\ids_{\ast}(t)
\end{eqnarray}
where $\wilde\ids_{\ast}(t)$ is the average number of distinct sites visited by a random walk generated by $\bH$ up to time $t$ \textendash\ cf. \eqref{eq:DV-average}.
As a result, our analysis in Appendix~\ref{app:lifshitz_proofs} gives
\begin{proposition}
[Asymptotic behavior for subcritical $\gamma<\gamma_{c}$]
\label{thm:Long_time_dynamics_subcritical}
If $z_\gamma>0$, we will have
\begin{equation}
\label{eq:thm_dynamics_subcrit}
\lim_{t\to+\infty} t^{-d/\alpha_{\eff}}
	\left[\log\ex\left[\left|\xi(0,t)\right|\right]+z_\gamma t\right]
	\leq -k(\alpha,d),
\end{equation}
where the effective pathloss exponent $\alpha_{\eff}$ is given by \eqref{eq:alpha-eff} and $k(\alpha,d)$ by \eqref{eq:app:k(a,d)_def}.
\end{proposition}
We demonstrate the tightness of the above inequality in a couple of cases.
First, assume that all initial powers are greater than $\sigma^2/z_\gamma$ (which itself is an upper bound for the elements of $\bp_V(0)$), so all elements in $\bxi_V(0)$ are positive.
Denoting with $\delta_{-}>0$ the minimum of all elements of $\bxi_V(0)$, we will have
\begin{eqnarray}
\label{eq:app:solutionFoschiniM_algo_bound3}
\ex\left[\left|\xi(0,t)\right|\right] &\geq & \delta_{-} e^{-z_\gamma t} \wilde\ids_{\ast}(t)
\end{eqnarray}
\begin{corollary}
If $\bp(t=0)\succ \sigma^2/z_\gamma\bu$, the inequality in \eqref{eq:thm_dynamics_subcrit} becomes an equality, i.e.
\begin{equation}
\label{eq:thm_dynamics_subcrit2}
\lim_{t\to+\infty} t^{-d/\alpha_{\eff}}
	\left[\log\ex\left[\left|\xi(0,t)\right|\right]+z_\gamma t\right]
	= -k(\alpha,d),
\end{equation}
\end{corollary}
The same can be shown when $\bp(0)$ has zero elements and hence $\bxi_{V}(0)\prec 0$ for finite $V$.
In the limit $V\to\infty$, the minimum of this vector will be zero;
that said, the operation of $\exp(-\bH_V t$ will project out such terms, so this issue does not make a difference.
Therefore, in this limit, the inequality \eqref{eq:app:solutionFoschiniM_algo_bound3} will hold for $\delta_{-}\leq 1/(z_\gamma+\lambda_{max})$, where $\lambda_{max}$ is the maximum eigenvalue of $\bH_0$. We therefore expect  that the equality in \eqref{eq:thm_dynamics_subcrit2} should be tight in general.

As a result of the above discussion we see that the timescale at which the system converges to its optimal vector $\bp^\ast$ is $t\sim z_\gamma^{-1}$ for $\gamma<\gamma_c$. It interesting to compare this timescale with the corresponding one at which an infeasible system with $\gamma>\gamma_c$ becomes unstable, that is the powers of the system become very large.
For concreteness, we will focus on the case where $\gamma_c|z_\gamma|\ll 1$ where we can make precise quantitative statements.
To that end, it will be more convenient to express the solution of \eqref{eq:FoschiniM_algo_matrix_form} in the form
\begin{flalign}
\label{eq:app:solutionFoschiniM_algo_matrix_form4}
\bp_{V}(t)
	&= e^{-(z_\gamma \bI+\bH_V)t} \bp(0) + \int_0^t \,ds\, e^{-(z_\gamma \bI+\bH_V)s}\bu
	\notag\\
	&= I_{1}(t) +I_{2}(t)
\end{flalign}
where $I_{1}$ and $I_{2}$ correspond to the two terms in the top line.
Taking the average over realizations and evaluating the element at $\bm=0$ in the infinite size  and $V$ limits, $I_{1}(t)$ will be bounded by $p_{\min} \wilde\ids_{\ast}(t) \geq I_1(t) \geq p_{\max} \wilde\ids_{\ast}(t)$, where $p_{\min}$ and $p_{\max}$ are the minimum/maximum values of the elements in $\bp(0)$, respectively.
The resulting integrand in the second term above can then be expressed as $\exp(-z_\gamma s)\wilde\ids_{\ast}(s)$.
For large times $t$, $\wilde\ids_\ast(t)\sim \exp(-k(\alpha,d)t^{d/\alpha_{eff}})$ so the integral may be approximated by
\begin{equation}
\label{eq:approx_integral}
\ex\big[I_{2}(t)\big]
	\sim
	\begin{cases}
	e^{-z_{\gamma}t - k(\alpha,d) t^{d/\alpha_{\eff}}}
		&\text{if $t\gg t^{\ast}$,}
	\\
	\bigoh(1)
		&\text{if $t\ll t^{\ast}$,}
	\end{cases}
\end{equation}
where $t^{\ast}$ is the solution of the equation $|z_\gamma| t^\ast-kt^{\ast d/\alpha_{\eff}}=0$.

This result can also be obtained by an asymptotic evaluation of the integral of the asymptotic expression of the integrand.
To do this, one only needs to bound the small time behavior of the integrand (where its approximate expression is not valid) and to control in a similar way the leading correction $o(t^{d/\alpha_{\eff}})$ to the asymptotic expression of $\wilde\ids_\ast(t)$.
Doing just that, we obtain:
\begin{proposition}
[Asymptotic behavior for supercritical $\gamma>\gamma_{c}$]
\label{thm:Long_time_dynamics_supercritical}
For $\gamma>\gamma_c$, such that $\gamma_c|z_\gamma|\ll 1$ we have
\begin{equation}
\label{eq:thm_dynamics_supercrit}
\log\ex\big[\vert p(0,t) \vert\big]
	=
	\begin{cases}
	\vert z_{\gamma} \vert t
		&\text{if $t\gg t^{\ast}$,}
	\\
	\bigoh(1)
		&\text{if $t\ll t^{\ast}$}.
	\end{cases}
%		= \left\{\begin{tabular}{lr}
%                                                    %\hline
%                                                    % after \\: \hline or \cline{col1-col2} \cline{col3-col4} ...
%                                                    $|z_\gamma| t$  & $t\gg t^\ast$ \\
%                                                    $\bigoh(1)$ & $t\ll t^\ast$ \\
%                                                    %\hline
%                                                  \end{tabular}\right.
\end{equation}
where the notation $\bigoh(1)$ refers to $|z_\gamma|$ and
\begin{equation}
\label{eq:t^ast}
t^{\ast} = \left[k(\alpha,d) / \vert z_{\gamma} \vert\right]^{\frac{\alpha_{\eff}}{\alpha_{\eff} - d}}
%  t^\ast = \left(\frac{k(\alpha,d)}{|z_\gamma|}\right)^{\frac{\alpha_{\eff}}{\alpha_{\eff}-d}}
\end{equation}
the effective pathloss exponent $\alpha_{\eff}$ is given by \eqref{eq:alpha-eff}
and $k(\alpha,d)$ by \eqref{eq:app:k(a,d)_def}.
\end{proposition}

\begin{remark*}
The above result shows that the characteristic time over which an infinite infeasible system becomes unstable is given by $t^\ast$.
This time for small $|z_\gamma|$ can be much larger than $|z_\gamma|^{-1}$.
\end{remark*}

%% file: Conclusions.tex
%----------------------------------------------------------------------
%%% CONCLUSIONS
%----------------------------------------------------------------------
% !TEX root = ./RandomPowerOptimization.tex

In this paper we studied the optimal power vector that achieves an \ac{SINR} target criterion in the presence of both randomness and interference.
In particular, we derived the statistics of the optimal power vector and the long-term behavior of the Foschini\textendash Miljanic power control algorithm \cite{Foschini1991_FMAlgorithm_PowerControl} in the presence of random erasures.
This was made possible by mapping the problem of power minimization in the presence of nonlinear \ac{SINR} constraints to the so-called Anderson impurity model which can be analyzed by studying random walks in a lattice with randomly placed traps.

Drawing tools and ideas from statistical physics, we calculated the average power and the variance of the optimal power vector by means of the \acf{CPA} approach, a method originally introduced in the study of disordered metals.
Despite the method's approximative nature, our results are fairly accurate over a wide range of parameters for the erasure density $\epr$ in the network and the users' target \ac{SINR} value $\gamma$;
on the other hand, the \ac{CPA} method fails to predict the infeasibility of power control in the system when the users' target \ac{SINR} exceeds a certain critical value.
To calculate the probability of the system becoming unstable, we then employed a different set of mathematical tools in order to calculate the low eigenvalue density of the random system.
Remarkably, the same tools also allowed us to estimate the tails of the power distribution under power control, thus obtaining a complementary outage criterion for networks with power-limited transmitters.
In all cases, our predictions for the system's instability probability and its large power tail behavior were confirmed by numerical simulations.
Finally, we calculated the average long-term behavior of the Foschini\textendash Miljanic power control algorithm in the presence of random erasures, and we showed that its rate of convergence exhibits nontrivial time-dependencies.

Summing up, we have found that approximate methods (like \ac{CPA}) provide good quantitative results for quantities related to \emph{bulk} properties of the system (such as the intra-sample average of the optimal power vector or its variance).
Nevertheless, rare events (such as instability or the occurrence of atypically large powers in the optimal power vector) are conditional on the appearance of large regions with no inactive transmitters. These regions are then responsible for the breakdown of the whole system, so our analysis focused on estimating the probability of observing such erasure-free regions.

We believe that the results (and insights) obtained in this paper regarding tail events may be applied to significantly more general network models.
For example, the probability that a finite-sized network can become infeasible may be approximated by the probability of occurrence of large regions of a given critical size with closely packed users.
Due to the size of the paper however, we decided not to present applications of these methods to specific situations, but to defer them instead to a future paper.

%\begin{enumerate}[Discussion]
%\item  Make contact with other concepts: (1) Localization - progressive ridance of unstable regions). (2) Off-diagonal disorder (3)fractional statistics of eigenfunctions
%\end{enumerate}
%\begin{enumerate}[Future Directions]
%\item Generalize to other diagonal disorder
%\end{enumerate}

%% file: App-CPA.tex
%----------------------------------------------------------------------
%%% APPENDIX: CPA
%----------------------------------------------------------------------
% !TEX root = ./RandomPowerOptimization.tex

In this appendix we will motivate the derivation of the \ac{CPA} equations applied in Section \ref{sec:CPA};
the interested reader can find more information on the method in \cite{Davies1963_CPA, Elliott1974_CPA_RevModPhys} and references therein.

Specifically, our aim will be to calculate the average resolvent operator $\ex[\bG(\lambda)]$ (\ref{eq:G_fun_CPA}).
Unfortunately, methods from random matrix theory cannot be applied here directly because the random matrix $\bV$ is diagonal (nevertheless, the end results will end up being related).
As such, the main idea behind \ac{CPA} is to replace the random matrix $\bV = V\bE$ in the resolvent operator $\bG$ with a constant diagonal matrix $\Sigma(\lambda)\bI$ so that the difference $\delta\bV=\bV-\bI \Sigma(\lambda)$ is ``small'' if we pick $\Sigma$ in the right way.%
\footnote{This assumption is correct for full random matrices, but only approximately so for diagonal random matrices $\bV$.}

We thus start by defining the matrix
\begin{equation}
\label{eq:G0_hat_app}
\hat\bG(\lambda)
	= \big[\left(\lambda-\Sigma(\lambda)\right)\bI-\bH_0\big]^{-1}
	= \bG_0(\lambda-\Sigma(\lambda))
\end{equation}
where $\bG_0$ is the resolvent operator in the absence of disorder:
\begin{equation}
\label{eq:G0_fun_app}
\bG_0(\lambda)
	= \left[\lambda \bI - \bH_{0}\right]^{-1}.
\end{equation}
The matrix $\bG$ can then be expressed as
\begin{flalign}
\label{eq:G_fun_Dyson}
\bG(\lambda)
	&= \big[{\hat \bG}(\lambda)^{-1}-\delta\bV\big]^{-1}
	\notag\\
	&= \hat\bG(\lambda) + {\hat \bG}(\lambda)\delta\bV\bG(\lambda)
	\notag\\
	&= \hat\bG(\lambda) + {\hat \bG}(\lambda)\bT{\hat \bG}(\lambda)
\end{flalign}
where the so-called scattering matrix $\bT$ is defined as
\begin{eqnarray}\label{eq:T_fun_def}
  \bT &=& \delta\bV\left[1-{\hat \bG}(\lambda)\delta\bV\right]^{-1}.
\end{eqnarray}
Up to this point everything is exact, and by expressing (\ref{eq:T_fun_def}) recursively and averaging over $\bV$ (in $\delta\bV$) we can obtain $\ex[\bT]$.
This could be plugged into  \eqref{eq:G_fun_Dyson} to obtain $\ex[\bG(\lambda)]$, but this is an impossible task in general.
On the other hand, if we assume that $\delta \bV$ is small, we may expect that the second term in the last equation will also be small on average.
The \ac{CPA} approach amounts to averaging over the randomness of a {\em single} random site $i$ and demanding that the corresponding diagonal element $T_{ii}$ of $\bT$ vanishes on average. Hence, it is an approximation which ``hides'' the effects of all other sites into $\Sigma$ and then reduces to a self-consistent single site problem. 
This somewhat obscure assumption leads to
\begin{equation}
\label{eq:T_fun_ii=0}
\ex \big[T_{ii}\big]
	= \ex \left[\frac{Ve_i-\Sigma(\lambda)}{1-(Ve_i-\Sigma(\lambda))g(\lambda)}\right]=0
\end{equation}
where
\begin{flalign}
\label{eq:g_ii_fun_def}
g(\lambda)
	&=\big[(\lambda-\Sigma(\lambda))\bI - \bH_{0}\big]_{ii}^{-1}
	\notag\\
	&= \int \frac{d\bk}{(2\pi)^d} \frac{1}{\lambda-\Sigma(\lambda)-\epsilon(\bk)}
\end{flalign}
is the (shifted) unperturbed resolvent operator evaluated at the $i$-th site (the second equality follows from the fact that the eigenvectors of $\bG_{0}$ are Fourier modes).

Rearranging terms in \eqref{eq:T_fun_ii=0} yields \eqref{eq:S_fun_CPA_eq};
together with \eqref{eq:g_ii_fun_def}, these two expressions constitute the CPA equations for this system, leading to expression \eqref{eq:Sigma_CPA2_eq} for $\Sigma(\lambda)$.
%As discussed in Section \ref{sec:CPA}, in our particular case, $\Sigma(\lambda)$ can be expressed as in \eqref{eq:Sigma_CPA2_eq}.

%% file: App-Outage.tex
%----------------------------------------------------------------------
%%% APPENDIX: OUTAGE
%----------------------------------------------------------------------
% !TEX root = ./RandomPowerOptimization.tex

%\renewcommand{\bx}{x}
%\renewcommand{\by}{y}
%\renewcommand{\bz}{z}

Our aim in this appendix will be twofold:
First and foremost, we seek to derive the low-energy asymptotic expressions \eqref{eq:thmLifsthitztails} for the \ac{IDS} of the disordered Hamiltonian matrix $\bH = (\bI-\bE) \bH_{0} (\bI-\bE)$
%of \eqref{eq:EME_def}
%which generates the erasure model \eqref{eq:EME_def} of \cite{Tulino2007_GaussianErasureChannel, Tulino2010_GaussianErasureChannel, Moustakas2012_PowerOptimizationErasures}.
in the large lattice limit $\vert\lattice\vert = \latsize^d \to\infty$. This will provide an approximation for the integrated density of eigenvalues for a large finite system, which will then be used to approximate the instability probability in Section \ref{sec:outage_probability}.
In so doing however, we will also provide the necessary tools that are required in Section \ref{sec:Dynamics} to estimate the long-term behavior of the Foschini\textendash Miljanic power control dynamics \eqref{eq:FoschiniM_algo_def}.
%\eqref{eq:FoschiniM_algo_matrix_form} will be derived

In a nutshell, our approach will be as follows:
\begin{enumerate}[1.]
\item
%in Section \ref{sec:sub1} Proposition \ref{prop:model-equivalence}
First, we will approximate the \ac{IDS} of the Hamiltonian $\bH$ of the erasure model \eqref{eq:EME_def} by the Anderson Hamiltonian $\bH_{V} = \bH_{0} + V\bE$ of \eqref{eq:H_def} for large $V$ and finite $\latsize$ (Section \ref{sec:sub1}).

\item
In Section \ref{sec:sub2}, we will derive the \ac{IDS} of $\bH$ in the large system limit by exchanging the limits $V\to\infty$ and $\latsize\to\infty$:
specifically, by working in the infinite system where $\bH$ and $\bH_{V}$ are viewed as infinite-dimensional operators (instead of as matrices of order $N = \latsize^{d}$), we will harvest the \acl{IDS} $\ids(\lambda)$ of $\bH$ from the density $\ids_{V}(\lambda)$ of $\bH_{V}$ by taking the limit $\lim_{V\to\infty} \ids_{V}(\lambda)$.

\item
To calculate $\lim_{V\to\infty} \ids_{V}(\lambda)$, we will take the Laplace transform $\wilde\ids_{V}(t) = \int_{0}^{\infty} e^{-t\lambda} \dd\ids_{V}(\lambda)$ of $\ids_{V}(\lambda)$ and express it as a Feynman\textendash Kac path integral over a random walk in $\Z^{d}$ with transition probabilities determined by $\bH_{0}$ (Section \ref{sec:sub3}).

\item
In Section \ref{sec:sub4}, we apply the analysis of \cite{Donsker1979_NumberDistinctSitesRW} to get an upper bound for the large $t$ behavior of $\wilde\ids(t)$ by averaging over the number of \emph{distinct} sites visited by the random walk.

\item
A matching lower bound for $\wilde\ids(t)$ is then obtained in Section \ref{sec:sub5} by using techniques discussed in \cite{Pastur_book1992_SpectraRandomOperators}.

\item
Finally, in Section \ref{sec:sub6}, we obtain the small $\lambda$ behavior of $\ids(\lambda)$ by  inverting the Laplace transform $\wilde\ids(t)$ for large $t$.
\end{enumerate}

In what follows, we will make this roadmap precise by encoding each step in a series of lemmas.

\subsubsection{Approximation of ${\cal N}^\lattice(\lambda)$ by ${\cal N}_V^\lattice(\lambda)$}
\label{sec:sub1}

First, to resolve any notational ambiguities, we will view $\bH_{0}$, $\bH = (\bI-\bE) \bH_{0} (\bI-\bE)$ and $\bH_{V} = \bH_{0} + V\bE$ as infinite-dimensional operators acting on $\ell^{2}(\Z^{d})$, and we will denote their restrictions to the lattice $\lattice = \Z_{L}^{d}$ by $\bH_{0}^{\lattice}$, $\bH^{\lattice}$ and $\bH_{V}^{\lattice}$ respectively.
With this in mind, we begin by showing that the spectrum of $\bH^{\lattice}$ can be approximated within $\bigoh(1/V)$ by that of the Anderson Hamiltonian $\bH_{V}^{\lattice}$.

More precisely, let $K = \dim\ker (\bI - \bE)$ denote the number of erased sites in the network model \eqref{eq:EME_def}.
Then, the spectrum of $\bH^{\lattice}$ will consist of $K$ zero eigenvalues (representing the erased lines of $\bI - \bE$) and $N-K$ non-negative eigenvalues comprising the \emph{effective} spectrum of $\bH$ over the range of $\bI - \bE$.
Similarly, for large $V$, Gershgorin's circle theorem shows that the spectrum of $\bH_{V}^{\lattice}$ will consist of $K$ large eigenvalues of order $\bigoh(V)$ and $N-K$ non-negative eigenvalues of order $\bigoh(1)$ which determine the stability behavior of the erasure model \eqref{eq:H_def}.%
\footnote{Simply note that the Gershgorin discs corresponding to the erased diagonal elements will be centered around $V$ and their radius will be of order $\bigoh(1)$.}

Obviously, for the erasure models \eqref{eq:H_def} and \eqref{eq:EME_def} to yield equivalent predictions, their effective spectra (defined as above) must agree in the limit $V\to\infty$.
Indeed, we have:

\begin{proposition}
\label{prop:model-equivalence}
For large $V>0$, the positive eigenvalues of $\bH_{V}^{\lattice}$ lie within $\bigoh(1/V)$ of the low-end eigenvalues of $\bH^{\lattice}$.
More precisely, the eigenvalues of $\bH^{\lattice}$ over the range of $\bI - \bE$ may be mapped bijectively to the eigenvalues of $\bH_{V}^{\lattice}$ that are of order $\bigoh(1)$, and the error of this bijection is at most $\bigoh(1/V)$.
\end{proposition}

\begin{IEEEproof}
By rearranging indices, the matrix $\bH_{V}^{\lattice} = \bH_{0}^{\lattice} + V \bE$ may be written as
\begin{equation}
\bH_{V}^{\lattice}
	=\left(
	\begin{array}{cc}
	\bA & \bB\\
	\bC & \bD + V \bI
	\end{array}\right),
\end{equation}
with the component blocks $\bA, \bB, \bC$ and $\bD$ all being independent of $V$.
Then, if $\bv = (\bv_{A}, \bv_{D})$ is the block decomposition of an eigenvector of $\bH_{V}^{\lattice}$ with eigenvalue $\lambda$, we will have:
\begin{equation}
\begin{aligned}
\bA \bv_{A} + \bB \bv_{D}
	&= \lambda \bv_{A},
	\\
\bC\bv_{A} + (\bD + V \bI) \bv_{D}
	&= \lambda \bv_{D}.
\end{aligned}
\end{equation}
Solving for $\bv_{D}$, we get $\bv_{D} = - \big[\bD + (V - \lambda) \bI\big]^{-1} \bC \bv_{A}$ and hence:
\begin{equation}
\label{eq:eigenvector-A}
\bA \bv_{A} - \bB\big[\bD + (V - \lambda) \bI\big]^{-1} \bC \bv_{A} = \lambda \bv_{A}.
\end{equation}
Coupled with the fact that $\bA,\bB,\bC$ and $\bD$ do not depend on $V$, Gershgorin's circle theorem shows that if $V$ is large enough, $K = \rank(\bE)$ eigenvalues of $\bH_{V}^{\lattice}$ will lie within $\bigoh(1)$ of $V$ while the $N-K$ remaining ones will be of order $\bigoh(1)$.
Thus, if $\lambda$ is an $\bigoh(1)$ eigenvalue, \eqref{eq:eigenvector-A} yields $\bA\bv_{A} = \lambda \bv_{A} + \bigoh(1/V)$, and our claim follows by noting that $\bH^{\lattice}$ may be written in the form $\bH^{\lattice} = \diag(\bA,\boldsymbol{0})$ after properly rearranging indices.
\end{IEEEproof}

\subsubsection{Exchanging the order of the limits $\latsize\to\infty$ and $V\to\infty$}
\label{sec:sub2}

We now turn to the definition of the \acl{IDS} of $\bH$ and $\bH_{V}$ viewed as random infinite-dimensional operators on $\ell^{2}(\Z^{d})$.
Fixing some finite cubic lattice $\lattice=\Z_{\latsize}^{d}$, let $\ids^{\lattice}(\lambda)$ (resp. $\ids_{V}^{\lattice}(\lambda)$) denote the so-called \emph{prelimit function} of $\bH$ (resp. $\bH_{V}$) on $\lattice$, namely the number of positive eigenvalues of $\bH^{\lattice}$ (resp. $\bH_{V}^{\lattice}$) not exceeding $\lambda$, normalized by the volume $\vert\lattice\vert = \latsize^{d}$ of $\lattice$;
formally, we let
\begin{equation}
\begin{aligned}
\ids^{\lattice}(\lambda)
	&= \latsize^{-d} \vert\{\lambda'\in\spec(\bH^{\lattice}): 0<\lambda'\leq\lambda\}\vert,
	\\
\ids_{V}^{\lattice}(\lambda)
	&= \latsize^{-d} \vert\{\lambda'\in\spec(\bH_{V}^{\lattice}): 0<\lambda'\leq\lambda\}\vert,
\end{aligned}
\end{equation}
where $\spec(\cdot)$ denotes the spectrum of the matrix in question.
It is then well-known that $\ids^{\lattice}$ and $\ids_{V}^{\lattice}$ converge vaguely%
\footnote{That is, in the weak$^{\ast}$ topology of Radon measures on $[0,+\infty)$.}
to a \emph{nonrandom} limit (see e.g. Theorem 4.4 in \cite{Pastur_book1992_SpectraRandomOperators}), i.e. there exist nonrandom densities $\ids$ and $\ids_{V}$ on $[0,+\infty)$ such that
\begin{equation}
%\label{eq:app:IDS_infte system}
\begin{aligned}
\ids(\lambda)
	&= \lim_{L\to\infty} \ids^{\lattice}(\lambda),
	\\
\ids_{V}(\lambda)
	&= \lim_{L\to\infty} \ids_{V}^{\lattice}(\lambda),
\end{aligned}
\end{equation}
at every continuity point of $\ids$ and $\ids_{V}$.
Accordingly, the limit cumulative density $\ids$ (resp. $\ids_{V}$) will be called the \emph{\acl{IDS}} of $\bH$ (resp. $\bH_{V}$).

On the other hand, Proposition \ref{prop:model-equivalence} shows that $\ids_{V}^{\lattice}$ converges to $\ids^{\lattice}$ as $V\to\infty$, so it is natural to expect that $\ids_{V}$ and $\ids$ are similarly related.
% i.e. $\lim_{V\to\infty} \ids_{V}^{\lattice}(\lambda) = \ids(\lambda)$ for every $\lambda\in(0,+\infty)$.
Indeed, the next lemma shows that the order of the limits $V\to\infty$ and $\latsize\to\infty$ can be exchanged, so $\ids$ can be harvested itself from the pointwise limit $\lim_{V\to\infty}\ids_{V}$:

\begin{lemma}
\label{lem.IDS.limit}
$\ids_{V}$ converges vaguely to $\ids$ as $V\to\infty$.
\end{lemma}

\begin{IEEEproof}
It obviously suffices to show that $\lim_{V\to\infty}\ids_{V}(\lambda) = \ids(\lambda)$ for all $\lambda\in(0,+\infty)$ at which $\ids$ is continuous.
To that end, let $\lambda>0$ be a continuity point of $\ids$;
then, by Proposition \ref{prop:model-equivalence}, there exists $\eps_{V}>0$ with $\eps_{V}\to 0$ as $V\to\infty$ such that
\begin{equation}
\label{eq:sandwich}
\ids_{V}^{\lattice}(\lambda - \eps_{V})
	\leq \ids^{\lattice}(\lambda)
	\leq \ids_{V}^{\lattice}(\lambda + \eps_{V})
\quad
\text{for all $\lambda$.}
\end{equation}
By a theorem of Craig and Simon \cite{Craig1983_LogHolderContinuityIDS}, the integrated density $\ids_{V}$ of $\bH_{V}$ will be continuous on $(0,+\infty)$.
Thus, letting $\latsize\to\infty$ in the sandwich above, we readily obtain $\ids_{V}(\lambda - \eps_{V}) \leq \ids(\lambda) \leq \ids_{V}(\lambda+\eps_{V})$ \textendash\ recall that the prelimit functions $\ids_{V}^{\lattice}$ and $\ids^{\lattice}$ converge vaguely to $\ids_{V}$ and $\ids$ respectively.
Our claim then follows by taking the limit $V\to\infty$ as in the proof of Lemma 4.6 in \cite{Pastur_book1992_SpectraRandomOperators}.
%our claim follows by taking the limit $V\to\infty$ and recalling that vague convergence is quantified on the continuity sets of $\lim_{V\to\infty} \ids_{V}$.
\end{IEEEproof}

\subsubsection{The Laplace transform of $\ids_{V}$}
\label{sec:sub3}

Our next step will be to calculate the Laplace transform $\wilde\ids_{V}$ of $\ids_{V}$, that is
\begin{equation}
\txs
\wilde\ids_{V}(t)
	= \int_{0}^{\infty} e^{-\lambda t} \dd \ids_{V}(\lambda).
\end{equation}
To that end, we will use the Feynman\textendash Kac path integral formula to express $\wilde\ids_{V}(t)$ as an integral over random walks:
%with transition probabilities determined by $\bH_{0}$
%as follows:
\begin{lemma}
\label{lem.FK}
Let $X(t)$ be a random walk on $\Z^{d}$ with generator $\bH_{0}$,
i.e.
%the transition kernel of hopping from $\bx\in\Z^{d}$ to $\by\in \Z^{d}$ after time $t\geq0$ is given by
\begin{equation}
\label{eq.transition}
%\prob(\by,\tau \vert \bx,0)
\prob\big(X(t) = \by \,\vert\, X(0) = \bx\big)
	\equiv \K(\bx,\by;t)
	= \be_\bx^{\top} \, e^{-t\bH_{0}} \be_\by,
\end{equation}
%where $\bH_{0}(x,y) = \langle \delta_{x}\vert\bH_{0}\vert\delta_{y}\rangle$ denotes the $(x,y)$-th element of $\bH_{0}$.
%i.e. the characteristic function of $X(t)$ when starting at $x\in\Z^{d}$ is given by
%\begin{equation}
%\ex^{x} e^{itX(t)} = \insum_{y\in\Z^{d}} e^{-t\bH_{0}}(x,y) e^{ity},
%\end{equation}
where $\be_\bx$, $\be_\by$ denote the basis vectors of $\ell^{2}(\Z^{d})$ corresponding to the sites $\bx,\by\in \Z^{d}$ \textendash\ i.e. $\be_\bx^{\top}\,  e^{-t\bH_{0}} \be_\by$ denotes the $(\bx,\by)$-th element of $e^{-t\bH_{0}}$.
Then:
\begin{flalign}
\label{eq.FK}
\wilde\ids_{V}(t)
	&= \ex_{\omega} \be_\b0^{\top}  e^{-t \bH_{V}} \be_\b0
	\notag\\
	&= \K(0,0;t)\cdot \ex_{\omega} \ex_{0,0}^{0,t}
	e^{-V \int_{0}^{t} \one\{\textup{$X(s)$ is erased}\} \dd s},
\end{flalign}
where the expectation $\ex_{\omega}$ is taken over the realizations of the erasure matrix $\bE$ of \eqref{eq:P(e)_def},
``\textup{$X(s)$ is erased}'' means that $\bE$ is equal to one at $X(s)$,
and $\ex_{\bx,0}^{\by,0}$ is the conditional expectation
\begin{equation}
\label{eq.cond.bridge}
\ex_{\bx,0}^{\by,t}\big[\cdot\big]
	= \ex\big[\cdot \vert X(t) = \by, X(0) = \bx\big].
\end{equation}
%denotes the conditional expectation over the law of $X$ with $X(0) = \bx$ and $X(t) = \by$.
\end{lemma}

To prove Lemma \ref{lem.FK}, we will first need an intermediate result (which will also be used in the proof of Propositions \ref{thm:Long_time_dynamics_subcritical}-\ref{thm:Long_time_dynamics_supercritical}):

%The first is a chain rule for the conditional expectation $\ex_{\bx,0}^{\by,t}\big[\cdot\big]$:
%
%\begin{lemma}
%\label{lem.chain}
%With notation as in Lemma \ref{lem.FK}, we have
%\begin{multline}
%\label{eq.chain}
%\ex_{\bx,0}^{\by,t} \big[F(X(\tau))\big]
%	=
%	\insum_{\bz} F(z)
%	\frac{\K(\bx,\bz;\tau) \K(\bz,\by;t-\tau)}{\K(\bx,\by;t)}
%%	\ex_{\bx,0}^{\by,\tau} \big[F(X(\tau))\big]
%\end{multline}
%for all $\tau\in[0,t]$ and for all $F\from\Z^{d}\to\R$.
%\end{lemma}
%
%\begin{IEEEproof}
%By the law of total expectation, we will have:
%\begin{flalign}
%\ex_{\bx,0}^{\by,t}
%	&\big[F(X(\tau))\big]
%	\notag\\
%	&= \insum_{\bz}
%%	\ex_{\bx,0}^{\bz,\tau} \big[F(X(\tau))\big]
%	F(z)
%	\cdot \prob_{\bx}\left(X(\tau) = \bz \,\midd\, X(t) = \by\right)
%	\notag\\
%	&= \insum_{\bz}
%%	\ex_{\bx,0}^{\bz,\tau} \big[F(X(\tau))\big]
%	F(z)
%	\cdot \frac
%	{\prob_{\bx} \left(X(t) = \by \,\midd\, X(\tau) = \bz\right) \prob_{\bx} \left(X(\tau) = \bz\right)}
%	{\prob_{\bx} \left(X(t) = \by\right)}
%	\notag\\
%	&= \insum_{\bz}
%%	\ex_{\bx,0}^{\bz,\tau} \big[F(X(\tau))\big]
%	F(z)
%	\cdot \frac{\K(\bz,\by;t-\tau) \K(\bx,\bz;\tau)}{\K(\bx,\by;t)},
%\end{flalign}
%where $\prob_{\bx}$ denotes the law of the random walk conditioned on $X(0) = \bx$, the third line follows from Bayes' law, and the last one from the stationarity of the increments of $X(t)$.
%%Our claim then follows by rearranging.
%\end{IEEEproof}
%With the chain rule \eqref{eq.chain} at hand, we obtain the following formula (which will also be used in the proof of Theorem \ref{thm:Long_time_dynamics}):

\begin{lemma}
\label{lem.propagator}
With notation as in Lemma \ref{lem.FK}, we have
\begin{equation}
\label{eq.propagator}
\be_\bx^{\top} e^{-t \bH_{V}} \be_\by
	= \K(\bx,\by;t)
	\cdot \ex_{\bx,0}^{\by,t} e^{- V \int_{0}^{t} \one\{\textup{$X(s)$ is erased}\} \dd s}
\end{equation}
%where $\ex_{\bx,0}^{\by,0}$ is now conditioned on $X(0) = \bx$ and $X(t) = \by$.
\end{lemma}

\begin{IEEEproof}
With $\bH_{V} = \bH_{0} + V\bE$, the Lie-Trotter product formula readily gives:
\begin{multline}
\be_\bx^{\top}  e^{-t\bH_{V}} \be_\by
	= \lim_{n\to\infty}
	\be_\bx^{\top} \left(e^{-t \bH_{0}/n} e^{-t\bE V/n}\right)^{n} \be_\by
	\\
	\qquad
	= \lim_{n\to\infty}
	\insum_{\bx_{1},\dotsc,\bx_{n-1}}
	\be_{\bx_{0}}^{\top}  e^{-t \bH_{0}/n} \be_{\bx_{1}}
	\;\dotsm\;
	\be_{\bx_{n-1}}^{\top}  e^{-t \bH_{0}/n} \be_{\bx_{n}}
	\\
	\times\exp\left(- tV/n \insum_{k=1}^{n} \one\{\textup{$\bx_{k}$ is erased}\}\right),
\end{multline}
where $\bx_{1},\dotsc,\bx_{n-1}\in\Z^{d}$ and $\bx_{0}  = \bx$, $\bx_{n} = \by$.
However, by the definition of the generator $\bH_{0}$ of $X$ and the definition of the conditional expectation over random walks \eqref{eq.cond.bridge},
the sum over $\bx_{0},\dotsc,\bx_{n}$ becomes
\begin{equation}
\label{eq.expsum}
\K(\bx,\by;t)
	\cdot \ex_{\bx,0}^{\by,t} \exp\left(-V \frac{t}{n} \insum_{k=1}^{n} \one\{\textup{$X(kt/n)$ is erased}\}\right).
\end{equation}
Hence, in the limit $n\to\infty$, we will have
\begin{equation}
\label{eq.expint}
\be_\bx^{\top} e^{-t\bH_{V}} \be_\by
	= \K(\bx,\by;t)
	\cdot \ex_{\bx,0}^{\by,t} e^{-V \int_{0}^{t} \one\{\textup{$X(s)$ is erased}\} \dd s}
\end{equation}
where the limit was moved under the expectation sign by applying the dominated convergence theorem and noting that the sum in \eqref{eq.expsum} converges a.s. to the integral in \eqref{eq.expint}.
\end{IEEEproof}

\begin{IEEEproof}[Proof of Lemma \ref{lem.FK}]
By the definition of $\ids_{V}$ and the weak convergence of the prelimits $\ids_{V}^{\lattice}$ to $\ids_{V}$,
we obtain
\begin{flalign}
\label{eq.exptrace}
\wilde\ids_{V}(t)
	&= \lim_{\latsize\to\infty} \int_{0}^{\infty} e^{-\lambda t} \dd \ids_{V}^{\lattice}(\lambda)
	\notag\\
	&= \lim_{\latsize\to\infty} \latsize^{-d} \ex_{\omega} \left[\tr \exp(-t \bH_{V}^{\lattice})\right]
%	\notag\\
%	&= \lim_{\latsize\to\infty} \ex_{\omega} \left\langle \b0 \midd  e^{-t \bH_{V}^{\lattice}} \midd \b0 \right\rangle
	= \ex_{\omega}\left[ \be_\b0^{\top}  e^{-t \bH_{V}} \be_\b0\right],
\end{flalign}
where $\b0$ has been chosen arbitrarily (recall that the randomness of $\bE$ is spatially homogeneous)%
\footnote{In the more precise language of \cite{Pastur_book1992_SpectraRandomOperators}, $\bE$ is \emph{metrically transitive}.}
%$\left\langle \bx\midd e^{-t \bH_{V}}\midd \by\right\rangle$ denotes the $(\bx,\by)$-th element of $e^{-\bH_{V}}$
and we have used the easily verifiable fact that the diagonal elements of $\bH_{V}$ are identically distributed.
Our assertion then follows by applying Lemma \ref{lem.propagator} with $\bx = \by = \b0$.
\end{IEEEproof}

Lemma \ref{lem.FK} shows that the contribution of (almost) every realization of the random walk $X(t)$ becomes exponentially small if the path spends a finite time on erased sites.
Hence, for a given realization of the erasure matrix $\bE$, the path integral of \eqref{eq.FK} will be dominated by paths that do not go through erasures;
more formally:
\begin{lemma}
\label{lem.distinct.sites}
With notation as in Lemma \ref{lem.propagator}, we have
\begin{equation}
\label{eq.sites.1}
\lim_{V\to\infty}
	\ex_{\omega}\left[
	\be_\bx^{\top}  e^{-t \bH_{V}} \be_\by\right]
%	\ex_{\bx,0}^{\by,t} e^{- V \int_{0}^{t} \one\{\textup{$X(s)$ is erased}\} \dd s}
	= \K(\bx,\by;t) \ex_{\bx,0}^{\by,t} \big[(1-\epr)^{\vert D(t) \vert}\big],
%	= \wilde\ids_{1}(t),
\end{equation}
where $D(t)\!\subseteq\!\Z^{d}$ is the set of sites visited by $X(t)$ up to time $t$.
%and
%\begin{equation}
%\label{eq.cond.open}
%\ex_{\bx,0}\big[\cdot\big]
%	= \ex\big[\cdot \vert X(0) = \bx\big].
%\end{equation}
\end{lemma}

\begin{IEEEproof}
By Lemma \ref{lem.propagator}, we readily get
\begin{flalign}
\label{eq.FK.exchanged}
\txs
\ex_{\omega} \left[ \be_\bx^{\top} e^{-t \bH_{V}} \be_\by \right]
	&= \K(\bx,\by;t) \ex_{\omega} \ex_{\bx,0}^{\by,t}
	e^{-V \int_{0}^{t} \one\{\textup{$X(s)$ is erased}\} \dd s}
	\notag\\
	&= \K(\bx,\by;t) \ex_{\bx,0}^{\by,t} \inprod_{r}\left[(1 - \epr) + \epr\cdot e^{-V s_{r}}\right],
\end{flalign}
where the product is taken over the sites $r\in D(t)$ visited by $X(t)$ up to time $t$, $s_{r}$ denotes the time spent by $X(t)$ at each \emph{distinct} site, and we have used Tonelli's theorem to exchange expectations in the first line.
Thus, for $V\to\infty$, we will have:
\begin{multline}
\lim_{V\to\infty} \ex_{\omega} \left[ \be_\bx^{\top} e^{-t \bH_{V}} \be_\by \right]
	\\
	= \K(\bx,\by;t) \lim_{V\to\infty} \ex_{\bx,0}^{\by,t} \inprod_{r\in D(t)} \left[(1 - \epr) + \epr\cdot\exp(-V s_{r})\right]
	\\
%	= \K(\bx,\by;t) \ex_{\bx,0}^{\by,t} \inprod_{r} (1 - \epr)
	= \K(\bx,\by;t) \ex_{\bx,0}^{\by,t} (1 - \epr)^{\vert D(t) \vert},
\end{multline}
where we used the dominated convergence theorem to take the limit $V\to\infty$ under the expectation in the second line (simply note that $(1 - \epr) + \epr\cdot\exp(-V s_{r})$ in \eqref{eq.FK.exchanged} is bounded by $1$).
\end{IEEEproof}

%Thus, if we use Tonelli's theorem to interchange the expectations in \eqref{eq.FK}, we will have
%\begin{flalign}
%\label{eq.FK.exchanged}
%\wilde\ids_{V}(t)
%	&= \ex_{0,0}^{0,t} \ex_{\omega} e^{-V \int_{0}^{t} \one\{\textup{$X(s)$ is erased}\} \dd s}
%	\notag\\
%	&\txs
%	= \ex_{0,0}^{0,t} \inprod_{r}\left[(1 - \epr) + \epr\cdot e^{-V s_{r}}\right],
%\end{flalign}
%where the product above is taken over the (random) set $D(t)$ of sites $y\in\Z^{d}$ visited by $X(t)$, and $s_{r}$ denotes the time spent by $X(t)$ at each \emph{distinct} site in $D(t)$.

Thanks to Lemmas \ref{lem.FK}, \ref{lem.propagator} and \ref{lem.distinct.sites}, the large $V$ limit of $\wilde\ids_{V}$ may be written in terms of the number of distinct sites visited by a random loop of $X(t)$ as follows:
\begin{equation}
\lim_{V\to\infty} \wilde\ids_{V}(t)
%	= \lim_{V\to\infty} \ex_{\omega} \left\langle 0 \midd e^{-t\bH_{V}} \midd 0 \right\rangle
	= \K(0,0;t) \cdot \ex_{0,0}^{0,t} \big[(1-\epr)^{\vert D(t) \vert}\big].
\end{equation}
The above shows that the limit $\wilde \ids_{\infty}(t) \equiv \lim_{V\to\infty} \wilde \ids_{V}(t)$ is well-defined;
hence, by the (vague) continuity of the Laplace transform (see e.g. Theorem 8.5 in \cite{Bhattacharya_book_ProbabilityTheory}), the convergence of $\wilde\ids_{V}$ to $\wilde\ids_{\infty}$ implies the vague convergence of the underlying integrated density $\ids_{V}$ to some limit density $\ids_{\infty}$ with Laplace transform $\wilde\ids_{\infty}$.
By Lemma \ref{lem.IDS.limit} and the uniqueness of the Laplace transform, it then follows that $\wilde\ids_{\infty}$ will simply be the Laplace transform of the \ac{IDS} of $\bH$, i.e.
\begin{equation}
\label{eq.ids.Laplace}
\wilde \ids(t)
	= \wilde \ids_{\infty}(t)
	= \K(0,0;t) \ex_{0,0}^{0,t} \big[(1-\epr)^{\vert D(t) \vert}\big].
\end{equation}
Thus, to obtain the asymptotic expression \eqref{eq:thmLifsthitztails} for $\ids(\lambda)$, we only need to derive the large $t$ behavior of $\wilde\ids(t)$ from \eqref{eq:DV-average} and then deduce the small $\lambda$ behavior of $\ids(\lambda)$ by inverting the Laplace transform.
We will achieve this by providing explicit bounds for \eqref{eq.ids.Laplace} that exhibit the same asymptotic behavior.

%In view of Lemmas \ref{lem.FK}, \ref{lem.propagator} and \ref{lem.distinct.sites}, we then get the following asymptotic expression for $\wilde\ids_{V}(t)$ in the limit $V\to\infty$:
%
%\begin{proposition}
%\label{prop.Laplace.sites}
%With notation as in Lemmas \ref{lem.FK} and \ref{lem.distinct.sites}, we have
%\begin{equation}
%\label{eq:DV-average}
%\lim_{V\to\infty} \wilde\ids_{V}(t)
%	= \ex_{0,0}^{0,t} \big[(1-\epr)^{\vert D(t) \vert}\big],
%\end{equation}
%where $\vert D(t) \vert$ is the number of distinct sites visited by the random walk $X(t)$ up to time $t$.
%\end{proposition}
%
%\begin{IEEEproof}
%The quantity $(1 - \epr) + \epr\cdot\exp(-V s_{r})$ in \eqref{eq.FK.exchanged} is bounded by $1$, so  the dominated convergence theorem allows us to exchange limits and expectations.
%We thus obtain
%\begin{flalign}
%\label{eq.distinct.sites}
%\lim_{V\to\infty} \wilde\ids_{V}(t)
%	&= \ex_{0,0}^{0,t} \inprod_{r} \lim_{V\to\infty} \left[(1 - \epr) + \epr\cdot\exp(-V s_{r})\right]
%	\notag\\
%	&= \ex_{0,0}^{0,t} \inprod_{r} (1 - \epr),
%\end{flalign}
%and our assertion follows.
%\end{IEEEproof}

\subsubsection{An upper bound for $\wilde\ids(t)$}
\label{sec:sub4}

First, let
\begin{equation}
\label{eq:DV-average}
\wilde\ids_{\ast}(t)
	= \ex_{0,0} \big[(1-\epr)^{\vert D(t) \vert}\big],
%	= \wilde\ids_{1}(t),
\end{equation}
where $\ex_{\bx,0}$ is the open-ended conditional expectation $\ex_{0,0}[\cdot] = \ex\big[\cdot \vert X(0) = 0\big]$.
The law of total expectation then yields:
\begin{multline}
\label{eq:IDS-lower-bound}
\wilde\ids_{\ast}(t)
	= \insum_{\by} \ex_{0,0}^{\by,t} \big[(1-\epr)^{\vert D(t) \vert}\big]
	\K(0,\by; t)
	\\
	\geq \K(0,0;t) \ex_{0,0}^{0,t} \big[(1-\epr)^{\vert D(t) \vert}\big]
	= \wilde\ids(t),
\end{multline}
so we are left to calculate the asymptotic behavior of \eqref{eq:DV-average}.

This can be done as follows:
first, let $S(t)$ denote the (a.s. finite) number of hops performed by $X$ over the interval $[0,t]$.
Then, conditioning the hop count $S(t)$ to some large $n\in\N$, we will use the calculations of \cite{Donsker1979_NumberDistinctSitesRW} for the number of \emph{distinct} sites $D(t)$ visited by a random walk to calculate the expectation of \eqref{eq:DV-average} conditioned on $S(t)$;
finally, to obtain \eqref{eq:DV-average}, we will average the result of this calculation over the hop count $S(t)$.

Since we have already averaged over the realizations of the erasure matrix $\bE$, the random walk $X(t)$ generated by $\bH_{0}$ will be spatially homogeneous.
As a result, the probability of hopping from $\bx$ to $\by$ given that $X$ does not remain at $\bx$ will be:
\begin{equation}
\label{eq:path_integral5_TransitionProb}
\Pi(\bx,\by)
	=\frac{\be_\bx^{\top} \bH_0 \be_\by}{\sum_{\bz\neq\b0} \be_\bx^{\top}  \bH_0 \be_\bz}
	\eqqcolon \gamma_c g(\bx,\by)
\end{equation}
Since $\Pi(\bx,\by)>0$ for every $\bx,\by\in \Z^d$, the random walk will be irreducible;
moreover, by the symmetry properties of $\bH_{0}$, it follows that $\Pi$ will actually be a function of the difference $\by - \bx$, i.e. $\Pi(\bx,\by) \equiv \Pi(\by-\bx)$.
Thus, to check that the criteria of \cite{Donsker1979_NumberDistinctSitesRW} apply to the random walk generated by $\bH_{0}$, we only need to calculate the characteristic function $\hat{\Pi}(\bq)$ of $\Pi$ for small values of $|\bq|$.
To that end, using \eqref{eq:low_q_eig_a>d+2} and \eqref{eq:low_q_eig_a<d+2}, we obtain
\begin{equation}
\label{eq:path_integral4_FTMarkovProb}
\hat{\Pi}(\bq)
	= 1 -
	\begin{cases}
	\gamma_{c} t_{2} \vert\bq\vert^{2} + \bigoh(|\bq|^4)
	&\text{if $\alpha>d+2$},\\
	\gamma_{c} t_{\alpha-d} \vert\bq\vert^{\alpha-d} + \bigoh(|\bq|^{2})
	&\text{if $d<\alpha<d+2$},
	\end{cases}
\end{equation}
or, more concisely, using the definitions in \eqref{eq:t-eff}, \eqref{eq:alpha-eff}, \eqref{eq:eps-eff}, $\hat{\Pi}(\bq)\approx 1-\gamma_c t_{\eff} \vert\bq\vert^{\alpha_\eff-d}$. Also, $\hat{\Pi}(\bq)=1$ if and only if $\bq=2\pi(k_{1},\ldots,k_{d})$, with integer $k_i$.
However, \eqref{eq:path_integral4_FTMarkovProb} shows that $\Pi$ lies in the domain of attraction of a Brownian process (for $\alpha \geq d+2$) and of a non-degenerate symmetric stable law of order $\alpha-d$ for $\alpha\in(d,d+2)$ \cite{Bouchaud1990_AnomalousDiffusion, Donsker1979_NumberDistinctSitesRW}.
Hence, by the analysis of \cite{Donsker1979_NumberDistinctSitesRW}, the expected value of \eqref{eq:DV-average} conditioned on the number of hops $S(t)$ will be asymptotically equal to:
\begin{equation}
\label{eq:DonskerVaradhanFormula1}
\ex_{0,0} \big[(1-\epr)^{\vert D(t) \vert} \:\big\vert\: S(t) = n\big]
	\sim \exp\left[-k(\alpha,d)\,\left(\gamma_c n\right)^{\frac{d}{\alpha_\eff}}\right].
\end{equation}
In the above equation, the coefficient $k(\alpha,d)$ is given by
\begin{equation}
\label{eq:app:k(a,d)_def}
k(\alpha,d)
	= \big[\log(1-\epr)\big]^{1-\frac{d}{\alpha_\eff}}
	\frac{\alpha_\eff}{\alpha_\eff-d}
	\left(\frac{(\alpha_\eff-d)t_{\eff}\eigmin}{d}\right)^{\frac{d}{\alpha_\eff}},
\end{equation}
where $\eigmin=\eigmin(\alpha,d)$ is the minimum eigenvalue of a $d$-dimensional ball of unit volume with Dirichlet boundary conditions, of the linear operator $\gen$ defined in \eqref{eq:generator} \cite{Donsker1979_NumberDistinctSitesRW}. As can be seen in \eqref{eq:generator} this operator is just the  Laplacian $\gen = -\nabla^2$ for $\alpha>d+2$, while in the case $d<\alpha<d+2$, it is the infinitesimal generator of a symmetric $(\alpha-d)$-stable process given in \eqref{eq:generator_sym_stable_proc}.%
\footnote{The fact that a ball minimizes the minimum eigenvalue is proven in \cite{Banuelos2010_SymmetrizationLevyProcesses}.}

In view of the above, to obtain the large $t$ behavior of $\wilde\ids_{\ast}(t)$, we only need to average the conditional expectation \eqref{eq:DonskerVaradhanFormula1} over the number of hops $S(t)$ that took place in $[0,t]$.
To that end, since $S(t)$ is Poisson distributed with parameter $t/\gamma_{c}$, we get
\begin{multline}
\wilde\ids_{\ast}(t)
	= \sum_{n=1}^\infty e^{-t/\gamma_c}\frac{t^n}{\gamma_c^n n!}
	\ex_{0,0} \left[(1-\epr)^{\vert D(t) \vert} \:\vert\: S(t) = n\right]
\end{multline}
For large $t$, the sum is dominated by large $n$, hence the asymptotic approximation of \eqref{eq:DonskerVaradhanFormula1} holds.
Approximating the sum by its maximum term for which $\gamma_c n^* = t$ we thus obtain
\begin{equation}
\label{eq:DonskerVaradhanFormula2}
%\wilde\ids(t)
\log\ex_{0,0} \left[(1-\epr)^{\vert D(t)\vert}\right]
	\sim -k(\alpha,d)\,t^{\frac{d}{\alpha_\eff}},
\end{equation}
so \eqref{eq:IDS-lower-bound} becomes:
\begin{equation}
\label{eq:DonskerVaradhanFormula3}
\log\wilde\ids(t)
	\leq \log\wilde\ids_{\ast}(t)
%\ex_{\bx,0}^{\by,0} (1-\epr)^{\vert D(t)\vert}
	\sim -k(\alpha,d)\,t^{\frac{d}{\alpha_\eff}}.
\end{equation}
%which provides an upper bound for $\log\wilde\ids(t)$.

\subsubsection{A Lower Bound for $\wilde\ids(t)$}
\label{sec:sub5}

To obtain a matching lower bound for \eqref{eq:DonskerVaradhanFormula3}, we will employ a slightly more elaborate variant of the methodology described in the main text.
Specifically, our approach will be based on Theorem 9.5 from \cite{Pastur_book1992_SpectraRandomOperators} which, in our notation, states that
\begin{equation}
\label{eq:thm_Pastur_lowerbound}
\ex_{\omega}\left[\be_\b0^{\top} e^{-t\bH_V} \be_\b0\right]
	\geq \frac{1}{\left|\lattice'\right|} \ex_{\omega}
	\left[e^{-t\bPsi^{\top}\bH_{V}\bPsi}\right]
\end{equation}
for every normalized vector $\bPsi\in\ell^{2}(\Z^{d})$ with finite support $\supp(\bPsi) = \lattice'\subseteq\Z^d$.
%$\left| \Psi\right\rangle$ is a real vector in $\Z^d$ with unit norm $\left\langle \Psi\left| \right.\Psi\right\rangle=1$, which has support $\wilde\Lambda$ with corresponding $\left| \wilde\Lambda\right|$.
%We base the analysis on a simple yet far-reaching theorem from \cite{Pastur_book1992_SpectraRandomOperators}, which in our notation can be written as follows:
%\begin{theorem*}[Theorem (9.5) in \cite{Pastur_book1992_SpectraRandomOperators}]
%\begin{equation}\label{eq:thm_Pastur_lowerbound}
%\ex\left[\left\langle 0\midd e^{-t\bH_V} \midd 0\right\rangle\right] \geq \frac{1}{\left| \wilde\Lambda\right|} \ex\left[e^{-t\left\langle \Psi\midd\bH_V\midd\Psi\right\rangle}\right]
%\end{equation}
%where $\left| \Psi\right\rangle$ is a real vector in $\Z^d$ with unit norm $\left\langle \Psi\left| \right.\Psi\right\rangle=1$, which has support $\wilde\Lambda$ with corresponding $\left| \wilde\Lambda\right|$.
%\end{theorem*}
We will thus have:
\begin{flalign}
\label{eq:LowerBound}
\wilde\ids(t)
	&\geq \frac{1}{\vert\lattice'\vert}
	e^{-t\bPsi^{\top} \bH_0\bPsi}
	\lim_{V\to\infty} \ex_{\omega}
	\left[e^{-t\bPsi^{\top}\bV\bPsi}\right]
	\notag\\
	&= \frac{1}{\vert\lattice'\vert}
	e^{-t\bPsi^{\top}\bH_0\bPsi}
	\lim_{V\to\infty} \prod_{\bx\in \lattice'}
	\left(1-\epr+\epr \cdot e^{-t V \vert\bPsi(x)\vert^{2}}\right),
\end{flalign}
where we have used the dominated convergence theorem to take the limit under the integral sign in the second line.

To make this last inequality as tight as possible, let $\bPsi$ be the eigenvector of the minimum eigenvalue of $\bH_{0}$ over $\lattice'$ with corresponding eigenvalue $\lambda(\lattice')$.
We will then have
\begin{equation}
\label{eq:LowerBound2}
\log\wilde\ids(t)
	\geq \sup\nolimits_{\lattice'} \left\{
	-t\lambda(\lattice')
	+ \vert\lattice'\vert \log(1-\epr)
	-\log\vert\lattice'\vert
	\right\},
\end{equation}
where the supremum is taken over all finite connected domains of $\Z^{d}$.
For large $t$, and in anticipation of the volume of $\lattice'$ being large, rescale all distances in the set $\lattice'$ by the length scale $R = \vert\lattice'\vert^{1/d}$ so that the rescaled set $\lattice_{R}' \equiv \{\by\in \R^d: \by=\bx/R \text{ for some } \bx\in \lattice'\}$ has unit volume.
As expected from dimensional analysis (and shown rigorously in \cite{Widom1963_ExtremeEIgenvaluesConvolutionOperators}), we will have $\lambda(\lattice') \sim R^{d - \alpha_{\eff}} \lambda(\lattice_{R}')$ where $\lambda(\lattice_{R}')$ is the minimum Dirichlet eigenvalue of $\gen$ over $\lattice'$.
%the operator $-\nabla^2$ if $\overline{\alpha}=2$ or, if $\overline{\alpha}=\alpha-d<2$, being the minimum eigenvalue of the infinitesimal generator of a symmetric ($\alpha-d$)-stable
%process, given in \eqref{eq:app:H0=Laplacian4}. Both operators act in $\wilde\Lambda_R$ with Dirichlet boundary conditions.
We thus get
\begin{equation}
\label{eq:LowerBound3}
\log\wilde\ids(t)
	\geq \sup
%	\nolimits_{\lattice_{R}',R}
	\left\{
	-t R^{d - \alpha_{\eff}} \lambda(\lattice_{R}')
	+R^{d} \log(1-\epr)
	-d\log R
	\right\},
\end{equation}
where the supremum is taken over all $\lattice_{R}'$ for fixed $R$ and over all $R$.
Therefore, by minimizing $\lambda(\lattice_{R}')$ over $\lattice_{R}'$ and then maximizing the RHS of \eqref{eq:LowerBound2} over $R$, we finally obtain the lower bound
\begin{equation}
\label{eq:LowerBound4}
\log\wilde\ids(t)
	\geq -k(\alpha,d)\,t\,^{d/\alpha_{\eff}}(1+o(1)),
\end{equation}
which is an asymptotic match for the upper bound \eqref{eq:DonskerVaradhanFormula3}.

\subsubsection{Harvesting $\ids(\lambda)$ from $\wilde\ids(t)$}
\label{sec:sub6}

From the matching exponential bounds \eqref{eq:DonskerVaradhanFormula3} and \eqref{eq:LowerBound4} above, we conclude that:
\begin{equation}
\label{eq:asymptN(t)_final}
\log \wilde \ids(t)
	\sim \log(1-\epr) k(\alpha,d)\,t\,^{d/\alpha_{\eff}}
	\quad
	\text{for large $t$,}
\end{equation}
with $k(\alpha,d)$ given from \eqref{eq:app:k(a,d)_def}.
%and $\overline{\alpha}=\min(2,\alpha-d)$.
We are thus left to invert the large $t$ behavior of the Laplace transform $\wilde\ids(t)$ to obtain the small $\lambda$ behavior of $\ids(\lambda)$;
to that end, Theorem 9.7 in \cite{Pastur_book1992_SpectraRandomOperators} readily yields
%which states that for the type of long time behavior of $\wilde\ids(t)$ given in the above equation, the following holds
\begin{flalign}
\label{eq:invLaplace_xform}
\log\ids(\lambda)
	&\sim \inf_t\left\{\lambda t + \log(1-\epr) k(\alpha,d)\,t\,^{d/\alpha_{\eff}} \right\}
	\notag\\
	&\sim \log(1-\epr)\left(\frac{\eigmin(\alpha,d)t_{\eff}}{\lambda}\right)^{\frac{d}{\alpha_{\eff} - d}}
	\quad
	\text{for small $\lambda$.}
\end{flalign}
%as $\lambda\to 0^+$.

%% file: App-ContApprox.tex
%----------------------------------------------------------------------
%%% APPENDIX: CONTINUUM APPROXIMATION
%----------------------------------------------------------------------
% !TEX root = ./RandomPowerOptimization.tex

In this appendix we will discuss briefly how the discrete operator $\bH_0$ can be approximated by a continuous one.
For simplicity, we will keep our discussion at an intuitive level;
for a rigorous treatment, the reader is instead referred to \cite{Widom1963_ExtremeEIgenvaluesConvolutionOperators}.

We will begin with the short-range interaction case $\alpha \geq d+2$ and assume that $\bH_{0}$ is defined over a large erasure-free region $\dom$ of  $\vert\dom\vert$ sites.
To that end, let $\scale$ be an arbitrary length scale which is much larger than the inter-site distance and much smaller than the effective radius of $\dom$, i.e. $1 \ll \scale \ll \vert\dom\vert^{1/d}$.
%Let us assume that the boundaries of the region are smooth on the scale of $L$, which is an arbitrary length $L$ much larger than the site distance.
Assume further that the boundary of $\dom$ is smooth when measured with balls of radius $\scale$,%
\footnote{By this, we mean that if we rescale everything by $|\dom|^{1/d}$, the boundary $\pd\dom$ of $\dom$ will have Hausdorff dimension $d-1$, and a cover of $\pd\dom$ by balls of radius $\scale/|\dom|^{1/d}$ will suffice to estimate its $(d-1)$-dimensional Hausdorff content as $|\dom|\to\infty$.
Alternatively, this means that, for large $|\dom|$, the boundary $\pd \dom$ of $\dom$ scaled down by $|\dom|^{1/d}$ is $m$-rectifiable by balls of size $\scale/|\dom|^{1/d}$ for all $m>d-1$).}
and that the optimal power vector $\bp^{\ast}$ vanishes at the boundary $\pd\dom$ of $\dom$.

Define now a function $\phi(\bx)$, $\bx\in\R^{d}$, such that $\phi(\bx)$ is equal to the value of $\bp$ at the site $\scale \bx$ whenever $\scale \bx\in \Z^{d}$ and $\phi$ interpolates smoothly between these values otherwise.
Then, letting $\be_{\bm}$ denote the basis vector corresponding to the site $\bm\in\dom$, we will have:
\begin{equation}
\label{eq:app:H0=Laplacian1}
\be_{\bm}^{\top} \bH_{0} \bp
	= \sum_{\bm'\in \dom} \big(\be_{\bm}^{\top} \bH_{0} \be_{\bm'}\big)\,\big(\be_{\bm'}^{\top} \bp\big)
	= \sum_{\bm'\in\dom} \big(\be_{\bm}^{\top} \bH_{0} \be_{\bm'}\big)\,\phi(\bx + \delta \bx)
\end{equation}
where $\bx = \bm/\scale$ and $\delta \bx = (\bm' - \bm)/\scale$.
%$\Delta \bm=\bm'-\bm$.
For large $\scale$, we may then expand $\phi$ to obtain:
\begin{equation}
\phi(\bx + \delta \bx)
	= \phi(\bx)
	+ \left(\delta \bx \!\cdot\! \nabla\right) \phi(\bx)
	+ \frac{1}{2}\left(\delta \bx \!\cdot\! \nabla \right)^{2} \phi(\bx)
	+ \bigoh(\scale^{-3}),
\end{equation}
%\begin{eqnarray}
%\label{eq:app:H0=Laplacian2}
%\hat{p}\left(\frac{\bm+\Delta\bm}{L}\right) &\approx& \hat{p}\left(\frac{\bm}{L}\right) + \frac{1}{L}\Delta\bm\cdot\nabla\hat{p}\left(\frac{\bm}{L}\right) \\ \nonumber
%&+& \frac{1}{2L^2}\left(\Delta\bm\cdot\nabla\right)^2\hat{p}\left(\frac{\bm}{L}\right) + o(L^{-2})
%\end{eqnarray}
We next plug this expression into \eqref{eq:app:H0=Laplacian1}. The first constant term ($\propto\phi(\bx)$) vanishes because each row (or column) of $\bH_0$ add to zero, while the second also vanishes, because $\sum_\bm (\bm-\bm')\be_{\bm}^{T} \bH_{0} \be_{\bm'}=0$. As a result, we obtain
\begin{equation}
\label{eq:app:H0=Laplacian3}
\be_{\bm}^{\top} \bH_{0} \bp
	= -\frac{t_{2}}{\scale^{2}} \nabla^{2} \phi(\bx) + \bigoh(\scale^{-3})
\end{equation}
with $t_2$ given by \eqref{eq:low_q_t2}.
%To obtain the above expression, we

On the other hand, for $\alpha<d+2$, $t_{2}$ is infinite due to the slow decay of the elements of the transition matrix $\bH_{0}$, so $\bH_{0}$ cannot be approximated by a differential operator in the sense of \eqref{eq:app:H0=Laplacian3}.
In this case, by following the reasoning of \cite{Donsker1979_NumberDistinctSitesRW} and \cite{Widom1963_ExtremeEIgenvaluesConvolutionOperators}, the leading order approximation to $\bH_{0}$ will be:
\begin{equation}
\label{eq:app:H0=Laplacian4}
\be_{\bm}^{\top} \bH_{0}\bp
	\approx \frac{t_{\alpha-d}}{\scale^{\alpha-d}}
	\int_{\lattice} \frac{2 \phi(\bx) - \phi(\bx+\bh) - \phi(\bx-\bh)}{\vert \bh \vert^{\alpha}}
	\dd \bh,
%	\left(\frac{\hat{p}\left(\frac{\bm}{L}+\by\right)+\hat{p}\left(\frac{\bm}{L}-\by\right)}{2}-\hat{p}\left(\frac{\bm}{L}\right)\right)
%	\frac{d^d\by}{|\by|^\alpha}
\end{equation}
with $\bx = \bm/\scale$ as before and $t_{\alpha-d}$ given by \eqref{eq:low_q_eig_a<d+2}, which is the generator of a $d$-dimensional symmetric stable process of degree $\alpha-d$ and can also be written as $\frac{t_{\alpha-d}}{\scale^{\alpha-d}}\big(- \nabla^{2} \big)^{(\alpha-d)/2}$ \cite{Banuelos2010_SymmetrizationLevyProcesses} (see discussion after \eqref{eq:generator}).

%% file: App-Tails.tex
%----------------------------------------------------------------------
%%% APPENDIX: TAILS
%----------------------------------------------------------------------
% !TEX root = ./RandomPowerOptimization.tex

In this Appendix our goal is to provide details on the lower and upper bounds on the exponential tails of the empirical power distribution of the optimal power vector discussed in Section \ref{sec:PowerDistribution}.

\subsection{A lower bound for the distribution of power in the network}
\label{app:Tails_lower_bound}

We start with \eqref{eq:lower-bound} and look for the minimum volume $V_p$ that can support the power $p$ at the origin. Since we are interested in large powers for $\alpha > d+2$, we will focus on large domains $V_{p}$; in addition, we anticipate domains whose boundaries are ``smooth'' at a length scale $\scale \gg 1$ which is sufficiently small compared to the effective size $E_{p} = V_{p}^{1/d}$ of the domain, and which will become irrelevant in the end (cf. Appendix \ref{app:cont_approx}).
%(in the sense that if we rescale everything by $E_{p}$, the boundary $\pd\dom_{p}$ of $\dom_{p}$ will have Hausdorff dimension $d-1$, and a cover of $\pd\dom_{p}$ by balls of radius $\scale/E_{p}$ will suffice to estimate its $(d-1)$-dimensional Hausdorff content in the limit $E_{p}\to\infty$).%
%\footnote{Alternatively, this means that, for large $E_{p}$, the boundary $\pd\dom_{p}$ of $\dom_{p}$ scaled down by $E_{p}$ is $m$-rectifiable by balls of size $\scale/E_{p}$ for all $m>d-1$).}
%As we shall see, the exact value of the length scale $\scale$ will be irrelevant in the end;
%(effectively, it defines the length scale in the problem);
In this way, the discrete equation $\bM\bp^{\ast} = (\bH_{0} + z_{\gamma}\bI) \bp^{\ast} =\sigma^{2}\bu$ which defines the optimal power vector $\bp^{\ast}$ may be approximated in the small $\scale/E_{p}$ limit by the \emph{stationary Klein\textendash Gordon equation} (sometimes referred to as the \emph{screened Poisson equation}):
\begin{equation}
\label{eq:KleinGordon}
\tag{KG}
-\nabla^2 \phi + \kappa_{\scale}^{2} \phi = 1,
\end{equation}
where
\begin{equation}
\label{eq:rescaled_phi}
\phi(\bx) = \frac{t_{2}}{\scale^{2} \sigma^2} \bp^{\ast}(\scale \bx)
	\quad
	\text{for $\scale \bx \in \Z^{d}$},
\end{equation}
and $\kappa_{\scale}^{2} = \kappa^{2} \scale^{2}$ with $\kappa^{2} = z_{\gamma}/t_{2}$.
In view of this, it suffices to determine the minimal domain $\dom\in\R^{d}$ for which the solution of the Dirichlet problem \eqref{eq:KleinGordon} with boundary conditions $\phi\equiv0$ on $\pd\dom$ has $\max_{x\in\dom} \phi(x) = t_2p/(\scale^{2}\sigma^{2})$ \textendash\ that is, $\bp^{\ast}_{\max} = p$.

This last problem may be reformulated as follows:
let $\phi$ be the solution to \eqref{eq:KleinGordon} with boundary conditions $\phi\vert_{\pd\dom} = 0$ for some (smooth) domain $\dom\subseteq\R^{d}$ containing $0$;
we then seek the domain for which $\phi(0)$ is maximal over all domains with unit volume $\vert\dom\vert=1$ (and, obviously, containing $0$).
On that account, let $G_{\dom}(\bx,\by)$ be the problem's Green's function, which is the solution to the unit impulse Dirichlet problem:
\begin{equation}
\label{eq:KleinGordon-Green}
\begin{aligned}
(\nabla^2 - \kappa_{\scale}^{2})\,G_{\dom}(\bx,\by)
	&= -\delta(\bx-\by)
	&\quad&\text{for all $\bx\in\dom$,}
	\notag\\
G_{\dom}(\bx,\by)
	&= 0
	&\quad&\text{for all $\bx\in\pd\dom$,}
\end{aligned}
\end{equation}
so that $\phi(\bx) = \int_{\dom} G_{\dom}(\bx,\by) \dd \by$.
Since the operator $\nabla^2 - \kappa_{\scale}^{2}$ is (uniformly) elliptic, an extension of Bandle's isoperimetric inequality for integrals of Green's functions readily gives \cite{Banuelos2010_SymmetrizationLevyProcesses}
\begin{equation}
\label{eq:isoperimetric}
\phi(0)
	\leq \sup_{\bx\in\dom} \int_{\dom} G_{\dom}(\bx,\by) \dd \by
	\leq \int_{\ball} G_{\ball}(0,\by) \dd \by,
\end{equation}
where $\ball$ is a ball of unit $d$-dimensional volume centered at $0$.
%
%The above shows that the maximal value for $\phi(0)$ is attained when $\dom=\ball$.
Hence, going back to the original problem of determining the minimal volume of a domain containing $0$ and giving rise to a solution $\phi$ of \eqref{eq:KleinGordon} with maximum value $t_{2} p/(\scale^{2} \sigma^2)$ at $0$,
we are left to determine the radius
$R_{\scale}(p)$
%$R_{\scale} = R_{p}/\scale$
of a ball centered at $0$ such that $\phi(0) = t_2p/(\scale^{2}\sigma^2)$ and $\phi\equiv0$ on its boundary.

To that end, spherical symmetry allows us to write \eqref{eq:KleinGordon} in the more convenient form
\begin{equation}
\label{eq:KleinGordon-radial}
\begin{aligned}
&\frac{1}{r^{d-1}} \frac{\pd}{\pd r} \left(r^{d-1} \frac{\pd\phi}{\pd r} \right) - \kappa_{\scale}^{2} \phi
	= -1,
	\\[2pt]
&\quad\phi(R_{\scale})=0.
\end{aligned}
\end{equation}
%which, after integrating, gives the expression
%\begin{equation}
%\phi(r) = \phi(0) - \frac{r^{2}}{2d}.
%\end{equation}
%The boundary condition $\phi(R_{d}^{2}) = 0$ then gives $R_{d} = \sqrt{2pd}$, so the required minimal volume $V_{p}$ will be:
%\begin{equation}
%\label{eq:minvol}
%V_{p}
%%	= (R_{d}/r_{d})^{d}
%	= K_{d} R_{d}^{d}
%	= C_{d} p^{d/2},
%\end{equation}
%where $K_{d} = \pi^{d/2}/\Gamma\left(\frac{d+2}{2}\right)$ represents the volume of the unit $d$-dimensional ball and
%$C_{d} = (2 d)^{d/2} K_{d}$.
Hence, focusing on the cases of interest $d=1$ and $d=2$, some calculus yields the radial solutions:
\begin{subequations}
\label{eq:KGSol}
\begin{align}
\label{eq:KGSol-1}
\phi(r)
	&= \kappa_{\scale}^{-2}
	\left[1 - \frac{\cosh(\kappa_{\scale} r)}{\cosh(\kappa_{\scale} R_{\scale})}\right]
	&\text{for $d=1$,}
	\\
\label{eq:KGSol-2}
\phi(r)
	&=\kappa_{\scale}^{-2}
	\left[1 - \frac{I_{0}(\kappa_{\scale} r)}{I_{0}(\kappa_{\scale} R_{\scale})}\right]
	&\text{for $d=2$,}
\end{align}
\end{subequations}
where
$I_{0}(x)$ is the $0$-th order hyperbolic Bessel function of the first kind.
Accordingly, with $\phi(0) = t_{2} p/(\sigma^{2} \scale^{2})$, we obtain the expressions
\begin{subequations}
\label{eq:Rp_app}
\begin{align}
\label{eq:Rp-1_app}
R_{p}
	&= \sqrt{\frac{t_2}{z_\gamma}} \arcosh\left(\frac{\sigma^2}{\sigma^2-z_\gamma p}\right)
	&\text{for $d=1$,}
	\\
\label{eq:Rp-2_app}
R_{p}
	&= \sqrt{\frac{t_2}{z_\gamma}} I_{0}^{-1} \left(\frac{\sigma^2}{\sigma^2-z_\gamma p}\right)
	&\text{for $d=2$,}
\end{align}
\end{subequations}
where $I_0^{-1}(y)$ is the function inverse of $I_0(x)$ and $R_{p} = \scale R_{\scale}(p)$ is the length in the original (unscaled) lattice \textendash\ so the final result does not depend on the length scale $\scale$, as expected.
We thus get:
\begin{subequations}
\label{eq:Vp}
\begin{align}
\label{eq:Vp-1}
V_{p}
	&= 2R_{p}
	\sim 2 \big(2 t_{2} p\big/\sigma^{2}\big)^{1/2}
	&\text{for $d=1$,}
	\\
\label{eq:Vp-2}
V_{p}
	&= \pi R_{p}^{2}
	\sim 4\pi t_{2} p/\sigma^2
	&\text{for $d=2$,}
\end{align}
\end{subequations}
where the asymptotic approximations are taken in the limit $z_\gamma p\to 0$ (and are exact for all $p$ in the critical case $z_\gamma=0$ which corresponds to $\gamma = \gamma_{c}$).
Proposition \ref{prop:lower-bound} then follows trivially.

\begin{remark*}
Extending the above analysis to $\alpha<d+2$, we see that we have to replace the stationary Klein-Gordon equation \eqref{eq:KleinGordon} with
\begin{equation}\label{eq:genKleinGordon_a<d+2}
\gen \phi+\kappa_{\scale}^{\alpha-d}\phi = 1
\end{equation}
where $\gen$ is given by \eqref{eq:generator_sym_stable_proc}, $\kappa_{\scale}^{\alpha-d}=z_\gamma a^{\alpha-d}/t_{\alpha-d}$  and $\phi(\bx)=t_{\alpha-d}\bp^\ast(ax)/(\sigma^2\scale^{\alpha-d})$. For fixed volume, the solution of the above differential equation with Dirichlet boundary conditions is maximized at $x=0$, when the domain is a ball  of radius $R_{\scale}$ centered at $x=0$ as in the case $\alpha>d+2$ discussed above \cite{Banuelos2010_SymmetrizationLevyProcesses}.
In this case, we obtain the radial solution:
\begin{equation}
\label{eq:KGSol_a<d+2}
\phi(r)
	= \kappa_{\scale}^{d-\alpha}
	\left[1 - \frac{{\overline \phi}(\kappa_{\scale} r)}{{\overline \phi}(\kappa_{\scale} R_{\scale})}\right]
\end{equation}
where ${\overline \phi}(\bx)$ is the non-increasing, radially symmetric solution of $\gen {\overline \phi}+{\overline \phi}=0$, with ${\overline \phi}(0)=t_{\alpha-d}p/(\scale^2\sigma^2)$.

In the critical case $\gamma=\gamma_c$, $\phi(r)$ satisfies $\gen \phi(\bx)=1$ with Dirichlet boundary conditions on the boundary of the ball of radius $R_{\scale}$.
By guessing that the solution of $\phi(\bx)$ is of the form $R_{\scale}^\beta {\overline \phi}(\bx/R_{\scale})$ we then find that $\beta=\alpha-d$.
As a result, we will have $V_p=\Omega_d R_p^d\propto p^{d/(\alpha+d)}$  and hence, for  $\gamma=\gamma_c$ and  $\alpha<d+2$, we conjecture that
\begin{equation}
\label{eq:lower-bound-final_alpha<d+2}
\pdist(p)
	\sim \exp(-c_{\alpha,d} p^{d/(\alpha+d)}),
\end{equation}
for some constant $c_{\alpha,d}$.
\end{remark*}

\subsection{A percolation-based upper bound}
\label{app:TailsUpperBound}

In this section, our goal will be to prove the upper bound \eqref{eq:upper-bound} for random networks where interference is only caused by nearest neighbors. Of course, in the one-dimensional case, this coincides with the Wyner model for which the lower bound obtained in the previous section is tight.
In the $2$-dimensional case, we consider four nearest neighbors per site, one for each of the unit steps in the $x$ and $y$ axes.
In this case, the cumulative power distribution $\pdist(p)$ may be written as
\begin{equation}
\label{eq:pdist-regions}
\pdist(p)
	= \insum_{\dom} \prob(\bp^{\ast}(0) > p \,\vert\, 0\in\dom)
	\cdot \prob(0\in\dom),
\end{equation}
where the event ``$0\in\dom$'' signifies that $0$ belongs to an erasure-free cluster $\dom\subseteq\Z^{2}$ whose boundary $\pd\dom$ is completely erased (i.e. $\bE=1$ on $\pd\dom$).
%and the indicator function $\one\{\bp^{\ast}(0) > p\}$ indicates whether the optimal power vector $\bp^{\ast}(0)$ exceeds $p$ at $0$ or not.
Thus, letting $V_{p}$ be the minimal volume which supports power $p$ at $0$, we obtain the upper bound:
\begin{equation}
\label{eq:upper-bound_app}
\pdist(p)
	\leq \insum_{\dom: \vert\dom\vert\geq V_{p}} \prob(0\in\dom)
	= F(V_{p}),
\end{equation}
where $F(V_{p})$ is the probability of $0$ belonging to an erasure-free cluster of size \emph{at least} $V_{p}$. The value of $V_p$ can be obtained from the discussion in the lower bound and is given by $V_p=\pi R_p^2$, where $R_p$ is the radius appearing in \eqref{eq:Rp_app}.

Since sites in $\Z^{2}$ are erased uniformly with probability $\epr$, $F(V_{p})$ may be viewed as the probability of $0$ belonging to a cluster of size at least $V_{p}$ in a site percolation model over $\Z^{d}$ with occupancy probability $1-\epr$ \cite{Grimmett1999_Percolation_book}.
As a result, the cumulative probability $F(V_{p})$ will be bounded from above by
\begin{equation}
\label{eq:site-bond}
F(V_{p}) \leq F_{\textrm{bond}}(V_{p}),
\end{equation}
where $F_{\textrm{bond}}$ now denotes the probability of $0$ belonging to an open cluster of size at least $V_{p}$ in an associated \emph{bond percolation} model with bonding probability $1-\epr$ \textendash\ see e.g. \cite[Sec.~1.6]{Grimmett1999_Percolation_book}.

Below the percolation threshold for $\Z^{2}$,%
\footnote{That is, for $1-\epr < p_{c}(d)$, where $p_{c}(d)$ is the supremum value of the bonding probability beyond which all open clusters are finite almost surely.}
it is well known that the probability of observing a cluster of size \emph{exactly} $V$ decays asymptotically as $P_{\textrm{bond}}(V) \sim e^{-\eta V}$ where $\eta \equiv \eta(\epr,d) > 0$ is a constant which depends only on the erasure probability $\epr$ and the dimensionality $d$ of the network (the bound $\eta \leq -\log(1-\epr)$ follows from the fact that $(1-\epr)^{V_p} \leq \pdist(p)$ and the above inequalities).
Combining all of the above, we thus obtain
\begin{equation}
\label{eq:site-bound}
\pdist(p)
	\leq F_{\textrm{bond}}(V_{p})
	\sim \frac{e^{-\eta V_{p}}}{1 - e^{-\eta}}
	= e^{-\eta V_{p} + \beta},
\end{equation}
whenever $1-\epr < p_{c}(d)$;
in particular, thanks to Kester's celebrated result that $p_{c}(2) = 1/2$ \cite{Grimmett1999_Percolation_book}, the asymptotic bound \eqref{eq:site-bound} will hold in $\Z^{2}$ for all $\epr \geq 1/2$.%
\footnote{The equality here follows from the fact that there is no percolation in the critical phase for $d=2$ \cite{Grimmett1999_Percolation_book}.}
This proves Proposition \ref{prop:upper-bound} and concludes our discussion.